\documentclass[12pt]{article}

\usepackage{amsmath,amssymb,euscript,array,cite}

\usepackage{graphicx}
\makeatletter
 
  \@addtoreset{equation}{section}
 \makeatother

\def\delbarslash{\,\,{\raise.15ex\hbox{/}\mkern-9mu {\bar\partial}}}

\newcommand{\be}{\begin{equation}}
\newcommand{\ee}{\end{equation}}
\newcommand{\bea}{\begin{eqnarray}}
\newcommand{\eea}{\end{eqnarray}}

\begin{document}
%zzz
\newcommand{\nd}[1]{/\hspace{-0.5em} #1}
\begin{titlepage}
\begin{flushright}
{\bf July 2005} \\ 
%DAMTP-\\
%SWAT-\\ 
hep-th/0507075 \\
\end{flushright}
\begin{centering}
\vspace{.2in}
 {\large {\bf Non-Critical Superstrings from Four-Dimensional Gauge Theory} 
}\\

\vspace{.3in}

Gaetano Bertoldi\\
\vspace{.1 in}
Department of Physics, University of Wales Swansea \\
Singleton Park, Swansea SA2 8PP, UK\\
\vspace{.1 in}
and \\
\vspace{.1 in}
Nick Dorey\\
\vspace{.1 in} 
DAMTP, Centre for Mathematical Sciences \\ 
University of Cambridge, Wilberforce Road \\ 
Cambridge CB3 0WA, UK \\
\vspace{.2in}
%and \\ 
%\vspace{.2in}
%
%
\vspace{.4in}
{\bf Abstract} \\

\end{centering}
We study large-$N$ double-scaling limits of $U(N)$ 
gauge theories in four dimensions. We focus on theories in a partially 
confining phase where an abelian subgroup $\hat{G}$ of the gauge group 
remains unconfined.   
Double-scaling is defined near critical points in the 
parameter/moduli space where states charged under $\hat{G}$ become
massless. In specific cases, we present
evidence that the double-scaled theory is dual to a non-critical
superstring background. Models studied include the 
$\beta$-deformation of ${\cal N}=4$ SUSY Yang-Mills which leads to a
non-critical string theory with sixteen supercharges. We also study 
${\cal N}=1$ SUSY Yang-Mills theory coupled to a 
single chiral superfield with a polynomial superpotential 
which leads to a related string theory with eight supercharges. 
In both cases the string coupling is small and the background is free
from Ramond-Ramond flux.

%\vspace{.05in}
%\baselineskip=.3in
\end{titlepage}
\section{Introduction and Overview}
\paragraph{}
The phenomenon of confinement plays a key role in the physics 
of the strong interactions. Obtaining an 
analytic description of the confining phase is an important goal 
which has, so far, remained elusive. 
One of the best hopes for progress in this direction is 
the large-$N$ expansion of $SU(N)$ gauge theory suggested many years
ago by 't Hooft \cite{TH1}. In particular, pure Yang-Mills theory 
is believed to be dual to a weakly-coupled string theory in this limit. 
The large-$N$ spectrum consists of infinite towers of stable 
glueballs corresponding to the excitations of a closed string.
The AdS/CFT correspondence \cite{AdS} provides examples of confining
models \cite{witconf,PS,KS,malnun} 
where the dual theory is a compactification of 
critical superstring theory in ten dimensions. 
Generally, however, the dual background contains Ramond-Ramond (RR) flux or
there are other obstacles to quantization of the string. In these
cases, progress is 
only possible in the supergravity limit where all but the lightest
glueball states decouple\footnote{For some models 
the spectrum can also be calculated for certain limits of large
quantum numbers \cite{BMN}.}. In this paper we will propose limits of 
certain confining theories which bypass this problem and lead to a
weakly-coupled dual string theory in a pure Neveu-Schwarz (NS)
background. The resulting backgrounds can also be understood as 
non-critical superstring theories of the type first introduced in 
\cite{KutSeib}. Our analysis involves a large-$N$ {\em double-scaling limit} 
closely related to those arising 
in the study of matrix models and also to the double-scaling limits of
Little String Theory \cite{LST,GK}. 
In the rest of this introductory section we will outline the main
ideas and results.  
\paragraph{}
In the usual 't Hooft large-$N$ expansion, the leading order
consists of an infinite sum of planar Feynman diagrams. Using the
double-line notation, individual
diagrams in this sum can have an arbitrary number of ``holes''
corresponding to closed index loops. In string theory constructions, 
where $SU(N)$ gauge theory describes the interactions of open strings 
living on the worldvolume of $N$ coincident D-branes, these holes are
simply the boundaries of the string world sheet which arise
naturally in open string perturbation theory. 
The basic idea of large-$N$ duality is that, by summing over all such
open string diagrams, a dual description in terms of 
closed strings is obtained. In spacetime, this amounts to
replacing the D-branes by their near-horizon geometry which
includes $N$ units of RR flux. In a large-radius limit where the RR
field-strength is small, we find that the holes have effectively been
filled in to give a continuous 
closed string world-sheet. What happens away from
this limit is less clear, as we do not know how to quantize the
string. Recent work \cite{OV1,OV2} offers interesting
hints that holes reappear in the closed string description as
``bubbles'' of a new phase of the worldsheet theory and that they are
directly related to the presence of RR flux. Another line of 
development, reviewed in \cite{TS}, suggests that 
the correct effective description of the world-sheet in the presence
of strong RR fields is actually discrete rather than
continuous\footnote{Such an 
effective description might emerge when 
a more traditional continuous description of the worldsheet, like that
suggested in \cite{OV1,OV2}, becomes strongly coupled. Further
discussion of these points is given in Section 8.}. 
For the basic case of the duality between string theory 
$AdS_{5}\times S^{5}$ and ${\cal N}=4$ 
SUSY Yang-Mills the continuous string is effectively 
replaced by a discrete spin chain when the background RR fields become
large \cite{Min}. This discretisation also suggests the presence of
world-sheet holes. 
\paragraph{}
Our main goal here is to find four-dimensional 
examples where the problems of RR flux and of the associated holes 
in the string worldsheet can be avoided. 
It is illuminating to begin by recalling 
the duality between matrix models and $c\leq 1$ non-critical string
theory (for a review see \cite{GZJ}). The matrix model has critical
points in its parameter space where the sum over diagrams of each
fixed genus diverges in a characteristic way and the standard $1/N$
expansion breaks down. Instead one may define an alternative
double-scaling limit where the couplings are tuned to their critical
values as $N\rightarrow \infty$ to obtain a finite contribution from 
each genus. The resulting genus expansion is controlled 
by a parameter $1/N_{eff}$ which is held fixed in the 
limit. In this context, the Feynman diagrams have a direct interpretation   
as discretisations of a dual string world-sheet. 
As we approach the critical point, the sum is
dominated by large diagrams with many holes. After an appropriate
rescaling, we obtain a 
continuum limit on a worldsheet of fixed size, where the average size of the
holes goes to zero. The result is a continuum string propagating in a 
linear dilaton background. The strong-coupling region is removed by
potential terms in the world-sheet action and the 
effective string coupling is identified with the double-scaling
parameter $1/N_{eff}$. 
\paragraph{}
In this paper we will define 
analogous double-scaling limits in four dimensions and argue that they
lead to pure NS backgrounds. We will make contact with several
features of the matrix model case, including the divergence in the sum
of planar diagrams, the appearance of a linear dilaton in the dual
string theory and the emergence of universality classes. More
generally, our results are consistent with an interpretation of
double-scaling as a continuum limit on the string world-sheet. 
The fact that such a limit yields a pure NS background also fits well 
with the association between RR flux and world-sheet holes
mentioned above.        
\paragraph{}
As in the case of matrix models, 
the first step is to find an appropriate critical point. 
The examples we will study here involve four-dimensional confining gauge
theories in a phase where 
an abelian subgroup of the gauge group remains unconfined and there is
no mass gap. One well-studied model \cite{DVPW,CDSW} 
which exhibits this behaviour is 
${\cal N}=1$ SUSY Yang-Mills with gauge group $G=U(N)$ coupled to an
adjoint chiral multiplet $\Phi$ with a polynomial
superpotential\footnote{This model is non-renormalisable and we will
  work with a fixed UV cut-off $M_{UV}$ in place.},  
\begin{equation}
W=\varepsilon {\rm Tr}_{N}\left[\frac{\Phi^{m+1}}{m+1}\,-\,
\sum_{l=1}^{m} \,g_{l}\frac{\Phi^{l}}{l} \right]   
\label{sp}
\end{equation} 
of degree $m+1$. In generic vacua, the $U(N)$ gauge symmetry is
confined down to a $U(1)^{m}$ subgroup denoted $\hat{G}$. 
The spectrum contains $m$ 
massless photons and their ${\cal N}=1$ superpartners. Despite the
absence of a mass gap, the Wilson loop evaluated in the fundamental
representation obeys an area law as it does in more conventional 
confining phases.  We will also study the $\beta$-deformation of ${\cal N}=4$ 
SUSY Yang-Mills which has a partially confining phase of exactly the 
same type \cite{nd1,nd2}. Beyond these specific
models, vacua in this phase are generic to all 
${\cal N}=1$ SUSY gauge theories coupled to adjoint matter with a 
superpotential. Such vacua also occur in 
non-supersymmetric models with adjoint scalars. The phenomena
described below should have interesting generalisations to this wider
class of models.   
\paragraph{}
To explain the relevance of these partially confining models, we begin
by describing some common features of their physics which will be
demonstrated in Section 3 below. 
First, as in pure Yang-Mills theory, the large-$N$ spectrum includes  
a tower of weakly-interacting glueballs corresponding to
excitations of a dual closed string. These states are neutral under
the unconfined gauge group $\hat{G}$ 
and have masses which remain fixed as $N\rightarrow \infty$.  
An important new ingredient is that these models also 
contain states which carry electric 
and/or magnetic charges under $\hat{G}$. For reasons which will be
explained in Section 3, we will refer to these states collectively as 
dibaryons. In the dual string theory, these dibaryons correspond to
wrapped/stretched D-branes.  In generic vacua their masses 
grow linearly with $N$ in accord with the identification $g_{s}\sim
1/N$. However, by varying 
the parameters/moduli we can also find special vacua where some dibaryons 
become massless. These are the critical points we will study. 
Near these special points standard 
large-$N$ counting arguments show that the 't Hooft $1/N$ expansion
breaks down. For the model (\ref{sp}), we will relate this directly to
a divergence in the sum of planar diagrams. These critical points can
also be related to the occurrence of singularities in the dual string
background where wrapped/stretched D-branes become massless. 
In the $\beta$-deformed theory, we will see that the region near the singular 
point in spacetime is replaced by an infinite throat with a linear
dilaton. The breakdown of closed string perturbation theory caused by
the growth of the dilaton 
corresponds to the breakdown of the $1/N$ expansion in the dual field
theory.      
\paragraph{}
As in the case of zero-dimensional matrix models, we can attempt to
define a double-scaling limit which yields a finite effective string 
coupling. In field theory, this corresponds to a limit where
couplings are tuned to their critical values as $N\rightarrow \infty$ 
in such a way that the mass $M_{B}$ of the lightest dibaryon remains 
constant. The tension $T$ of the confining string, which sets the
glueball mass is also held fixed. 
As noted above, the conventional 't Hooft large-$N$ limit leads to
a free theory of colour singlet states with all interactions suppressed
by powers of $1/N$. The key characteristic of the 
double-scaling limit studied here is the emergence of a 
decoupled sector of states with residual interactions 
controlled instead by the double
scaling parameter $1/N_{eff}\sim \sqrt{T}/M_{B}$. 
For both models  studied in detail in this
paper, we will argue that this interacting sector has a 
tractable dual description 
as a double-scaled Little String Theory or, via holography
\cite{holog,GKP}, as a non-critical superstring theory. 
Apart from providing, in these special cases, a solvable limit of
confining gauge theory, these results also suggest a new approach to the 
non-perturbative definition of string theory in linear dilaton
backgrounds. 
\paragraph{}
In the case of the $\beta$-deformed theory, a double-scaling 
limit of exactly the type described above 
was considered in \cite{nd2}. In the following we will review
the relevant results of \cite{nd2} adding several new details and 
physical interpretation. We will then go on to describe our new results
for the model with superpotential (\ref{sp}). 

\subsection{Review of the $\beta$-deformation}
\paragraph{}
The $\beta$-deformation of ${\cal N}=4$ SUSY Yang-Mills with gauge
group $SU(N)$ is dual to a 
certain deformation of IIB string theory on $AdS_{5}\times S^{5}$ 
\cite{AKY,lunin}. 
The model has a partially confining branch for
special values of the deformation parameter. 
Generic vacua on this branch 
correspond to configurations with NS fivebranes located at different radial
positions in $AdS_{5}$. Each NS5 is also wrapped on a (topologically
trivial) toroidal submanifold of $S^{5}$.
The general features of the large-$N$ spectrum
described above are visible in the string dual.   
Electric (magnetic) dibaryons are dual to 
toroidally-wrapped D3 branes (D-strings), stretched between the 
NS5 branes. Critical points with massless dibaryons correspond to 
configurations with coincident NS fivebranes. In these vacua, the
theory flows to ${\cal N}=4$ SUSY Yang-Mills in the IR, with gauge
group $SU(m)$ for $m$ coincident fivebranes. 
\paragraph{} 
For sufficiently large 't Hooft coupling, the 
critical points described above can be studied using supergravity. 
Adapting earlier work by Polchinski and Strassler \cite{PS}, the 
supergravity dual was found in \cite{nd2}. The dual geometry
interpolates smoothly between an asymptotically AdS region in the UV 
and the near-horizon geometry of coincident NS5 branes wrapped on
$T^{2}$. The wrapped NS5 branes also carry a total of 
$N$ units of D3 brane charge on their worldvolume. 
As usual the near-horizon region includes a semi-infinite
throat with a linear dilaton. Although the full background includes 
non-zero Ramond-Ramond fields, these die off rapidly as we enter the 
throat leaving only the fields of the standard NS fivebrane solution. 
\paragraph{}
The singularity corresponding to coincident IIB NS5 branes in 
asymptotically flat ten-dimensional spacetime has been 
studied from many points of view. In particular, the system has a
decoupling limit leading to a six-dimensional 
Little String Theory (LST) \cite{LST}. In this limit, the 
asymptotically flat region of the solution decouples leaving a throat 
which is infinite in both directions. The IIB theory on this infinite
throat background is holographically dual to the Little String Theory
living on the fivebranes \cite{holog}. In the present case we have coincident 
NS5 branes embedded in an asymptotically AdS spacetime. The novelty
here is that the {\em full} geometry, including both the asymptotic region
and the fivebranes, is holographically dual to the four-dimensional 
gauge theory we started with. What remains is to reinterpret known
aspects of fivebrane physics in terms of the four-dimensional theory.          
\paragraph{}
The large-$N$ expansion breaks down at critical
points where dibaryons become massless. This is directly visible in
the string dual where the growth of the dilaton leads to the breakdown
of closed string perturbation theory. However, as for
other linear dilaton backgrounds, we can cure this problem by
deforming the theory in a way which eliminates the strong coupling
region. For NS5 branes in flat space, such a deformation was 
identified by Giveon and Kutasov \cite{GK}. 
These authors argued that separating 
the fivebranes slightly in their transverse directions has the effect 
of replacing the infinite tube in the string dual with a semi-infinite
cigar where the dilaton reaches a maximum value at the tip of the
cigar. More precisely, they consider a double-scaling limit where 
the separation $r_{0}$, between the fivebranes and the asymptotic string
coupling, $g_{s}$, both go to zero with the ratio 
$r_{0}/g_{s}$ and $\alpha'$ held fixed. 
This ensures that the masses $M_{D1}\sim r_{0}/g_{s}\alpha'$ 
of stretched D-strings remain constant 
in the limit. The maximum value of the effective string coupling
attained at the tip of the cigar is then 
$g_{cigar}\sim 1/M_{D1}\sqrt{\alpha'}$.
\paragraph{}
Following \cite{nd2}, we can now reinterpret the double-scaling limit 
of Giveon and Kutasov in the dual gauge theory. As usual taking the
limit $g_{s}\rightarrow 0$ with $\alpha'$ fixed corresponds to a 't
Hooft large-$N$ limit. The new ingredient is that the moduli of the
partially confining branch (corresponding to the separation of the
fivebranes) are scaled so as to keep the masses of the lightest 
electric and magnetic dibaryons fixed. In this limit, the asymptotically
AdS region of the geometry is separated from the region near the
tip of the cigar by a throat of length $\sim \log N$. 
In the dual field theory this
corresponds to the decoupling of two different sectors of the
large-$N$ Hilbert space. One sector, corresponding to states localised
near the tip of the cigar, consists of infinite towers of glueballs with
residual interactions controlled by the coupling $1/N_{eff}=g_{cigar}$. 
The formula for this coupling and the 
effective string tension in terms of field
theory parameters is given in eqns 
(\ref{final1},\ref{final2}) below. When $1/N_{eff}$ is small, 
this sector of the theory is described by
tree-level string theory on an exactly solvable background. 
This in turn is equivalent to a six-dimensional double-scaled
LST with sixteen supercharges compactified on $T^{2}$. The emergence
of two additional compact directions in the dual field theory is
related to the phenomenon of deconstruction \cite{nd2}.
The other sector of the theory, corresponding to the asymptotically
AdS region, becomes free as $N\rightarrow \infty$.   
\paragraph{}
The type of decoupling described above is quite
novel from a field theoretic point of view. Normally two sectors of states 
decouple only when there is a large hierarchy between the
corresponding mass scales. We emphasize that there is no such
hierarchy in the present case: the masses of typical states in the
two sectors are essentially the same. The two sectors decouple 
because they are separated by an infinitely long throat in the 
dual geometry. The decoupling is therefore related to the approximate
locality of the bulk theory which is usually hard to understand from
the boundary side of the correspondence. 
An obvious question is whether the phenomena described above are 
somehow special to the $\beta$-deformation of ${\cal N}=4$ SUSY 
Yang-Mills. A more interesting possibility is that non-trivial
double-scaling limits can be defined for other partially confining theories 
which have critical points in their parameter/moduli spaces. 
In this paper, we will show that 
the model with superpotential (\ref{sp}) provides an interesting new
example of this type.

\subsection{A new example} 
\paragraph{}
The ${\cal N}=1$
theory with superpotential (\ref{sp}) is dual to Type II string theory on a
certain non-compact Calabi-Yau manifold with non-zero RR fluxes 
\cite{intv}. In this context, the electric and
magnetic dibaryons of the field theory correspond to D-branes wrapped
on particular topologically non-trivial cycles which each have zero
units of RR flux. 
At a critical point, one or more of these cycles shrink to zero
size, leading to a singular geometry, 
and the corresponding dibaryons become massless. 
Although various singularities are possible we focus on a particular
case which has a straightforward interpretation in the dual field
theory. The singularity occurs when the
F-term couplings take the values $g_{1}=2\Lambda^{m}$, with $g_{l}=0$ for
$l>1$ \cite{eguchi,bert}. 
Here $\Lambda$ is the RG invariant scale of the gauge theory. 
In this case, the dual Calabi-Yau is described by the polynomial
equation, 
\begin{equation}
u^{2}+v^{2}+w^{2}+\varepsilon^{2}z^{m}(z^{m}-4\Lambda^{m})=0
\label{gencon1} 
\end{equation}
which exhibits a generalised conifold singularity at the origin
$u=v=w=z=0$. Unlike the $\beta$-deformed case, there is no obvious
limit where the dual string geometry (\ref{gencon1}) can be studied
using supergravity. Instead we will analyse the critical point using
field theory methods. 
\paragraph{}
The low-energy F-term effective action for the model (\ref{sp}) 
has been determined by Dijkgraaf and Vafa \cite{DVPW}. 
As we review in Section 3, the resulting F-term action includes 
2-, 3- and 4-point interaction vertices for certain single-trace
operators which arise as components of ${\cal N}=1$ chiral
superfields. As usual these quantities are dominated by 
planar Feynman diagrams in the 't Hooft large-$N$ 
limit. A remarkable feature of the results of Dijkgraaf and Vafa is
that they provide the first known examples where the sum over planar
diagrams can be evaluated exactly in four dimensions. In fact 
these special F-term
quantities are given exactly by a sum over planar diagrams even at
finite $N$. 
The key simplification exploited by Dijkgraaf and Vafa
is that the planar diagrams in question effectively reduce to those of 
an auxiliary zero-dimensional complex matrix model \cite{DVPert1}.
As in other examples related to matrix models, 
the existence of the critical point at $g_{1}=2\Lambda^{m}$
(with $g_{l}=0$ for $l>1$) reflects the finite radius of
convergence of the sum over planar diagrams. 
As we approach the critical point this sum diverges. 
Near the critical point we define a 
large-$N$ double-scaling limit, which tames this divergence, leaving 
a theory with finite (but non-zero) interactions between
colour-singlet states of fixed mass\footnote{The reader may wonder why
  a similar analysis is not possible for the $\beta$-deformed theory
  described in the previous section. The answer is that, in that case, 
the F-term action on the partially confining branch is actually 
trivial because of the
  constraints of conformal invariance and global symmetries. To avoid
  confusion, note that the matrix model analysis of \cite{DHK,ben,DH} 
was applied to the Coulomb branch of the model where the F-terms
are non-trivial.}. In Section 5, we argue that the unique limit
with this property involves taking $N\rightarrow 
\infty$, $\varepsilon\rightarrow \infty$
and $\delta=g_{1}-2\Lambda^{m}\rightarrow 0$ with the following 
quantities held fixed,  
\begin{eqnarray}
 \tilde{\varepsilon}=\varepsilon/N \qquad{} 
\Delta=\varepsilon\delta^{\frac{m+2}{2m}} & \,\,\, {\rm and} \,\,\, & 
\tilde{g}_{l}=g_{l}/\delta^{\frac{m-l+1}{m}} \qquad{} {\rm  
for}\,\,\,\, l=2,\ldots, m 
\label{dscale0}
\end{eqnarray}
As in the conventional 't Hooft large-$N$ limit, 
the dynamical scale $\Lambda$ is also held fixed. The relation of this
limit to the known double-scaling limits of the associated matrix
model is discussed in Section 5.   
\paragraph{}
At the level of the F-term effective action, we can demonstrate
some interesting and suggestive properties of the 
resulting double-scaled theory.  In particular, we find that the
theory has a sector which remains interacting 
in the double-scaling limit . This sector contains
\footnote{Here $[k]$ denotes the greatest integer less than or equal
  to $k$.} $p=[(m-1)/2]$ massless $U(1)$ vector
multiplets of ${\cal N}=1$ supersymmetry 
and $p$ neutral chiral multiplets. The effective superpotential for these
chiral multiplets goes to zero in the double-scaling limit and, in
particular, their masses vanish in this limit. 
The interactions between these degrees 
of freedom are controlled by a coupling $1/N_{eff}\sim 1/\Delta$ 
which is held fixed in the double-scaling limit. 
In contrast F-term interactions with the remaining degrees
of freedom of the theory are suppressed by powers of $1/N$.
\paragraph{}
Like the case of the $\beta$-deformed theory, the critical point 
described above occurs when 
electric and magnetic dibaryons become massless. 
These states correspond to D-branes in the dual string theory wrapped
on cycles which vanish at the critical point. 
The resulting divergences in the F-term action can be 
interpreted as a consequence 
of the additional massless degrees of freedom. On the other hand, the
masses of these states scale like $1/g_{s}\sim N$ in generic vacua 
away from the critical point. Hence it is natural 
to interpret the double-scaling limit as a limit where the
couplings approach their critical values as $N\rightarrow \infty$ so
that the dibaryon masses are held fixed. 
\paragraph{}
On the string theory side, this again corresponds to a limit of the type
considered by Giveon and Kutasov. 
In fact, the case of a generalised conifold singularity of the form 
\begin{equation}
U^{2}+V^{2}+W^{2}+Z^{m}=0
\label{uvw}
\end{equation}
arising in a Type IIB Calabi-Yau geometry {\em without flux} was considered in 
\cite{GK}. The decoupling limit $g_{s}\rightarrow 0$, yields a 
Little String Theory with eight supercharges 
which flows to the ${\cal N}=2$ Argyres-Douglas (AD)
fixed-point\footnote{In the special case $m=2$, (\ref{uvw})
  describes the standard conifold and the dual theory is IR free.} 
of type $a_{m-1}$ in the IR. 
The resulting LST is holographically dual to an infinite throat with a
linear dilaton which describes the IIB string propagating in the region 
near the singularity. As usual the singularity can be resolved by
blowing up three-cycles so that wrapped D3 branes have non-zero masses
$M_{D3}$.    
A double-scaling limit is then obtained by scaling the resolution
parameters, so that the masses of these state remains fixed as
$g_{s}\rightarrow 0$. As for the NS fivebrane theory discussed above, 
the effect of double-scaling is to replace the infinite throat in the
holographic dual by a semi-infinite cigar. The effective string
coupling then reaches a maximum value $g_{cigar}\sim
1/(M_{D3}\sqrt{\alpha'})$ at the tip of the cigar.   
\paragraph{} 
In the present case we are interested in the non-compact Calabi-Yau
geometry (\ref{gencon1}) which includes $n$ units of flux through each
of $m$ non-intersecting three-cycles. The presence of flux breaks the 
${\cal N}=2$ supersymmetry of the Calabi-Yau compactification down to
an ${\cal N}=1$ subalgebra. As described above, the dual field theory
has a double-scaling limit with a decoupled sector 
where residual interactions are controlled by the parameter 
$1/N_{eff}\sim 1/\Delta$. Inspired by the example of the 
$\beta$-deformed theory, we would like to propose that the interacting
sector is dual to the double-scaled Little String Theory which arises in the 
Giveon-Kutasov limit of the generalised conifold singularity 
without flux. As before the effective string coupling $g_{cigar}$
is identified with the double-scaling parameter 
$1/N_{eff}$ and $1/(2\pi \alpha')$ is identified with the tension of the
confining string. The identification of string theory and gauge
parameters is given in more detail in Eqns (\ref{ident6}) below.    
\paragraph{}
The most obvious objection to this proposal is
that the double-scaled LST in question preserves eight supercharges
corresponding to ${\cal N}=2$ supersymmetry in four dimensions. In
contrast, the gauge theory with superpotential (\ref{sp}) has only
${\cal N}=1$ SUSY. Our claim is that supersymmetry is enhanced to
${\cal N}=2$ only in the decoupled sector and that the enhancement 
occurs specifically in the double-scaling limit. We will argue that
supersymmetry enhancement in the decoupled sector occurs at all energy
scales not just in the IR. 
This is similar to the
enhancement of ${\cal N}=1$ SUSY to ${\cal N}=4$ which occurs in the 
the decoupled sector of the $\beta$-deformed theory. 
The F-term calculations described above provide evidence in favour of
this conclusion. In particular, the superpotential for the fields 
in the decoupled sector vanishes in the double-scaling limit and the
resulting massless field content described above is consistent with
the presence of $p=[(m-1)/2]$ massless vector multiplets of ${\cal N}=2$
supersymmetry. More generally we compute the exact ${\cal N}=1$ F-terms in the
low-energy effective action on both sides of the proposed
duality and show that they agree precisely. Note that, for the case
$m=2$ of the ordinary conifold, we have $p=0$ and the interacting
sector has no massless fields. In this special case, we therefore have much
less evidence in favour of our conjecture.  
\paragraph{}
The supergravity analysis of the $\beta$-deformed theory reviewed
above immediately suggests a heuristic 
interpretation of our proposed correspondence. The full
string background dual to the gauge theory with superpotential
(\ref{sp}) is some complicated warped geometry with strong RR fluxes. 
The generalised conifold singularity appears when various flux-free
cycles vanish. As in the case without flux, the region near the
singularity is replaced by an infinite throat with a linear dilaton 
\cite{Comments}. 
Although RR fields are present in the bulk, these fields decay as we
move down the throat. In the double-scaling limit, the throat is
replaced by a cigar whose length becomes infinite as $N\rightarrow
\infty$. String states localised near the tip of the cigar
decouple from the bulk and have the same world-sheet description as in
the case without flux. As in the $\beta$-deformed case, many details
of the UV theory are lost when the bulk region decouples. This is even
more striking in the present case where the dual field theory is 
non-renormalisable and has an explicit UV cut-off which is held fixed in
the limit. In the decoupled sector, all dependence on the cut-off 
is absorbed onto the two independent parameters of double-scaled LST. 
We believe that this type of behaviour, which is surely impossible at
finite $N$, is an interesting new aspect of large-$N$ field theory
which should have other applications.  
\paragraph{}
The results presented in this paper have some overlap with earlier
work. The idea that the standard 
large-$N$ expansion breaks down near the
critical points in the moduli/parameter space of four-dimensional
gauge theories has been emphasized in a series of papers by Ferrari 
\cite{ferr2,ferr3,ferr1}, 
which also contain proposals for double-scaling limits. 
In particular, one of the limits considered in \cite{ferr1} is closely 
related to the $m=2$ case of our analysis. 
Large-$N$ limits leading to dual string backgrounds without RR flux
have also been discussed in \cite{AMV} (see also \cite{DV2}). 
The field theory and the 
specific vacuum studied in this reference also coincide with the $m=2$ case 
discussed below. However, the specific limit considered in \cite{AMV}
is quite different from the double-scaling limit considered here 
(this is discussed further in Section 6.2). The 
four-dimensional interpretation of
double-scaling in the Dijkgraaf-Vafa matrix model has also been
discussed before in \cite{DV2,ferr1}. 
\paragraph{}
The rest of the paper is organised as follows. Section 2 provides
several preliminaries for the analysis of the model (\ref{sp}). In
particular, we discuss the vacuum structure of the theory and its
classification in terms of order parameters. This is mainly a review
of  parts of \cite{CSW}. We also review the F-term effective action of 
\cite{DVPW}. 
In Section 3, we present an analysis of the 
spectrum and interactions of the model based on large-$N$ counting. 
We also review basic features of the 
dual string theory background of \cite{intv}. Section 4
identifies the critical point in parameter space and exhibits the
breakdown of the $1/N$ expansion. Double-scaling limit is formulated
in Section 5 and our calculations of the F-terms in this limit are
described. A more detailed description
of the calculations is given in an Appendix. Our proposal for a 
duality between the model 
(\ref{sp}) and the double-scaled LST of the generalised 
conifold is formulated in 
Section 6.  A more detailed review of the $\beta$-deformed
theory and its double-scaling limit is
provided in Section 7. Finally, a discussion 
of our results is given in Section 8. 
    
\section{The Model}
\paragraph{}
In this paper we will mainly be concerned with gauge theories in a
phase where confinement is partial in the sense that an abelian
subgroup $\hat{G}$
of the original gauge group $G$ remains unconfined and the low-energy
theory contains massless gauge fields\footnote{More precisely, 
we consider theories with gauge group $G$ which have an unconfined 
gauge symmetry, $\hat{G}$, at low energies. At least in
one regime of parameters (the weak-coupling regime discussed below)
it is clear that $\hat{G}$ is a subgroup of the original gauge group
$G$. More generally, $\hat{G}$ and $G$ might not be related in a simple way.}. 
We begin by reviewing a 
family of models which exhibits this behaviour in generic vacuum states. 
As in Section 1 above, 
we will consider an ${\cal N}=1$ supersymmetric gauge theory with
gauge group $G=U(N)$. The matter content of the theory is an ${\cal
  N}=1$ vector multiplet, with chiral field strength 
$W_{\alpha}$, and a single adjoint chiral multiplet $\Phi$. The theory 
has a classical superpotential of the form, 
\begin{equation}
W=\varepsilon {\rm Tr}_{N}\left[\frac{\Phi^{m+1}}{m+1}\,-\,
\sum_{l=1}^{m} \,g_{l}\frac{\Phi^{l}}{l} \right]   
\label{sp1}
\end{equation} 
One obvious objection to this model is that it is non-renormalisable 
and requires embedding in some larger theory with a sensible UV
definition. For our purposes it will be sufficient to introduce an
explicit UV cut-off at some scale $M_{UV}$ larger than the other mass 
scales in the problem. 
\paragraph{}
The classical vacua of the theory 
are determined by the stationary points of the
superpotential (\ref{sp1}). For generic values of the couplings, we
find $m$ stationary points at the zeros of, 
\begin{equation}
W'(x)=\prod_{l=1}^{m}(x-x_{l})
\label{df}
\end{equation}
The classical vacua are configurations where each of the
$N$ eigenvalues of $\Phi$ takes one of the $m$ values, $\{x_{l}\}$, for
$l=1,2,\ldots,m$. 
Thus vacua correspond to partitions of $N$ 
where $N_{l}\geq 0$ eigenvalues take the value $x_{l}$ with
$N_{1}+N_{2}+\ldots N_{m}=N$. 
\paragraph{}
Provided $N_{l}\geq 2$ for all $l$, the classical low-energy gauge 
group in such a vacuum is, 
\begin{equation}
\hat{G}_{cl}=\prod_{l=1}^{m} U(N_{l}) \simeq \prod_{l=1}^{m}
U(1)_{l}\times SU(N_{l})
\label{ghat}
\end{equation}
The massless modes in this vacuum comprise an ${\cal N}=1$
vector multiplet of $\hat{G}_{cl}$. 
The classical spectrum also includes massive states which carry
electric charges under the unbroken $U(1)$ factors in the low-energy gauge 
group. In particular there are both vector and chiral multiplets 
transforming in the 
$({\bf N}_{r},\bar{\bf N}_{s})$ of $SU(N_{r})\times SU(N_{s})$ with charges 
$(+1,-1)$ under $U(1)_{r}\times U(1)_{s}$. There are also massive
states transforming in the adjoint representation of each non-abelian
factor which are neutral under the unbroken $U(1)$s. 
The masses of all these states 
are set by the dimensionful parameters in the superpotential 
(\ref{sp1}). We will 
denote the lightest such mass-scale $M$. Finally, the classical theory
also includes 't Hooft-Polyakov monopoles which are magnetically
charged under pairs of $U(1)$ factors in $\hat{G}_{cl}$. The masses of these
states are of order $\tilde{M}=M/g_{0}^{2}$ where $g_{0}$ is the bare 
gauge coupling. 

\subsection{Weak-coupling analysis}
\paragraph{}
Quantum corrections drastically modify 
the properties of the classical vacuum state described above. The 
UV gauge theory is asymptotically free and the gauge
coupling runs with energy scale. The resulting dynamics is
easiest to analyse in a regime where the running
coupling $g^{2}(\mu)$ is small at the scale $\mu=M$ set by the
parameters in the superpotential. Following previous authors, we will refer
to this as the weak-coupling regime\footnote{This is a misnomer
because, of course, the coupling is only really weak in the UV.}. 
In terms of the RG invariant scale $\Lambda$ associated with the 
running of the $U(N)$ gauge coupling in
the UV theory, the weak-coupling regime corresponds to $\Lambda<<M$. 
In this regime, the low-energy theory at scales far below 
$M$ is ${\cal N}=1$ SUSY Yang-Mills with gauge group 
$\hat{G}_{cl}=\prod_{l=1}^{m}U(1)_{l}\times SU(N_{l})$ and we can 
draw on the standard facts about the dynamics of this theory. 
In particular, each non-abelian factor in $\hat{G}_{cl}$ is 
asymptotically free and runs to strong coupling in the IR
below the scale, 
\begin{equation}
\Lambda_{(l)}\sim M\exp\left(-\frac{8\pi^{2}}{3N_{l}g^{2}(M)}
\right)
\label{laml}
\end{equation} 
The resulting strong coupling dynamics confines all states charged under the
non-abelian factors of $\hat{G}_{cl}$ and generates a mass gap of order 
$\Lambda_{(l)}$ in the $l$'th factor. The unconfined low-energy
gauge-group is $\hat{G}=U(1)^{m}=\prod_{l=1}^{m}U(1)_{l}$ and the only
remaining massless states are $m$ abelian multiplets of ${\cal N}=1$ 
supersymmetry. 
\paragraph{}  
Another effect of the strong-coupling dynamics 
is to produce non-zero gluino condensates in each non-abelian
factor of $\hat{G}_{cl}$. If we denote 
as $W_{\alpha l}$, the chiral field strength of the $SU(n_{l})$
vector multiplet in the low-energy theory, we can define a corresponding 
low-energy glueball superfield 
$S_{l}=-(1/32\pi^{2})
\langle {\rm Tr}_{N_{l}}(W_{\alpha l}W^{\alpha}_{l}) \rangle$ in each
factor. Non-perturbative effects generate a superpotential of the form 
\cite{VY,CSW}, 
\begin{equation}
W_{eff}(S_{1},\ldots,S_{m})=\sum_{l=1}^{m}N_{l}\left(S_{l}
\log(\Lambda_{(l)}^{3}/S_{l})+S_{l}\right) + 2\pi i \sum_{l=1}^{m} b_{l}S_{l}
\label{vy}
\end{equation}
where $b_{l}$ are integers to be discussed below. Note that this
simple additive formula for the superpotential is only valid in the 
weak-coupling regime $\Lambda<<M$
where the dynamics of the different non-abelian factors in
$\hat{G}_{cl}$ are effectively decoupled. Assuming that the low-energy
dynamics is correctly described by an effective action for the fields $S_{l}$,
we can find the vacuum states of the theory by minimizing $W_{eff}$ to
obtain, 
\begin{equation}
\langle S_{l} \rangle  = 
\Lambda_{(l)}^{3}\exp\left(\frac{2\pi i}{N_{l}}b_{l}\right)
\label{cond}
\end{equation}
The phase of each gluino condensate depends on an integer
$b_{l}$ defined modulo $N_{l}$ which labels inequivalent supersymmetric
vacua. Thus the quantum theory has a total of $N_{1}N_{2}\ldots N
_{l}$ 
distinct SUSY vacua corresponding to the 
single classical vacuum we started with. 
\paragraph{}
As first discussed by Cachazo, Seiberg and Witten (CSW) \cite{CSW}, 
the physics of the 
partially confining phase depends sensitively on the integers $b_{l}$ 
appearing in the phase of the gluino condensate (\ref{cond}). 
In the more familiar case of theories with a mass gap, the possible 
confining phases were classified by 't Hooft \cite{TH2} in terms of
the electric 
and magnetic quantum numbers of states which condense in the vacuum. 
Conventional confinement is associated with the condensation of
magnetic degrees of freedom, while various oblique confining 
phases occur when dyonic states condense. In pure 
${\cal N}=1$ SUSY Yang-Mills with gauge group $SU(n)$ the
corresponding integer $b=0,1,2,\ldots,n-1$, appearing in the phase of the
gluino condensate, determines which 
kind of confinement is realised in each vacuum. Precisely one
value of $b$ is associated with ordinary confinement and
the remaining values lead to oblique confinement. In this case the 
$n$ vacua are also related by a spontaneously broken discrete
symmetry. As the theory contains no dynamical 
charges in the fundamental representation, the different phases can 
only be distinguished by introducing
external probes.  
In the present case
with several non-abelian factors, the theory contains states carrying
electric and magnetic charge under pairs of unconfined $U(1)$s. In this case 
the physics of each vacuum depends non-trivially on differences in the mode of 
confinement between different factors even in the absence of
external probes. As there are no states charged under the central
$U(1)$, the physics only depends on the differences of
the integers $b_{l}$. Following \cite{CSW} we adopt a convention where 
$b_{1}=0$. 

\subsection{Order parameters}
\paragraph{}
For a theory containing only fields in the adjoint representation of
the gauge group, the natural order parameter for confinement is the
Wilson loop, 
\begin{equation}
{\cal W}_{R}=\left\langle {\rm Tr}_{R}  P\exp\left(-\oint_{C} A \right)
\right\rangle
\label{wilson}
\end{equation} 
where $C$ is a closed contour and 
the trace is evaluated in some representation $R$ of the gauge
group $G$. The signal for confinement of external charges tranforming in the
representation $R$ of $G$ is that the corresponding Wilson line decays
exponentially with the area enclosed by the contour $C$. For an
$SU(n)$ gauge theory in a conventional confining phase, this area law 
holds when $R$ is the fundamental representation ${\bf n}$ and also for tensor
products of $r<n$ copies of the fundamental representation 
${\bf n}^{r}={\bf n}\otimes{\bf n}\otimes \ldots\otimes{\bf n}$. 
For $r=n$, this tensor product contains a singlet, the baryon, which
is not confined and consequently the area law does not hold. More generally,
following CSW \cite{CSW}, it is useful to introduce the 
{\em confinement index}, which is an integer $t$ defined as 
the lowest value of $r$ such that the Wilson loop evaluated in the 
representation ${\bf n}^{r}$ {\em does not} exhibit the area law. 
Because the $n$-fold product always contains a singlet, the
confinement index is naturally defined modulo $n$.  
\paragraph{}
For the vacua of the $U(N)$ theory considered above, the 
confinement index is easy to compute in the case 
where the mode of confinement in each non-abelian factor is the same 
\cite{CSW}. This corresponds to the choice $b_{1}=b_{2}=\ldots=b_{m}=0$ for the
integers defined in the previous section which label the vacua. 
The fundamental representation ${\bf N}$ of $SU(N)$ transforms as,
\begin{equation}
({\bf N}_{1},{\bf 1},\ldots,{\bf 1})\oplus ({\bf 1}, 
{\bf N}_{2},\ldots,{\bf 1})\oplus\ldots\oplus({\bf 1}, {\bf 1},
\ldots,{\bf N}_{m}) 
\end{equation}
under $\prod_{l=1}^{m}SU(N_{l})\subset \hat{G}_{cl}$. If we take 
a tensor product of $r$ copies of this representation a singlet will
occur if $r$ is equal to any of the integers $N_{l}$. Thus a singlet
also occurs for $r=l_{1}N_{1}+l_{2}N_{2}+\ldots l_{m}N_{m}$ for any
integers $l_{l}$. The smallest non-zero value of $r$ modulo $N$ 
which can be obtained in this way is equal to the greatest common
divisor of the $N_{l}$. Thus we find that the confinement index $t$ 
is equal to the greatest common divisor of the 
integers $N_{l}$ with $l=1,2,\ldots, m$. 
\paragraph{}
The above discussion takes into account the possibility of the
screening of external charges by pair creation of chromoelectric
charges. 
To extend this analysis to the more general vacuum where the integers 
$b_{l}$ are not all equal, we also need to consider magnetic
screening. This is discussed in detail in Section 2.1 of \cite{CSW} 
and we will only quote the final result: the confinement index in a general 
vacuum is equal to the greatest common divisor of the $N_{l}$ and the  
$b_{l}$. We will give a string theory derivation of this result in
Section 3.3 below.  
\paragraph{}
So far our discussion of confinement has been restricted to the 
weak coupling regime $\Lambda<<M$. However, 
the behaviour of the Wilson loop can only change if we
encounter a phase transition. Thus we can interpret the confinement
index $t$ as a universal quantity characterising the partially confining
phase. Another such quantity is the rank $r=m$ of the low-energy
unconfined gauge group $\hat{G}$. The investigation of \cite{CSW} showed
that, as the parameters of the superpotential are varied it is
possible to interpolate smoothly between classical vacua
corresponding to different partitions of $N$. Thus the filling fractions 
$N_{l}/N$ characterising a particular classical vacuum are not
universal in the sense described above. The same applies to the
individual integers $b_{l}$ which determine the mode of confinement in
each non-abelian factor. In contrast CSW
found that the confinement index $t$ and the rank $r$ of the
unconfined gauge group can only change on passing through a singular 
point where two branches of the moduli space intersect. Other more
subtle order parameters which distinguish different branches were also
introduced in \cite{CSW}.
\subsection{The F-term effective action}
\paragraph{}
In the weak-coupling regime $\Lambda<<M$, we saw that 
the low-energy effective theory contains an ${\cal N}=1$ vector multiplet of 
the unbroken classical gauge group $\hat{G}_{cl}=\prod_{l=1}^{m}
U(N_{l})$. From each factor $U(N_{l})\simeq U(1)_{l}\times SU(N_{l})$ 
in $\hat{G}_{cl}$, we obtain a 
chiral field-strength superfield $W_{\alpha l}$. 
Interactions between fields living in different factors are mediated 
by bifundamental diquarks of mass $M$. 
Provided we remain at weak coupling these interactions are effectively
suppressed by powers of the small parameter $\Lambda/M$. More generally,
we can obtain corrections to this limit by integrating out the massive
degrees of freedom to obtain an effective action for the
vector multiplets $W_{\alpha l}$. In principle, a second step would
then be to quantize this action to obtain a complete
effective action which also takes into account the strong-coupling
gauge dynamics at scales of order $\Lambda$. 
\paragraph{}
Although this goal is far
too ambitious for generic observables, theories with ${\cal N}=1$
supersymmetry contain a special subsector of observables which are
constrained by holomorphy in complex parameters and background
fields. These observables are determined by the F-terms 
in effective action. Such terms include the effective superpotential for
the glueball superfields $S_{l}=
-(1/32\pi^{2}){\rm Tr}_{N_{l}}[W_{\alpha l}W^{\alpha}_{l}]$, 
whose weak-coupling form was given in Eq (\ref{vy}) above. 
In a remarkable series of papers \cite{DV2,DV3,DVPW}, Dijkgraaf and
Vafa determined the exact F-terms in the effective action for
$W_{\alpha l}$ obtained by integrating out the massive degrees of freedom. 
With some plausible additional assumptions their approach also
yields the full F-term effective action including the effects of gauge
dynamics. The resulting F-term effective action 
can be written in terms of the glueball superfields
$S_{l}$ and $m$ additional superfields 
$w_{\alpha l}=(1/4\pi){\rm Tr}_{N_{l}}[W_{\alpha l}]$, for
$l=1,2,\dots, m$ which
contain as components 
the massless photons and their ${\cal N}=1$ superpartners. The 
general form of the action is,   
\begin{equation} 
{\cal L}_{F}={\rm Im}\left[\int d^{2}\,\theta\, W_{eff}\right] =
{\rm Im}\left[\int d^{2}\,\theta \, W_{eff}^{(0)}+W_{eff}^{(2)}\right]
\label{fterm0}
\end{equation}
with, 
\begin{eqnarray} 
W_{eff}^{(0)}   & = 
& \sum_{l}\, N_{l}\frac{\partial{\cal F}}{\partial S_{l}} +2\pi i
      \tau_{0}\sum_{l}S_{l}+2\pi i \sum_{l}b_{l}S_{l} \nonumber \\ 
W_{eff}^{(2)}   & = 
& \frac{1}{2}\sum_{k,l} 
\frac{\partial^{2}{\cal F}}{ \partial S_{k} \partial S_{l}}
\,w_{\alpha k}w^{\alpha}_{l} 
\label{fterm1}
\end{eqnarray}
To avoid any confusion we will refer to $W_{eff}$ as the total 
superpotential and $W^{(0)}_{eff}$ on its own as the glueball 
superpotential. The function ${\cal F}(S_{1},S_{2},\ldots,S_{m})$ 
is known as the prepotential.

%%%%%%%%%%%%%%%%%%%%%%%%%%%%%%%%%%%%%%%%
\vspace{1.0cm}
\begin{figure}[ht]
\begin{center}
\input{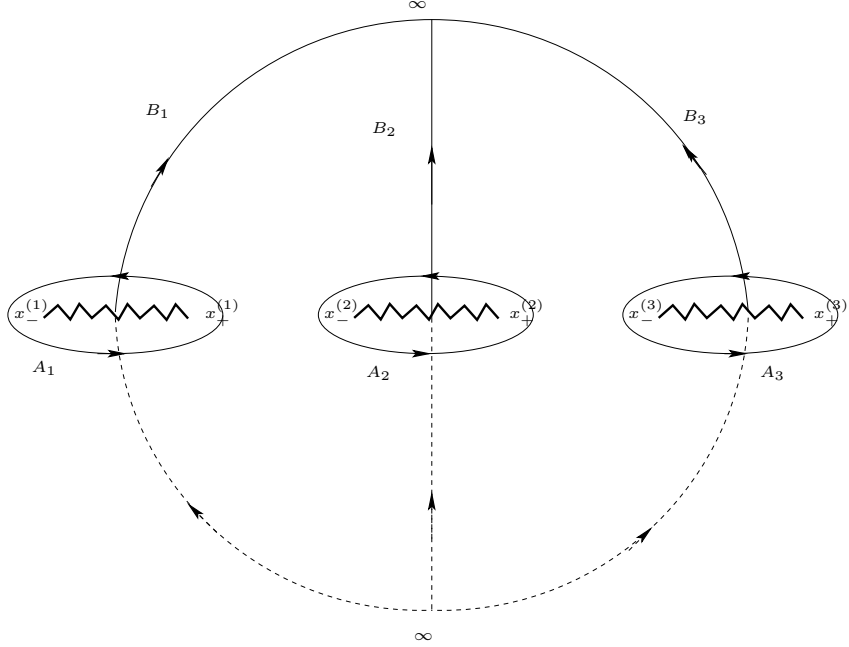}
\caption{ \label{fig1} 
The spectral curve $\Sigma$ for $m=3$ }
\end{center}
\end{figure}
\vspace{0.5cm}
%%%%%%%%%%%%%%%%%%%%%%%%%%%%%%%%%%%%%%%%
\paragraph{}      
The exact F-terms are governed by
the corresponding spectral curve $\Sigma$, 
\begin{equation}
y^{2}=W'(x)^{2}-f(x)=\varepsilon^{2}\prod_{l=1}^{m}(x-x_{+}^{(l)})
(x-x_{-}^{(l)})
\label{spectral}
\end{equation}
where $f(x)=\sum_{l=1}^{m} \kappa_{l}x^{l-1}$ depends on $m$
parameters $\kappa_{l}$ which will be determined below. The curve is a
double-cover of the complex $x$ plane branched over the $2m$ points 
$x^{(l)}_{\pm}$ for $l=1,2,\ldots,m$, with a marked point at infinity
on both sheets. 
The two sheets of the cover are joined along $m$ branch cuts
$C_{l}$. We chose the cuts so that $C_{l}$ runs from $x^{(l)}_{+}$ to 
$x_{-}^{(l)}$ for $l=1,2,\ldots m$. We also introduce  
the following basis of one-cycles
$A_{k}$ and $B_{k}$ with intersections, 
\begin{eqnarray}
A_{k} \circ A_{l} = B_{k} \circ B_{l}=0 & \qquad{} & 
A_{k} \circ B_{l}=\delta_{kl} 
%\oplus \odot
\label{intersect}
\end{eqnarray}
The cycle $A_{k}$ surrounds the cut $C_{k}$. 
The (non-compact) dual cycle $B_{k}$ runs from the point at infinity 
on the lower sheet to the same point on the upper sheet, passing from
the lower to the upper sheet by traversing the cut $C_{k}$. The
resulting non-compact Riemann surface has genus $m-1$. 
The surface $\Sigma$ and its one-cycles are shown in
Figure \ref{fig1} for the case $m=3$. The solution is then
determined by the periods of the differential $ydx$
according to,   
\begin{eqnarray}
S_{l}=\,\oint_{A_{l}}\, ydx & \qquad{} & S^{D}_{l}=
\frac{1}{2\pi i}\frac{\partial{\cal F}}{\partial S_{l}}=\,\oint_{B_{l}}\, ydx 
\label{ssd}
\end{eqnarray}
In particular, the first equation in (\ref{ssd}) effectively
determines the $m$ parameters $\kappa_{l}$ in terms of the $S_{l}$. The
second equation in (\ref{ssd}) can then be used to express ${\cal
  F}$ as a function of the glueball superfields. This in turn
determines the full F-term action (\ref{fterm0},\ref{fterm1}). 
The abelian gauge couplings appearing in (\ref{fterm1}) are determined by the
matrix of second partial derivatives of ${\cal F}$ which can be
thought of as a generalised period matrix for the non-compact Riemann
surface $\Sigma$.   
\paragraph{}
As we are interested in moving away from the weak coupling
limit, the definitions of the chiral superfields $S_{l}$ and
$w_{\alpha l}$ used above are not really satisfactory because they are only
invariant under the classically unbroken subgroup $\hat{G}_{cl}$, and
not under the full $U(N)$ gauge symmetry of the
 theory. A very elegant 
gauge-invariant definition of these observables, which reduces to the
definition given above at weak-coupling, was suggested in
\cite{CDSW}. The definition involves the closed contours, $A_{l}$, in the
complex $x$-plane introduced above. It is easy to check that each
contour $A_{l}$ surrounds exactly one critical point, $x_{l}$, 
of the classical superpotential $W(x)$. We can then define,  
\begin{eqnarray}
S_{l} & = & - \frac{1}{2\pi i}\oint _{A_{l}} \, dx\, \frac{1}{32\pi^{2}} 
{\rm Tr}_{N}\left[ \frac{W_{\alpha}W^{\alpha}}{x- \Phi}\right]
\nonumber \\ 
w_{\alpha l} & = & \frac{1}{2\pi i}\oint _{A_{l}} \, dx 
\, \frac{1}{4\pi} {\rm Tr}_{N}\left[ 
\frac{W_{\alpha}}{x- \Phi}\right]  
\label{gidef}
\end{eqnarray}

\section{The Large-$N$ limit}
\paragraph{}
In this Section we will consider the large-$N$ behaviour 
of the theory in its partially confining phase. We begin 
by briefly reviewing the standard discussion of 
$SU(N)$ Yang-Mills theory with and without quarks (see for example 
\cite{man}). In subsequent subsections we adapt these standard counting
arguments to the case of partial confinement, discuss the F-term
effective action at large-$N$ and finally discuss the dual string
theory.   

\subsection{A review of large-$N$ counting}
\paragraph{}
We will begin 
by reviewing the standard large-$N$ analysis of 
pure Yang-Mills theory with gauge group $SU(N)$. 
A similar discussion applies to any (totally) confining theory 
with all fields in the adjoint representation of the gauge group. 
To describe the large-$N$ limit, it is convenient to rescale the fields so
that the bare gauge coupling $g_{0}^{2}$ appears only via an overall
prefactor of $1/g_{0}^{2}$ in front of the action.      
The 't Hooft limit then corresponds to 
taking $N\rightarrow \infty$ and $g_{0}^{2}\rightarrow 0$ with the 
combination $g_{0}^{2}N$ held fixed. 
In terms of renormalisation group invariant
quantities, this corresponds to a limit where $N\rightarrow \infty$
with the dynamical scale $\Lambda$ held fixed. Thus the mass gap of 
the theory remains constant in the limit. 
\paragraph{}
The spectrum of pure Yang-Mills theory consists of colour-singlet glueballs 
with masses of order $\Lambda$. At large-$N$, the corresponding 
single-particle states are created by gauge-invariant single trace
operators acting on the vacuum. With the conventions described above,
where $N$ appears as an overall prefactor in front of the action,
single-trace operators create single-particle states with amplitude of
order one.  
For example scalar glueballs are created with amplitude of order one 
by the operator $\hat{O}={\rm Tr}_{N}[F_{\mu\nu}F^{\mu\nu}]$.  
Connected correlation functions of such operators obey the scaling rule, 
\begin{eqnarray}
\langle \hat{O}(x_{1})\hat{O}(x_{2})\ldots\ldots
\hat{O}(x_{L})\rangle & 
\qquad{} \sim 
\qquad{} N^{2-L}
\label{scaling}
\end{eqnarray}
This means that glueball masses, which are determined by the
exponential decay of two-point functions, remain fixed in the
large-$N$ limit, while decay widths corresponding to three-point
functions go like $1/N$. On shell $2\rightarrow 2$ scattering 
amplitudes are extracted from four-point functions which scale like
$1/N^{2}$. As $N\rightarrow \infty$, we therefore expect a free theory 
with an infinite tower of stable glueballs. 
This is exactly the behaviour expected from a closed string 
theory with string coupling $g_{s}\sim 1/N$. 
\paragraph{}
Although the supersymmetric models considered in this paper contain 
only adjoint fields, it will be useful to extend our discussion of 
pure $SU(N)$ Yang-Mills
theory by adding massive quarks in the fundamental representation. 
In this case the 
glueball spectrum described above will be accompanied by massive
mesons and baryons. Standard large-$N$ counting indicate that mesons
have masses of order one and $L$-point interactions which scale like 
$N^{1-L/2}$. Thus mesons, like glueballs, become free in the large-$N$
limit. In a dual string theory description with string coupling
$g_{s}\sim 1/N$ mesons are naturally identified as excited states of
an open string. Mixing between mesons and glueballs is also suppressed
in the large-$N$ limit. Baryons have very different 
properties at large-$N$ \cite{witbar1}. Their masses grow linearly
with $N$ reflecting the presence of $N$ constituent quarks, while
their shape and size remain fixed at large-$N$. A key feature which
will play an important role in the following is that the interactions
of baryons, both with mesons and with glueballs, are {\em not} suppressed in
the large-$N$ limit. In particular the interactions of glueballs with 
baryons are governed by matrix elements of single trace operators such as 
$\hat{O}={\rm Tr}_{N}[F_{\mu\nu}F^{\mu\nu}]$ between one-baryon
states. Applying standard counting rules we find,    
\begin{eqnarray}
\langle B| \hat{O}(x_{1})\hat{O}(x_{2})\ldots\ldots
\hat{O}(x_{L})|B'\rangle & 
\qquad{} \sim 
\qquad{} N^{1-L}
\label{scalingb}
\end{eqnarray}   
Thus the three-point coupling between two baryons and a single
glueball corresponds to the case $L=1$ and therefore scales like 
$N^{0}$. A similar coupling for two baryons and a single meson grows
like $N^{\frac{1}{2}}$. All of these features are easily understood
from the perspective of the dual string theory where glueballs and
mesons are identified with the excitations of closed and open strings
respectively. In this context baryons should be thought of as
D-particles on which open strings can end \cite{witbar2}. The leading contribution to
the matrix element (\ref{scalingb}) corresponds to a disc amplitude
coupling the D-brane to $L$ closed string states which scales as
$g_{s}^{L-1}\sim N^{1-L}$.
 
\subsection{Partially confining theories at large-$N$}
\paragraph{} 
How does the above story change when we have a theory in the partially
confining phase? The obvious difference is the presence in the
spectrum of massless gauge bosons of the unconfined gauge group 
$\hat{G}=U(1)^{m}$. All the large-$N$ limits we will consider in this paper 
are ones where the rank $m$ of the low-energy gauge group is held
fixed. Heuristically, 
the presence of the low-energy gauge symmetry is only significant 
if there are states in the spectrum which carry the corresponding 
charge. In the absence of light charged states the gauge bosons decouple in
the IR. In the following we will see that the spectrum does contain 
states which are electrically and/or magnetically charged under 
$\hat{G}$. However, in generic vacua, these states become very massive 
at large $N$ (somewhat like the baryons of large-$N$ QCD described in
the previous subsection) 
and we will argue that the remaining spectrum of
$\hat{G}$-neutral states resembles that of a standard large-$N$ gauge 
theory.       
\paragraph{}
We will now describe the large-$N$ limit of the partially confining
phase in more detail. 
As before we take the $N\rightarrow \infty$ with $g_{0}^{2}N$ fixed, which
corresponds to holding the dynamical scale $\Lambda$ fixed.  
Again, it is convenient to normalize the bare action so that
a power of $1/g_{0}^{2}$ appears as a prefactor for the vector 
multiplet kinetic terms. We also adopt conventions where the same prefactor
appears in front of the kinetic term for the chiral multiplet. 
Finally, to ensure that the remaining terms in the action 
also scale linearly with $N$, we also take the parameter $\varepsilon$
appearing in the superpotential (\ref{sp1}) to infinity with 
$\tilde{\varepsilon}=\varepsilon/N$ held fixed. The remaining
superpotential couplings $g_{l}$ are held fixed. The only remaining 
choice is the scaling of the filling fractions, $N_{l}/N$, which
characterise the vacuum. We will restrict our attention to the 
simplest possibility, where the filling fractions approach 
constant values as $N\rightarrow \infty$. 
\paragraph{}
To analyse the resulting spectrum, it 
is simplest to start our discussion by considering the weak-coupling
regime $\Lambda<<M$. In this case the effective theory at energy
scales below $M$ is ${\cal
 N}=1$ SUSY Yang-Mills with gauge group $\hat{G}_{cl}=
\prod_{l=1}^{m} U(1)_{l}\times SU(N_{l})$. 
The large-$N$ limit defined above involves 
scaling each $N_{l}$ like $N$ and also holds the corresponding dynamical
scales $\Lambda_{(l)}$ fixed.  Thus our chosen limit reduces at low
energies to the standard 't Hooft limit of the ${\cal N}=1$ SUSY 
Yang-Mills theory in each non-abelian factor
of $\hat{G}_{cl}$. The theory in the $SU(N_{l})$ factor 
has a mass gap of order $\Lambda_{(l)}\sim N^{0}$
and the spectrum includes an infinite tower of weakly-interacting 
stable glueballs with masses which remain fixed as $N\rightarrow 
\infty$. As before, the spectrum 
also includes the $m$ massless photons of the unconfined gauge-group 
$\hat{G}=U(1)^{m}$.  
\paragraph{}
The remaining degrees of freedom of the full $U(N)$ gauge theory 
acquired masses of order $M$ in the classical theory. These include 
states $Q_{rs}$ which transform in the $({\bf N}_{r},{\bf N}_{s})$
representation of $SU(N_{r})\times SU(N_{s})\subset \hat{G}_{cl}$ 
and carry electric charges $(+1,-1)$ under the unconfined abelian subgroup 
$U(1)_{r}\times U(1)_{s}\subset \hat{G}$. From the point of view 
of the low-energy gauge-group factors $SU(N_{r})$ and $SU(N_{s})$ 
these states are heavy diquarks with large bare masses of order $M>>\Lambda$. 
The abelian subgroups 
$U(1)_{r}$ and $U(1)_{s}$ correspond to a gauged version of the 
corresponding baryon number symmetries.    
At low-energies, the diquarks will be confined to form singlets of 
$SU(N_{r})\times SU(N_{s})$, 
including dimesons $Q_{rs}\bar{Q}_{rs}$ and dibaryons
$Q_{rs}Q_{rs}\ldots Q_{rs}$ where the minimum number, $N_{rs}$, 
of diquarks needed to form a
colour-singlet is the lowest common multiple of the integers 
$N_{r}$ and $N_{s}$.
\paragraph{} 
In the weak-coupling limit, 
the properties of the dimesons and dibaryons can be understood using the
standard large-$N$ analysis of QCD with massive quarks reviewed above. 
In particular, as the dynamics of different non-abelian factors are 
effectively decoupled, their contributions to dibaryon and dimeson
masses are additive. Thus 
we deduce that these states have masses of order $N$ and $N^{0}$
respectively. The three-point coupling of any two dibaryons charged under 
a $U(1)_{l}\subset \hat{G}$ with a single glueball of $SU(N_{l})$ 
is of order $N^{0}$. The only subtlety that arises concerns dimeson 
interactions. In particular, a three-point self-coupling for dimesons of 
$SU(N_{r})\times SU(N_{s})$ gains a suppression factor 
$\sim 1/\sqrt{N}$ from each factor and thus scales like $1/N$ just
like the three-point glueball self-couplings. Similar counting leads
to the conclusion that the large-$N$ properties of dimesons are
essentially the same as those of glueballs\footnote{This reflects the 
fact that the theory contains only adjoint fields, which means that
(in the absence of D-branes) the dual string theory contains only
closed strings. In particular both glueballs and dimesons correspond
to closed string excitations.}.                
\paragraph{}
The classical theory also contained 
magnetic monopoles of mass $\tilde{M}\sim M/g^{2}_{0}$. 
If we focus on the vacuum with $b_{1}=b_{2}=\ldots=b_{m}=0$ 
where confinement is due to magnetic condensation in each factor, 
these states will remain unconfined. However their masses will be
renormalised in the quantum theory. 
We can estimate the renormalised mass by replacing the bare
coupling $g_{0}^{2}$ by the running coupling $g^{2}(\mu)$ evaluated 
at the scale $\mu=M$ of classical symmetry breaking. Further renormalisation 
can occur due to strong coupling effects in the IR which produce 
additive correction of order $\Lambda_{(l)}$. As 
$g^{2}(M)$ goes like $1/N$ in the large-$N$ limit defined above, we deduce
that the masses of magnetically charges states grow linearly with
$N$. As discussed above, the masses of electrically charged dibaryons
also grow with $N$ as $N\rightarrow \infty$. There are other interesting
parallels between the large-$N$ behaviour of magnetic monopoles and of
dibaryons \cite{witbar1}. Monopoles are classical field configurations 
whose size and shape are independent of $g_{0}^{2}$ and thus of $N$. 
Interactions between elementary quanta and these classical solitons also
remain unsuppressed as $N\rightarrow \infty$. In the following we will
see that magnetic monopoles and dibaryons appear on an equal footing
in the dual string theory as wrapped/stretched D-branes. We will also
come across dyonic particles corresponding to bound-states of these
objects, as well as interesting 
low-energy electric-magnetic dualities which permute all of these states. 
For ease of reference, we will sometimes refer to the $\hat{G}$-charged states
collectively as dibaryons. 
\paragraph{}
So far our discussion of the large-$N$ spectrum has been limited 
to the weak-coupling limit
$\Lambda<<M$. Moving away from this limit, we can no longer 
rely on the low-energy description in terms of a $\hat{G}_{cl}=
\prod_{l=1}^{m}U(N_{l})$ gauge theory. In general states can only 
be distinguished 
by their quantum numbers under the true low-energy gauge group 
$\hat{G}=U(1)^{m}$. Thus, as we increase the ratio $\Lambda/M$ the
$\hat{G}$-neutral glueballs and dimesons described above will mix 
and it no longer makes sense to distinguish between them. 
On the other hand the electric and magnetic dibaryons 
are still distinguished by their charges under $\hat{G}$.
In the following we will see that the following qualitative 
aspects of the large-$N$ spectrum described above persist beyond the
weak-coupling coupling limit, 

{\bf 1:} States neutral under $\hat{G}$ have masses of order $N^{0}$
and interactions (with other neutral states) suppressed by powers of
$1/N$. 

{\bf 2:} States carrying electric and/or magnetic charges under
$\hat{G}$ have masses which grow linearly (or faster) with $N$. 

{\bf 3:} Interactions between $\hat{G}$-charged and $\hat{G}$-neutral
states remain unsuppressed at large-$N$. 

\subsection{F-terms and the large-$N$ limit}
\paragraph{}
In this subsection we will consider the exact 
F-term effective action (\ref{fterm0},\ref{fterm1}) of Dijkgraaf and Vafa 
in the context of the large-$N$ limit.  
The effective action is written in terms of chiral superfields $S_{l}$
and $w_{\alpha l}$ which are defined as gauge-invariant single-trace  
operators in (\ref{gidef}). 
It will also be convenient to define component fields for each of
these superfields, 
\begin{eqnarray}
S_{l} & = & s_{l}+\theta_{\alpha}\chi^{\alpha}_{l}+ \ldots \nonumber \\ 
w_{\alpha l}& = & \lambda_{\alpha l} +\theta_{\beta}f^{\beta}_{\alpha
  l}+ \ldots 
\label{comp}
\end{eqnarray}
The component fields,  $s_{l}$ and $f_{l}$ 
are bosonic single trace operators while  
$\chi_{l}$ and $\lambda_{l}$ are fermionic single trace operators. 
In the large-$N$ limit, these operators should 
create bosonic and fermionic
colour-singlet single particle states respectively. 
It is instructive to consider the interaction vertices for these
fields contained in the F-term effective action. Expanding 
(\ref{fterm0}) in components we find terms like,  
\begin{equation} 
{\cal L}^{(2)}_{F}= \int\, d^{2}\theta\, W_{eff}^{(2)} \supset 
V^{(2)}_{ij}f^{i}_{\alpha\beta}f^{\alpha\beta j}  + 
V^{(3)}_{ijk} \chi^{i}_{\alpha}f^{\alpha\beta j}\lambda^{k}_{\beta}
+V^{(4)}_{ijkl}\chi^{i}_{\alpha}\chi^{\alpha j}
\lambda^{k}_{\beta}\lambda^{\beta l} 
\label{vijk}
\end{equation}
where, 
\begin{equation}
V^{(L)}_{i_{1}i_{2}\ldots i_{L}} =\left\langle 
\frac{\partial^{L} {\cal F}}{\partial S_{i_{1}} \partial S_{i_{2}}
\ldots\partial S_{i_{L}}}\right \rangle 
\label{vp}
\end{equation}
for $L=2,3,4$. 
We will also consider the 
$2$-point vertex coming from the
glueball superpotential $W_{eff}^{(0)}$,
\begin{equation}
{\cal L}^{(0)}_{F}= \int\, d^{2}\theta\, W_{eff}^{(0)} \supset 
H^{(2)}_{ij}\chi_{\alpha}^{i}\chi^{\alpha j}
\end{equation}
where 
\begin{equation}
H^{(2)}_{ij}=\left\langle 
\frac{\partial^{2} W_{eff}^{(0)}}{\partial S_{i} \partial S_{j}}
\right \rangle 
\end{equation}
The matrix $H^{(2)}_{ij}$ therefore 
effectively determines the masses of the chiral multiplets $S_{l}$.   
\paragraph{}
In the large-$N$ limit, the leading
contribution to these vertices should come from
planar Feynman diagrams. By the standard counting arguments reviewed in
Section 3.1, the planar contribution to the L-point vertex $V^{(L)}$
should scale like $N^{2-L}$. Similarly the two-point vertex $H^{(2)}$
scales like $N^{0}$. For these F-terms, the form of the perturbation series 
is also constrained by holomorphy and dimensional analysis \cite{DVPW,CDSW}. 
To simplify the analysis, we will focus on 
a particular submanifold in parameter space defined by 
$g_{1}={\cal U}$, $g_{l}=0$ for $l>1$,
where the superpotential (\ref{sp1}) preserves a ${\bf Z}_{m}$
symmetry. In this case, the 
series of planar contributions to the three-point
coupling has the general form, 
\begin{equation}
V^{(3)}_{ijk}\sim \frac{1}{\varepsilon {\cal U}^{\frac{m+1}{m}}} 
\, \sum_{r=0}^{\infty} C^{(ijk)}_{i_{1}i_{2}\ldots
  i_{r}}\hat{S}_{i_{1}}\hat{S}_{i_{2}}\ldots \hat{S}_{i_{r}}
\label{series1}
\end{equation}  
where $\hat{S}_{l}=(S_{l}/\varepsilon {\cal
  U}^{\frac{m+1}{m}})$. The $r$'th term in the series (\ref{series1})
  comes from a sum of planar diagrams with $r+2$ loops. 
To keep track of the $N$ dependence recall 
that $\varepsilon \sim N$, ${\cal  U}\sim N^{0}$ and that the 
expectation value of a single-trace operator $S_{l}$ scales like $N$. 
Thus we confirm that the 
 planar contribution to $V^{(3)}_{ijk}$ scales like $1/N$ as 
expected for a vertex coupling three single-trace operators.  
\paragraph{}
The exact formulae (\ref{fterm1},\ref{spectral},\ref{ssd}) of
Dijkgraaf and Vafa 
effectively determine the coefficients $C^{(ijk)}$ in the expansion 
(\ref{series1}). 
From our point of view these results
have two remarkable features. The first is simply that, for 
these special F-term quantities, the full sum over planar diagrams can
be performed. This is the first example where this is possible in an
interacting four-dimensional field theory. In particular, such a
resummation for generic observables in the same theory is still far out of
reach. The simplification can be
understood as a cancellation of four-dimensional kinematic factors in
each planar diagram reducing the calculation to a sum over the 
planar Feynman diagrams of an associated zero-dimensional matrix model.  
The second feature is that the results are given exactly by the
planar contribution even for finite $N$. Thus there are no $1/N$
corrections to the leading order result. This feature relies
heavily on holomorphy and is therefore also special to the F-term
observables. The absence of corrections at any order in the $1/N$
expansion\footnote{This fact is usually emphasized by introducing a new
variable $\hat{N}$ for the dimension of 
the matrices of the associated matrix model which is different from
$N$. The sum over zero-dimensional planar diagrams is then generated
by considering an $\hat{N}\rightarrow \infty$ limit with $N$ fixed.}
means that the Dijkgraaf-Vafa results hold equally well at finite $N$. 
In this paper we are mainly interested in the large-$N$ limit and 
will mainly exploit Dijkgraaf and Vafa's results as providing examples where
the leading order can be computed exactly. 
\subsection{Large-$N$ string dual}
\paragraph{}
A large-$N$ string theory dual for the model with generic 
superpotential (\ref{sp1}) has been proposed by Cachazo, Intriligator 
and Vafa \cite{intv}. 
In their construction the $U(N)$ gauge theory in 
question is realised by wrapping $N$ D5 branes on topologically 
non-trivial two-spheres in a certain  
Calabi-Yau three-fold. In particular, the second homology group of
the three-fold contains a basis of $m$ non-intersecting 
two-cycles each homeomorphic to $S^{2}$. Labelling these two-spheres as
$S^{2}_{l}$ for $l=1,2,\ldots,m$, the classical vacuum with low energy 
gauge group $\hat{G}_{cl}=\prod_{l=1}^{m}U(N_{l})$ is realised by 
wrapping $N_{l}$ D5 branes on $S^{2}_{l}$ for each $l$.  
The couplings $\{g_{l}\}$ 
appearing in the superpotential (\ref{sp1}) correspond to geometrical 
parameters controlling the complex structure of the Calabi-Yau
manifold. As noted above, the theory described by the superpotential 
(\ref{sp1}) is non-renormalisable, and the string theory construction 
implicitly provides some UV completion of the model. In this case the 
UV theory is related to the six-dimensional theory living on the five-branes. 
Unwanted additional degrees of freedom appear 
in the form of Kaluza-Klein modes of mass $\sim 1/R$ 
arising from compactification of the
six-dimensional theory on an $S^{2}$ of radius $R$. As in other examples 
\cite{malnun}, $R$ needs to be {\em large} in string units ensure 
a large hierarchy between the UV scale $M_{UV}\sim R^{-1}$ and the 
scale $\Lambda$ of four-dimensional gauge dynamics\footnote{For the
  theory of $N$ D5 branes wrapped on a two-sphere of radius $R$ we
  find that $\Lambda/M_{UV}\sim \exp(-4\pi^{2}/Ng^{2}_{UV})$ where the UV
  gauge coupling $g^{2}_{UV}$ is proportional to $\alpha'/R^{2}$}.  
\paragraph{}
The large-$N$ duality of \cite{intv}, corresponds to a geometric
transition in which the D5 branes disappear and are replaced by 
flux. More precisely, the two-spheres $S^{2}_{l}$ on which the branes
are wrapped are blown down to zero size, while $m$ non-intersecting 
three-spheres $S^{3}_{l}$, are blown up to finite size. The resulting
Calabi-Yau geometry is defined by the complex equation, 
\begin{equation}
u^{2}+v^{2}+w^{2}+W'(x)^{2}-f(x)=0
\label{geom3}
\end{equation}
Here $f(x)$ is the same function appearing in the definition 
(\ref{spectral}) of the spectral curve $\Sigma$. 
Indeed the three-fold (\ref{geom3}) can be thought of as fibration of
an $a_{1}$ singularity over the Riemann surface $\Sigma$. In this
context, the homology three-sphere $S^{3}_{l}$ corresponds to the product
of the unique compact two-cycle of the $a_{1}$ singularity, with the
one-cycle $A_{l}$ on $\Sigma$ defined above.  The resulting
closed string background includes $N_{l}$ units of Ramond-Ramond
three-form flux through this cycle. 
In the absence of flux, the low-energy theory contains an 
abelian vector multiplet of ${\cal N}=2$ SUSY associated with each  
three-cycle $S^{3}_{l}$. The scalar component of each vector 
multiplet is the holomorphic volume of the corresponding three-cycle.  
Including RR flux through the basis three-cycles introduces a
superpotential which breaks the ${\cal N}=2$ supersymmetry down to an 
${\cal N}=1$ subalgebra. As in the field theory, the remaining
massless fields comprise $m$ abelian 
vector multiplets of ${\cal N}=1$ SUSY.     
\paragraph{}     
In principle, the geometric transition described above provides  
a dual closed string theory formulation of the gauge theory in its
partially confining phase. Unfortunately, the resulting geometry is
complicated because of the presence of strong background RR fields due to
the flux through the three-spheres. The corresponding
solutions of the supergravity field equations are unknown and would
probably be irrelevant anyway due to the presence of large $\alpha'$ 
corrections. For this reason, applications of the geometric transition 
have mainly been confined to determining F-terms which are insensitive 
to the presence of background flux. 
\paragraph{}
Despite the problems described above, we may still make a few qualitative 
remarks about the dual string theory and compare it with our
expectations of the large-$N$ field theory. The first point, emphasized in 
\cite{vafa}, is that the large-$N$ duality implied by the geometric
transition is of a standard form. In particular, the dual string
coupling $g_{s}$ scales like $1/N$ as anticipated by 't Hooft, while 
the string mass scale $M_{s}=1/\sqrt{\alpha'}$ is identified with 
the dynamical scale $\Lambda$ of the gauge theory. Thus, as in other
confining gauge theories, we should identify the excited states of the
dual string with the tower of stable glueballs which is
expected at large $N$. 
\paragraph{}
Other features of the large-$N$ gauge theory
discussed above are also visible in the string theory dual. In
particular, we can identify states which are electrically and/or 
magnetically charged with respect to the unconfined $U(1)$'s in the
low-energy gauge group. These states correspond to wrapped D3 branes
around some of the non-contractible three-cycles. 
If a three-cycle has $n$ units 
of threeform flux, a wrapped D3 brane by itself is not consistent with 
charge conservation \cite{witbar2}. Instead, well-known selection
rules demand that $n$ fundamental strings must end on the wrapped
brane. The resulting object corresponds to a baryon vertex in the 
dual field theory. The basis cycle 
$S^{3}_{l}$ introduced above carries $N_{l}$ units of threeform flux. 
A D3 brane wrapped around $S_{l}$, corresponds to the baryon vertex 
of the $SU(N_{l})$ factor in the low-energy gauge group 
$\hat{G}_{\rm cl}$. However a dibaryon of $SU(N_{r})\times SU(N_{s})$
can be realised by wrapping $N_{rs}/N_{r}$ D3 branes on $S^{3}_{r}$ and 
$N_{rs}/N_{s}$ anti-D3 branes on $S^{3}_{s}$ 
where, as above, $N_{rs}$ is the lowest common 
multiple of $N_{r}$ and $N_{s}$. To accomodate the selection rule
described above, we must also join the two sets of wrapped branes with 
$N_{rs}$ fundamental strings. The mass of the resulting dibaryon then
scales like $N_{rs}$ as expected from field theory. The 
large-$N$ scaling of glueball-dibaryon
interactions obtained by field theory arguments in the previous
section also follows immediately we identify the dibaryon as a 
D-brane.   
\paragraph{}
In addition to the basis cycles $S^{3}_{l}$ there 
are also additional compact cycles in the geometry. These include
linking cycles with 
intersection numbers $(+1,-1)$ with respect to pairs of the basis 
cycles $S^{3}_{l}$. A D3 brane wrapped on a cycle $S^{3}_{r,s}$ linking 
$S^{3}_{r}$ and
$S^{3}_{s}$ corresponds to a monopole carrying magnetic charges
$(+1,-1)$ under $U(1)_{r}\times U(1)_{s}\subset \hat{G}$. 
As before such a wrapped brane is only allowed to exist in isolation
if there is no flux through the linking cycle. As shown in \cite{CSW}, 
the number of units of flux through each linking cycle is
determined by the integers $b_{l}$ which measure the 
differences between the modes of confinement in each vacuum. In
particular, the number of units of flux through $S^{3}_{r,s}$ is equal to 
$b_{r}-b_{s}$. Thus a wrapped brane is allowed only if $b_{r}=b_{s}$  
which is true only in a vacuum where the modes of confinement in 
the non-abelian factors $SU(n_{r})$ and $SU(n_{s})$ are the same. 
This is simply the condition that the corresponding 
magnetic (or dyonic) state remains unconfined in the dual field
theory. The masses of the wrapped D-branes scale like 
$1/g_{s}\sim N$ giving the expected large-$N$ scaling for magnetic monopoles.  
The fact that dibaryons and magnetic monopoles both correspond to
wrapped D-branes also accounts for the similarities between the two
types of states noted above.   
\paragraph{}
Finally it is also straightforward to check the formula of \cite{CSW} 
for the confinement index $t$ in a generic vacuum quoted in the previous
Section. The integer $t$ is the smallest positive integer such that an
external charge in the 
representation of $U(N)$ formed from the product of $t$ copies of the
fundamental representation can be screened. In the string dual 
such an external charge can be formed from $t$ fundamental strings 
ending on the boundary (ie at infinity). Screening will occur if 
we can find a configuration of wrapped branes with 
finite energy on which these strings can end. To achieve this we wrap 
$k_{l}$ D3 branes around each basis cycle $S^{3}_{l}$ and $m_{l}$ D3
branes around the linking cycle $S^{3}_{l,l+1}$. Here negative values
of $k_{l}$ or $m_{l}$ correspond to anti-D3 branes. In addition to 
the $t$ external strings ending on 
the branes we can also join branes wrapping different cycles with 
additional fundamental strings. This can be done provided that we can
choose the integers $k_{l}$ and $m_{l}$ such that,
\begin{equation}
\sum_{l=1}^{m} \, k_{l}N_{l}+ m_{l}b_{l}=t \,\,\,\, {\rm mod}\, N
\end{equation} 
The smallest positive integer $t$ satisfying this requirement is 
the greatest common divisor of the $N_{l}$ and the $b_{l}$. 
Thus we find the expected value of the confinement index \cite{CSW}.    

\section{A Critical  Point}
\paragraph{}
So far we have discussed the theory at generic points in a partially
confining phase. In these generic vacua, the masses of all states
charged under the unconfined gauge group $\hat{G}$ grow linearly (or faster)
with $N$. As $N\rightarrow \infty$ these charged states decouple
leaving a conventional large-$N$ spectrum of $\hat{G}$-neutral
glueballs. In particular, the standard scaling law (\ref{scaling}) for
the correlators of single trace operators holds. Equivalently,
glueballs are free at $N=\infty$ and their interactions can be
calculated in a $1/N$ expansion. In this Section 
we will consider critical points in the parameter space where this 
simple picture breaks down.  
\paragraph{}
As we move around in the space of complex parameters the masses of
the electrically charged dibaryons and of the magnetic 
monopoles will change. A natural possibility to
consider is that of singular points where one or more of these states
becomes massless. In the context of the
string dual described in the previous section, the charged states
correspond to D3 branes wrapped around non-contractible three-cycles 
in the geometry, specifically those cycles which do not have RR flux 
through them. The states in question become massless when the
corresponding cycles shrink to zero size. These states are not BPS and 
we do not know their exact masses. However ${\cal N}=1$ 
F-terms in these theories can be computed exactly for any values of
the parameters and these include the low-energy $U(1)$ couplings. 
As in theories with ${\cal N}=2$ supersymmetry, the
presence of light charged states can be detected via the 
resulting logarithmic running of these couplings.

%%%%%%%%%%%%%%%%%%%%%%%%%%%%%%%%%%%%%%%%
\vspace{1.0cm}
\begin{figure}[ht]
\begin{center}
\input{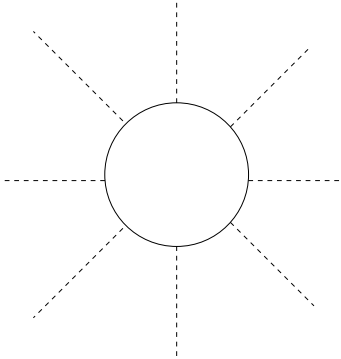}
\caption{ \label{figloop} 
A one-loop contribution to an L-point interaction
vertex (for $L=8$).}
\end{center}
\end{figure}
\vspace{0.5cm}
%%%%%%%%%%%%%%%%%%%%%%%%%%%%%%%%%%%%%%%%

\paragraph{}
Another consequence
of light charged states is that the simple picture of the large-$N$
limit decribed above no longer holds. This is particularly clear when we
recall that interactions between glueballs and $\hat{G}$-charged states remain
unsuppressed in the large-$N$ limit. Noting the
result of the previous section that the three-point coupling $V$ of a
single glueball to two charged states is of order $N^{0}$, we may 
estimate the contribution of loops of charged states to
the interaction vertices of $L$ glueballs. A one-loop contribution to
the $L$-point interaction vertex with
fermionic charged states of mass $M_{B}\sim N$ running around the 
loop is shown in Figure \ref{figloop}. Without taking into account the
index structure of the vertices or the constraints of supersymmetry we
will obtain a rough estimate of this contribution which will suffice
to illustrate the main points. We evaluate 
the loop integral with a UV
cut-off $M_{UV}\sim N^{0}$. The behaviour at large-$N$ is
given by,  
\begin{equation}
V^{L}\int_{p^{2}<M^{2}_{UV}} d^{4}p\, {\rm tr} 
\left(\frac{\gamma_{\mu}p^{\mu}+M_{B}}{p^{2}+M_{B}^{2}}\right)^{L}
\sim V^{L} M_{B}^{L}\int_{0}^{M_{UV}^{2}} \frac{x
  dx}{(x+M_{B}^{2})^{L}}\sim M_{UV}^{4}\left(\frac{V}{M_{B}}\right)^{L}
\label{loop}
\end{equation} 
As $M_{B}\sim N$, this is consistent with the standard large-$N$
counting rule which says that each additional glueball adds one power
of $1/N$. However, we can also see that the loop contribution (\ref{loop}) will
diverge as we approach a critical point where $M_{B}$ vanishes. 
The divergence grows worse as we increase $L$, violating standard
large-$N$ scaling. This is our first indication that the conventional 
$1/N$ expansion breaks down at the critical point. 
\paragraph{}
As mentioned above, the presence of critical points where
new massless charged states appear can be detected using the exact
F-term action of Dijkgraaf and Vafa. In this paper, we will 
focus on critical points where the dual string background 
(\ref{geom3}) develops a generalised conifold singularity of the form 
(\ref{uvw}). In the following, we will relate the 
violation of large-$N$ scaling
described above to the well-known breakdown of 
closed string perturbation theory at the singularity.    
In the light of the relationship between the spectral
curve $\Sigma$ and the dual Calabi-Yau, the critical point 
corresponds to a point
where $\Sigma$ develops a singularity of the form $y^{2}=x^{m}$. 
Such points have been studied 
previously by several authors \cite{ferr1, eguchi, bert, shih}. 
The particular case relevant to our present study was considered in 
\cite{bert}.  
\paragraph{}
To identify and analyse the critical point, it is
useful to  proceed via two steps, 

{\bf 1} Off-shell critical points can be identified as points in
field space with coordinates where the F-term effective action
(\ref{fterm1}) becomes singular. This corresponds to finding special
values of the gluino superfields $\{S_{1},S_{2},\ldots,S_{m}\}$ 
where the spectral curve $\Sigma$ degenerates. 

{\bf 2} Having found the off-shell critical point, the second step is
to find a supersymmetric vacuum where the criticality is realised
on-shell. In other words we must find a supersymmetric vacuum where
the critical values of the gluino superfields are attained as vacuum
expectation values. 

\begin{figure}[ht]
\begin{center}
\input{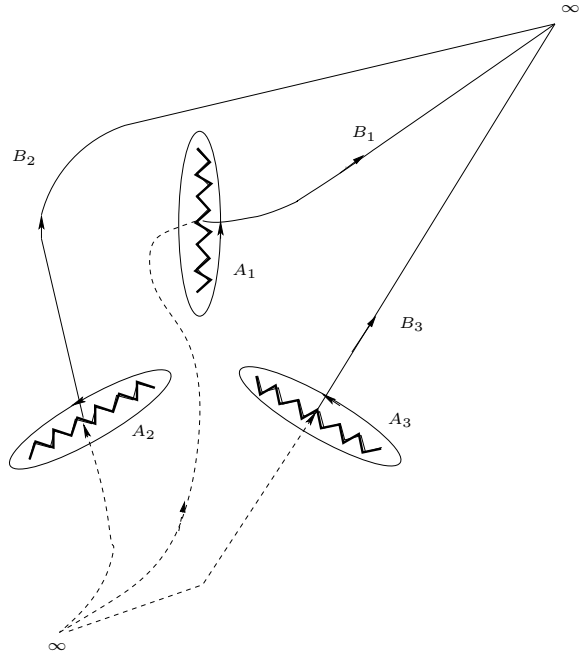}
\caption{ \label{fig2} The $A$ and $B$ cycles for $m=3$. 
}
\end{center}
\end{figure}

To exhibit the off-shell critical point,  
we will focus on the special case $g_{1}={\cal U}$ with
$g_{l}=0$ for $l>1$, where the superpotential 
(\ref{sp1}) has a  ${\bf Z}_{m}$ symmetry. It is also convenient to
restrict our attention to a particular submanifold in the field
space which respects this symmetry: we set 
$S_{l}=S\exp(2\pi il/m)$, for $l=1,2,\ldots,m$. 
This in turn is equivalent to the choice $\kappa_{l}=0$ for all
$l>1$, for the coefficients of the polynomial $f(x)$
in (\ref{spectral}). Ultimately, these choices will be realised on-shell
by stationarising the glueball superpotential.  
\paragraph{}
With these choices the spectral curve becomes, 
\begin{equation}
y^{2}=\varepsilon^{2}(x^{m}-{\cal U})^{2}-\kappa_{1}
\label{spec2}
\end{equation}
with branch points which lie at, 
\begin{equation}
x_{\pm}^{(l)}= 
\left( {\cal U}\pm \frac{\sqrt{\kappa_1}}{\varepsilon}\right)^{\frac{1}{m}}
e^{\frac{2\pi il}{m}} 
\end{equation}
The corresponding Riemann surface $\Sigma$ is shown for the case $m=3$
in Figure \ref{fig2}.

%%%%%%%%%%%%%%%%%%%%%%%%%%%%%%%%%%%%%%%%
\vspace{1.0cm}
\vspace{0.5cm}
%%%%%%%%%%%%%%%%%%%%%%%%%%%%%%%%%%%%%%%%
\paragraph{}
The critical point of interest occurs 
for $\sqrt{\kappa_{1}}=\pm {\cal U}\varepsilon$. 
Choosing the positive
sign, we see that the $m$
branch points $x^{(l)}_{-}$, for $l=1,2,\ldots,m$, collide at the
origin and the curve becomes 
\begin{equation}
y^{2}=\varepsilon^{2}x^{m}(x^{m}-2{\cal U})
\label{spec8}
\end{equation}
which has the desired singularity of the form $y^{2}\sim x^{m}$ at the origin.
Using the first equation in (\ref{ssd}), 
we find that the 
gluino superfields take the values $S_{l}=S_{cr}\exp(2\pi il/m)$ with 
\begin{equation}
S_{cr}=d_{m}\varepsilon {\cal U}^{\frac{m+1}{m}}
\end{equation}
at the critical point. The overall numerical prefactor is given by, 
\begin{equation}
d_m = - \frac{\Gamma(\frac{m+2}{2m})}
{\Gamma(\frac{1}{2m}) \Gamma(\frac{3}{2}+\frac{1}{2m})} \ .
\end{equation} 
\paragraph{}
The next step is to find SUSY vacua in which the
critical value of the glueball superfields is attained as a vacuum
expectation value. To do this we must minimise the full effective
superpotential, 
\begin{equation}
W^{(0)}_{eff}=\sum_{l=1}^{m} N_{l}\frac{\partial {\cal F}}{\partial S_l}
+ 2 \pi i \tau_{0} \sum_{l=1}^{m} S_l + 2\pi i \sum_{l=1}^{m} b_l S_l
\label{sp3}
\end{equation}
As explained above the integers $N_{l}$ determine the pattern of classical
symmetry breaking while the integers $b_{l}$ correspond to a choice of
vacuum in the quantum theory. 
The problem of minimising the superpotential (\ref{sp3}) is fully 
equivalent to that of finding a factorisation of the associated 
${\cal N}=2$ curve \cite{intv} of the form, 
\begin{equation}
y^{2}_{{\cal N}=2}=P_{N}(x)^{2}-4\Lambda^{2N}=H_{N-m}(x)^{2}y^{2}
\label{neq2}
\end{equation}
where $P_{N}$ and $H_{N-m}$ are polynomials of degree $N$ and  $N-m$ 
repectively and, as above, $y$ is given by equation (\ref{spectral}) which
defines the ${\cal N}=1$ spectral curve $\Sigma$. Once a factorisation
has been found, the values of $N_{l}$ and $b_{l}$ in the corresponding
vacuum state are then determined by the periods of the holomorphic
differential, 
\begin{equation}
T=\left\langle {\rm
  Tr}_{N}\left[\frac{dx}{x-\Phi}\right]\right\rangle=\frac{P_{N}'(x)}{
\sqrt{P_{N}(x)^{2}-4\Lambda^{2N}}}\, dx
\label{tdef}
\end{equation}
as,  
\begin{eqnarray}
N_{l}=\,\oint_{A_{l}}\, T & \qquad{} & -\tau_{0}-b_{l}=\oint_{B_{l}}\, T 
\label{nT}
 \end{eqnarray}
Note that the periods of $T$ around the compact cycles $A_{l}$ and 
$B_{k}-B_{l}$ are integer-valued as expected. This follows immediately
from the relation \cite{CSW},
\begin{equation}   
T= - d \log 
\left(P_{N}(x) - \sqrt{ P_{N}(x)^{2}-4\Lambda^{2N} }\right)
\end{equation}
In the language of the dual string theory, 
the periods of $T$ around any given one-cycle on $\Sigma$ measure 
the number of units of RR three-form flux through the corresponding
three-cycle of the Calabi-Yau geometry (\ref{geom3}). The first
relation in (\ref{nT}) confirms that there are $N_{l}$ units of flux
through the basis three-cycles $S^{3}_{l}$. The second relation tells
us that the RR flux through the compact linking cycles discussed in
Section 3.4 is determined by the integers $b_{l}$.  
\paragraph{}
The problem of factorising the curve to find an on-shell realisation
of the critical behaviour described above was solved in
\cite{eguchi,bert}. The desired factorisation can only be obtained
when $N$ is divisible by $m$. In this case we set 
$\kappa_{1}=4\varepsilon^{2}\Lambda^{2m}$ in 
(\ref{spec2}) with $\kappa_{l}=0$ for $l>1$ as before. Thus the
spectral curve becomes, 
\begin{equation}
y^{2}=\varepsilon^{2}(x^{m}-{\cal U})^{2}-4\varepsilon^{2}\Lambda^{2m}
=\varepsilon^{2}(x^{m}-{\cal U}-2\Lambda^{m})(x^{m}-{\cal
  U}+2\Lambda^{m})
\label{spec3}
\end{equation}
The on-shell curve becomes singular at ${\cal U}=\pm
 2\Lambda$ and coincides with the critical curve
 (\ref{spec8}). In particular, choosing the positive sign we find the curve,  
\begin{equation}
y^{2}=\varepsilon^{2}x^{m}(x^{m}-4\Lambda^{m})
\label{degen}
\end{equation}
with the singularity $y^{2}\sim x^{m}$ at the origin. 
\paragraph{}
The values of the integers $N_{l}$ and $b_{l}$ corresponding to this
vacuum can then be computed
using (\ref{nT}). In this way we obtain 
$N_{1}=N_{2}=\ldots N_{m}=n$ where $N=mn$ and 
$b_{1}=b_{2}=\ldots b_{m}=0$. 
The first condition means that the
unbroken classical gauge group in this vacuum 
$\hat{G}_{cl}$ is $U(n)^{m}$ and the second implies that the mode of
confinement in the non-abelian part of each factor is the same. 
By the formula of section 2, the confinement index in this vacuum is 
given by $t=n$. It is striking that the critical curve 
(\ref{spec8},\ref{degen}) can only be realised on-shell in this
specific vacuum. In particular, the restriction on $b_{l}$ is
essential to ensure that the charged states which become massless at
the critical point are present in the theory as unconfined particles. 
\paragraph{}
In the context of ${\cal N}=2$ supersymmetry,
the singular curve (\ref{degen}) corresponds to a special point on the
Coulomb branch of ${\cal N}=2$ SUSY Yang-Mills theory with gauge group 
$SU(m)$ where a number of mutually non-local electric and magnetic
charges become simultaneously massless (for $m>2$). In this vacuum the theory
flows to the $a_{m-1}$ Argyres-Douglas superconformal fixed point \cite{AD}. 
The present context is somewhat different. We are considering a $U(N)$ 
gauge theory in a partially confining phase with ${\cal N}=1$ supersymmetry. 
The degeneration of the spectral curve which
controls the coupling constants of the low-energy gauge group 
$\hat{G}=U(1)^{m}$ is consistent with the presence of the same set of
massless electric and magnetic charges which appear at the
Argyres-Douglas fixed point. In our case these states are identified
as electric dibaryons and their magnetic (and dyonic) counterparts as
discussed above. It has been proposed \cite{eguchi} that the resulting
${\cal N}=1$ theory flows to the {\em same} ${\cal N}=2$ 
superconformal fixed point in the IR. Evidence in favour of this
includes the spectrum of chiral operator dimensions, essentially 
determined by the curve (\ref{degen}), which matches that 
of the Argyres-Douglas CFT. Although we believe this proposal is 
plausible we will not assume it to be true in the following. 
\paragraph{}
The above points are further illuminated by considering the behaviour
of the dual string geometry at the critical point. For ${\cal
  U}=2\Lambda^{m}$ (\ref{geom3}) becomes    
\begin{equation}
u^{2}+v^{2}+w^{2}+\varepsilon^{2}x^{m}(x^{m}-4\Lambda^{m})= 0 
\label{geom2}
\end{equation}     
which exhibits a generalised conifold singularity of type $a_{m-1}$ at 
the point $u=v=w=x=0$. The case of ${\cal N}=2$ supersymmetry
described above corresponds to considering IIB strings on the
singular Calabi-Yau geometry (\ref{geom2}) {\em without flux}. 
The low-energy effective action for string theory in
this background includes the ${\cal N}=2$ Argyres-Douglas
SCFT living in the remaining four flat dimensions of the ten
dimensional spacetime. As above, the massless electric and magnetic
states correspond to D3 branes wrapped on three-cycles which vanish at
the critical point. 
The resulting four-dimensional superconformal theory can be decoupled from
gravity by taking the limit $g_{s}\rightarrow 0$. We will review the
standard analysis of this flux-free case in more detail in Section 6.1 below.  
\paragraph{}
As in our field theory discussion, the present context is somewhat
different. In particular the geometry includes non-trivial RR flux
through certain three-cycles. To be precise, as we have $N_{l}=n=N/m$ for
all $l$, the basis three-cycles
which project to the one-cycles $A_{l}$ on $\Sigma$ each have exactly 
$n$ units of flux. This flux introduces a superpotential for the
moduli of the Calabi-Yau which breaks the spacetime supersymmetry to 
${\cal N}=1$ in four dimensions. On the other hand, 
as the integers $b_{l}$ vanish in our chosen
vacuum, there is no flux through the compact three-cycles which project
to compact one-cycles of the form $B_{k}-B_{l}$ on $\Sigma$. In
particular, this means that there is no flux through the cycles which
vanish at the critical point\footnote{This can also be understood by
  noting that a vanishing cycle with a
  non-zero quantised flux automatically leads to to a divergence in the RR
  field strength. This would correspond  to a divergence in the
  superpotential and its derivatives which is not consistent with a
  SUSY vacuum for finite values of the couplings.}. As mentioned above 
this means that the theory contains unconfined 
charged states corresponding to D3 branes wrapped on these cycles. 
This in turn is consistent with the appearance at the critical point 
of the same set of massless charges associated with the ${\cal
  N}=2$ Argyres-Douglas fixed point as suggested above.     

\section{The Double-Scaling Limit}
\paragraph{}
In the previous section we gave a general argument that the $1/N$ expansion
should break down at critical points where the electric and magnetic 
dibaryon become massless. We can now illustrate this breakdown by
considering the behaviour of the planar diagram expansion (\ref{series1}) for 
the three-point coupling $V^{(3)}$ as we approach the 
critical point. As above we begin by considering the
off-shell behaviour on the submanifold in field space defined by
setting $S_{l}=S\exp(2\pi il/m)$, for $l=1,2,\ldots,m$. 
In this case (\ref{series1}) simplifies to, 
\begin{equation}
V^{(3)}_{ijk}=\frac{1}{\varepsilon{\cal U}^{\frac{m+1}{m}}}\, 
\sum_{r=0}^{\infty} c^{(r)}_{ijk} \left(\frac{S}
{\varepsilon{\cal U}^{\frac{m+1}{m}}}\right)^{r}
\label{series2}
\end{equation}
with coefficients $c^{(r)}_{ijk}$ coming from summing the planar
diagrams with $r+2$ loops. 
The exact answer for the sum (\ref{series2}) is then determined by
equations (\ref{vp}), (\ref{ssd}) and (\ref{spec2}). Near the critical
point we obtain
\footnote{The formulae (\ref{scr1}) and (\ref{scr2}) below are 
meant to indicate the leading behaviour of the 
most divergent elements of the tensor $V^{(3)}_{ijk}$. We will give much
more precise results in the next Section.}, 
\begin{eqnarray}
V^{(3)}_{ijk} &\sim & \frac{1}{\varepsilon{\cal U}^{\frac{m+1}{m}}} 
\left(1- \frac{1}{d_{m}}\left(\frac{S}
{\varepsilon{\cal U}^{\frac{m+1}{m}}}\right)\right)^{-1} \nonumber \\ 
& \sim & \frac{1}{S-S_{cr}}
\label{scr1}      
\end{eqnarray}
As above we can realise this behaviour on-shell by setting 
$\kappa_{1}=4\varepsilon^{2}\Lambda^{2m}$. The critical behaviour is
then attained at ${\cal U}=2\Lambda^{m}$. Near this point in parameter
space, (\ref{scr1}) is equivalent to, 
\begin{equation}
V^{(3)}_{ijk}\sim \frac{1}{\varepsilon\Lambda^{\frac{m}{2}}
({\cal U}-2\Lambda^{m})^{\frac{m+2}{2m}}}
\label{scr2}
\end{equation}
\paragraph{}
These results indicate that the series of planar diagrams
(\ref{series1}) has a finite radius of convergence and, in particular,
it diverges at the critical point $S_{l}=S_{cr}\exp(2\pi i l/m)$. 
This type of behaviour is very familiar in the context of matrix
models and is hardly surprising given that the planar diagrams
contributing to the F-term couplings reduce to those of an auxiliary matrix
model. However, the important point is that the divergence reflects
the presence of new light states at the critical point with couplings
to the glueballs which remain unsuppressed in the large-$N$ limit. 
This means that similar divergences should also occur in generic 
correlation functions of single trace operators. 
\paragraph{}
In the context of zero dimensional matrix models, the type of critical
behaviour described above can be used to define a double-scaling
limit (for a review see \cite{GZJ}). Near the critical
point, the sum of planar diagrams is dominated by large diagrams with
many vertices. Related divergences also occur in the sum over diagrams of
each fixed genus corresponding to contributions at each order in the
$1/N$ expansion. By taking $N\rightarrow \infty$ and simultaneously
tuning the parameters to approach the critical point, one can cure all
of these divergences simultaneously and obtain a coherent sum over
large diagrams of each genus. The sum over different genera 
is then controlled by a renormalised expansion parameter $\kappa$
which is held fixed as $N\rightarrow \infty$. A characteristic
feature of the resulting double-scaled theories is the existence of 
universality classes associated with each critical point. After taking
the limit, detailed information about the definition of the
theory is lost and seemingly different models become equivalent.    
\paragraph{}
Double-scaled matrix models also have an interesting interpretation in terms of
a dual string theory. In this context individual Feynman diagrams 
correspond to discretisations of the string world-sheet. The
double-scaling limit is then equivalent to a continuum limit on 
the worldsheet. In many cases, this yields a 
continuum non-critical string theory.  Typically 
the dilaton grows in the Liouville direction 
but this growth is cut-off by the presence
of a potential term in the world sheet action (the ``Liouville
wall''). The resulting effective 
string coupling is identified  
with the double scaling parameter $\kappa$. In the following, we 
will define a corresponding double-scaling limit for our 
four-dimensional theory where very similar features emerge. 
\paragraph{}
In the standard large-$N$ 't Hooft limit discussed in Section 3, the
interaction vertices of $L$ single trace fields, go to zero like $N^{2-L}$.  
Motivated by the above discussion, we can instead define a large-$N$ 
double-scaling limit where these interactions are held fixed. 
As before we take $N\rightarrow \infty$ with $\varepsilon\sim N$ and 
$\Lambda$ fixed. The new ingredient is that we simultaneously take the
limit ${\cal U}\rightarrow 2\Lambda^{m}$ with, 
\begin{equation}
\Delta=\varepsilon({\cal U}-2\Lambda^{m})^{\frac{m+2}{2m}} 
\end{equation}
held fixed. From (\ref{scr2}), we see that the three-point vertex
$V^{(3)}_{ijk}\sim (1/\Lambda^{m/2}\Delta)$ remains fixed in this
limit. We will analyse the effect of this limit on the rest of the F-term 
effective action below. After allowing for the generalisation (\ref{dscale})
discussed below, this is the unique limit which leads to a finite
non-zero interaction between colour-singlet states and also keeps the
mass scale of the glueball spectrum fixed. As discussed above, the
breakdown of the $1/N$ expansion can also be understood as a
consequence of the presence of light $\hat{G}$-charged states near the
critical point. In particular the loop contribution (\ref{loop}) 
suggests that the effective coupling for glueball interactions is 
proportional to $1/M_{B}$ where $M_{B}$ 
is the mass of the lightest charged dibaryon. Combining these
observations, we conclude that the double-scaling limit proposed above 
also has the effect of keeping $M_{B}$ constant. In the following,
several checks of the consistency of this conclusion will emerge. 
\paragraph{}
In the context of zero dimensional matrix models, one of the key
features of the double-scaling limit is that 
it yields finite non-zero contributions at each genus. This should also be true
in four-dimensions if the limit defined above holds the dibaryon mass
fixed as we claim. 
Ideally one could check this at each order by considering higher genus
corrections to the F-term effective action. However the
effective superpotential is saturated by planar diagrams and it
receives no $1/N$ corrections. On the other hand, such corrections do
appear when the four-dimensional field theory is coupled to gravity
and a background field-strength for the graviphoton is turned on. 
These corrections are in turn related to the higher genus corrections 
to the free energy of the auxiliary matrix model. Although we will not
discuss this in detail here, the above discussion suggests that the
limit defined above should correspond to a double-scaling limit of the 
auxiliary matrix model which yields a finite contribution at each
genus. This is hard to verify directly, as very
little is known about higher genus corrections for the multi-cut
configurations of the complex matrix model which are relevant 
to the critical point of interest. On the other hand, the conjectured
duality between matrix models and topological string theory developed
in \cite{DV2,DV3,DVPW} relates these corrections to the higher-genus 
contributions to the B-model on the Calabi-Yau geometry
(\ref{geom3}). Possible tests of our proposal along these lines will
be discussed elsewhere. Previous work \cite{DV2,DVPert2,ferr1} 
on double-scaling limits of the Dijkgraaf-Vafa matrix model and its
four-dimensional interpretation has focussed on cases which can be
analysed using known results for one- and two-cut solutions of the
hermitian matrix model. The analysis of the two-cut solution given in 
the Appendix of \cite{DVPert2} should be relevant to the special case
$m=2$ of our analysis. 
\paragraph{}
It will also be useful to generalise the double-scaling limit slightly
by restoring
the full set of superpotential couplings $g_{l}$ appearing in
(\ref{sp1}) for $l=1,2,\ldots, m$. As before the critical point lies
at $g_{1}={\cal U}=2\Lambda^{m}$ with $g_{l}=0$ for $l>1$. 
We approach the critical point by taking the limit 
$N\rightarrow\infty$ and $\varepsilon \rightarrow \infty$  with 
$\delta=(g_{1}-2\Lambda^{m})\rightarrow 0$
and $g_{l}\rightarrow
 0$. Apart from $\Lambda$, the quantities held fixed are, 
\begin{eqnarray}
 \tilde{\varepsilon}=\varepsilon/N \qquad{} 
\Delta=\varepsilon\delta^{\frac{m+2}{2m}} & \,\,\, {\rm and} \,\,\, & 
\tilde{g}_{l}=g_{l}/\delta^{\frac{m-l+1}{m}} \qquad{} {\rm  
for}\,\,\,\, l=2,\ldots, m 
\label{dscale}
\end{eqnarray}
In this more general limit, the three-point function has the behaviour, 
\begin{equation}
V^{(3)}_{ijk}\sim \frac{1}{\Lambda^{\frac{m}{2}}\Delta}\, 
v^{(3)}_{ijk}(\tilde{g}_{l})    
\end{equation}
The tensor $v^{(3)}_{ijk}$ is a function of the additional 
double-scaling parameters $\tilde{g}_{l}$ which will be determined
explicitly below. 
\paragraph{}
In order to understand the behaviour of the theory in the
double-scaling limit described above, it is convenient to exploit the  
electric-magnetic duality invariance of the F-term action
(\ref{fterm1}). In particular, the ${\cal N}=1$ superspace 
action $W_{eff}^{(2)}$ for the low-energy abelian vector multiplets 
$w_{\alpha l}$ is invariant under the $Sp(2m; {\bf Z})$ low-energy S-duality
transformations which are familiar from the study of ${\cal N}=2$
supersymmetric gauge theory \cite{SW1}. As we review below, this 
invariance extends to the full
F-term action with an appropriate transformation law for the integers
$N_{l}$ and $b_{l}$ appearing in $W_{eff}^{(0)}$. For technical reasons we
will restrict our attention to the case of odd $m$ setting $m=2p+1$. 
We will comment on the case of even $m$ at the end of this Section.   
\paragraph{}
In terms of the spectral curve $\Sigma$, these electric-magnetic duality
transformations correspond to our freedom in choosing a basis of
homology cycles with the canonical intersection properties
(\ref{intersect}). Starting from the basis of cycles, $A_{l}$ and
$B_{l}$, with $l=1,2,\ldots, m$, defined in the previous section, we
can find a new canonical basis with cycles $\tilde{A}_{l}$ and
$\tilde{B}_{l}$ where, 
\begin{equation}
\left(\begin{array}{c} \tilde{B} \\   \tilde{A}
\end{array}\right)= M\cdot  \left(\begin{array}{c} B \\   A
\end{array}\right)  
\end{equation}
for $M\in Sp(2m;{\bf Z})$. 
\paragraph{}
We will choose to work in a very specific basis which we now
describe. We define basis cycles, 
\begin{eqnarray}
\tilde{A}_{l}=A_{-}^{(\kappa)} & \qquad{} & \tilde{B}_{l}=B_{-}^{(\kappa)} 
\qquad{} l=\kappa=1,2,\ldots,p \nonumber \\ 
       =A_{+}^{(\kappa)} & \qquad{} &\,\,\,\,\,\,\,\, =B_{+}^{(\kappa)} 
\qquad{} l=p+\kappa=p+1,p+2,\ldots,2p \nonumber \\ 
       =A_{\infty} & \qquad{} & \,\,\,\,\,\,\,\, =B_{\infty} 
\qquad{} l=2p+1 \nonumber 
\end{eqnarray} 
Apart from having canonical intersections the new basis cycles also
have the following properties, 

{\bf 1} The cycles $A_{-}^{(\kappa)}$ and $B_{-}^{(\kappa)}$ for
$\kappa=1,2,\ldots,p$ are compact cycles which vanish at the critical
point. 

{\bf 2} The cycles $A_{+}^{(\kappa)}$ and $B_{+}^{(\kappa)}$ for
$\kappa=1,2,\ldots,p$ are compact cycles which have zero intersection
with all the vanishing cycles. 

{\bf 3} The compact cycle $A_{\infty}=A_{1}+A_{2}+\ldots+A_{m}$
corresponds to a large circle in the complex $x$-plane enclosing all
of the branch cuts. $B_{\infty}$ is a non-compact cycle which
necessarily has zero intersection with all the vanishing cycles. 

%%%%%%%%%%%%%%%%%%%%%%%%%%%%%%%%%%%%%%%%
\vspace{1.0cm}
\begin{figure}[ht]
\begin{center}
\input{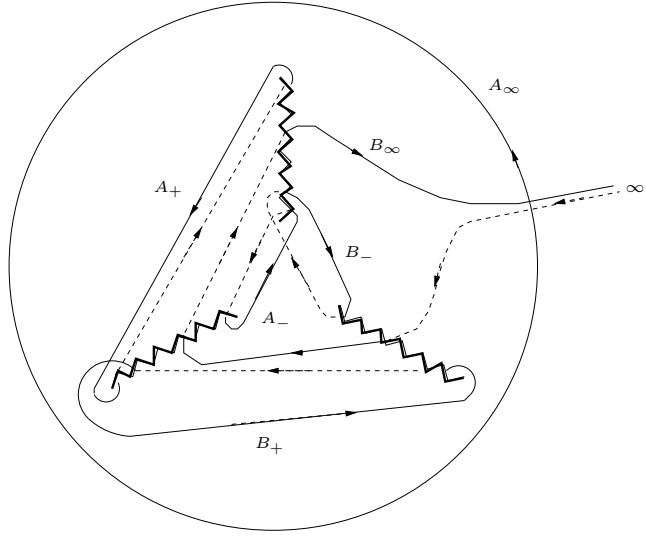}
\caption{ \label{fig3}
The new basis for $m=3$}
\end{center}
\end{figure}
\vspace{0.5cm}
%%%%%%%%%%%%%%%%%%%%%%%%%%%%%%%%%%%%%%%%

The new basis is illustrated for the case $m=3$ in Figure \ref{fig3}. 
The corresponding $Sp(6;{\bf Z})$ transformation is given in the
Appendix. It should
be clear by inspection that such a choice is possible for any odd
value of $m$. In terms of the dual string theory, the $Sp(2m;{\bf Z})$
transformation corresponds to a change of basis for the third homology
group of the Calabi-Yau manifold (\ref{geom3}). In the new basis, the
compact three-cycles corresponding to $A^{(\kappa)}_{-}$ and 
$B^{(\kappa)}_{-}$, for $\kappa=1,2,\ldots,p$ 
all vanish at the generalised conifold singularity, while the
remaining three-cycles are non-vanishing.  
\paragraph{}
The change of basis described above induces a non-trivial
transformation of the glueball superfields $S_{l}$ and the conjugate 
variables $S^{D}_{l}$. These quantities are given in (\ref{ssd}) above
as $A$- and $B$-periods of the holomorphic differential $ydx$. Thus we
have, 
\begin{equation}
\left(\begin{array}{c} \tilde{S}^{D} \\   \tilde{S}
\end{array}\right)= M\cdot  \left(\begin{array}{c} S^{D} \\   S
\end{array}\right)  
\end{equation} 
where, 
\begin{eqnarray}
\tilde{S}_{l}=\,\oint_{\tilde{A}_{l}}\, ydx & \qquad{} & 
\tilde{S}^{D}_{l}=\frac{1}{2\pi i}\frac{\partial\tilde{\cal F}}
{\partial \tilde{S}_{l}}=\,\oint_{\tilde{B}_{l}}\, ydx 
\label{ssd2}
\end{eqnarray}
This transformation allows us to 
rewrite the F-term effective action in terms of the new fields, 
\begin{eqnarray}
\tilde{S}_{l} & = & -\frac{1}{2\pi i}
\oint _{\tilde{A}_{l}} \, dx\, \frac{1}{32\pi^{2}} 
{\rm Tr}_{N}\left[ \frac{W_{\alpha}W^{\alpha}}{x- \Phi}\right]
\end{eqnarray}
The resulting action for the dual vector multiplets 
$\tilde{w}_{\alpha l}$ is determined by a dual prepotential 
${\tilde {\cal F}}(\tilde{S}_{1},\ldots,\tilde{S}_{m})$ which is
defined by the second equality in (\ref{ssd2}). The corresponding 
contribution to the superpotential is,  
\begin{equation}
W_{eff}^{(2)}    = 
 \frac{1}{2}\sum_{k,l} 
\frac{\partial^{2}\tilde{\cal F}}{ \partial \tilde{S}_{k} \partial 
\tilde{S}_{l}}
\,\tilde{w}_{\alpha k}\tilde{w}^{\alpha}_{l} 
\label{fterm1b}
\end{equation}
After expanding the action in
component fields as in (\ref{vijk}), we find $L$-point
interaction vertices,  
\begin{equation}
\tilde{V}^{(L)}_{i_{1}i_{2}\ldots i_{L}} =\left\langle 
\frac{\partial^{L} \tilde{\cal F}}{\partial \tilde{S}_{i_{1}} \partial 
\tilde{S}_{i_{2}}
\ldots\partial \tilde{S}_{i_{L}}}\right \rangle 
\label{vp2}
\end{equation}
for $L=2,3,4$. 
\paragraph{}
To rewrite the glueball superpotential $W_{eff}^{(0)}$ 
in terms of the new variables, we need
to consider the action of $Sp(2m;{\bf Z})$ on the integers $N_{l}$ and
$b_{l}$ corresponding to our choice of vacuum. The
quantities $N_{l}$ and $N^{D}_{l}=-\tau_{0}-b_{l}$ are given by periods of the
holomorphic differential $T$ and therefore transform as,  
\begin{equation}
\left(\begin{array}{c} \tilde{N}^{D}\\   \tilde{N}
\end{array}\right)= M\cdot  \left(\begin{array}{c} N^{D} \\   N
\end{array}\right)  
\label{sp2m}
\end{equation} 
where,
\begin{eqnarray}
\tilde{N}_{l}=\,\oint_{\tilde{A}_{l}}\, T & \qquad{} & 
\tilde{N}^{D}_{l}= -\tilde{b}_{l}=\,\oint_{\tilde{B}_{l}}\, T 
\, 
\qquad{}
l=1,\ldots,2p
\label{nT3}
\end{eqnarray} 
\begin{eqnarray}
\tilde{N}_{2p+1}=\,\oint_{\tilde{A}_{\infty}}\, T = N & \qquad{} & 
\tilde{N}^{D}_{2p+1}= - \tilde \tau_0 -\tilde{b}_{2p+1}
=\,\oint_{\tilde{B}_{\infty}}\, T 
\, 
\end{eqnarray} 
The bare coupling transforms as $\tilde \tau_0 = \tau_0 + q$ for some
integer $q$, which corresponds to an irrelevant shift of the bare
vacuum angle by a multiple of $2\pi$. 
The glueball superpotential has the form of a simplectic product
between period vectors and is therefore 
invariant under the $Sp(2m;{\bf Z})$ duality transformation. In the new
basis it can be written as, 
\begin{eqnarray} 
W_{eff}^{(0)}   & = 
& \sum_{l=1}^{2p}\, 
\left(
\tilde{N}_{l}\frac{\partial{\tilde{\cal F}}}{\partial 
\tilde{S}_{l}} 
+ 2\pi i \tilde{b}_l \tilde{S}_{l}
\right)
+ \tilde N_{2p+1} \frac{\partial{\tilde{\cal F}}}{\partial 
\tilde{S}_{2p+1}}  
+ 2\pi i ( \tilde \tau_0 + \tilde{b}_{2p+1} )
\tilde{S}_{2p+1}
\label{sp7}
\end{eqnarray}
The complete effective superpotential in the new basis is given by the
sum of (\ref{fterm1b}) and (\ref{sp7}).  
\paragraph{}
We now present our results for the behaviour of the F-term 
couplings in the double-scaling limit. Details of the calculation are
given in the Appendix.  
We begin by considering the two point coupling $\tilde{V}^{(2)}_{ij}$ which
determines the matrix of low-energy abelian gauge couplings. As above,
this object is a generalised period matrix for the (non-compact)
Riemann surface $\Sigma$. In the double-scaling limit, we find that the period
matrix assumes the following block-diagonal structure, 
\begin{equation}
\tilde{V}^{(2)}= \left( \begin{array}{cc} \Pi^{-} & 0 \\ 0 & \Pi^{+}
\end{array} \right)
\end{equation}
Here $\Pi^{-}$ is a $p\times p$ matrix-valued function of the moduli
$\tilde{g}_{l}$ where, as above, $p=(m-1)/2$. This matrix can be
identified with the period matrix of the {\em reduced spectral curve}
$\Sigma_{-}$. The latter is a compact Riemann surface of genus $p$
described by the complex equation, 
\begin{equation}
{\cal Y}^{2}={\cal X}^{m}-\sum_{l=2}^{p}\tilde{g}_{l}{\cal X}^{l-1} - 1  
\label{sigmaminus}
\end{equation}
\paragraph{}
The remaining $q\times q$ block $\Pi^{+}$ (with $q=m-p$) can be
thought of as the generalised period matrix of the non-compact Riemann
surface $\Sigma_{+}$ described by the curve, 
\begin{equation}
y^{2}=\varepsilon^{2}(x^{m}-4\Lambda^{m})
\label{sigmaplus}
\end{equation}
%%%%%%%%%%%%%%%%%%%%%%%%%%%%%%%%%%%%%%%%
\vspace{0.5cm}
\begin{figure}[ht]
\begin{center}
\input{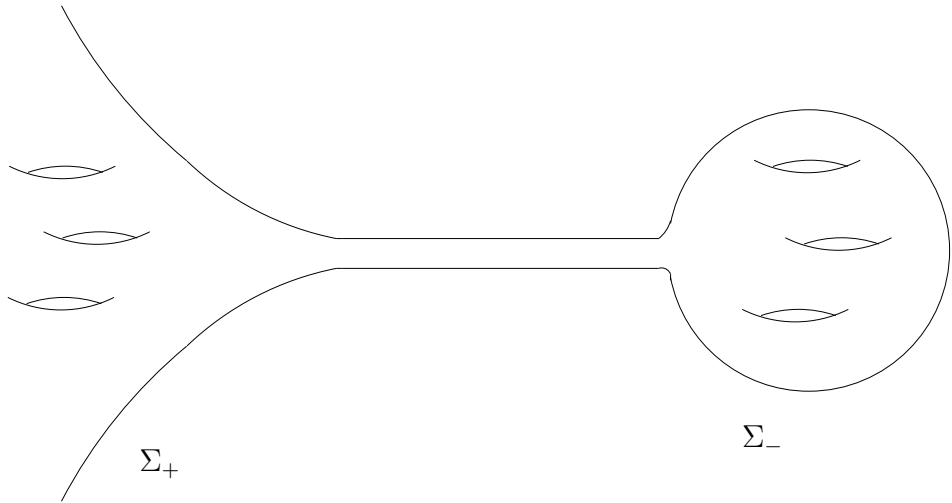}
\caption{The degeneration of the spectral curve $\Sigma$ in the
  double-scaling limit.}
\end{center}
\end{figure}
\vspace{0.5cm}
%%%%%%%%%%%%%%%%%%%%%%%%%%%%%%%%%%%%%%%%
\paragraph{}
The block diagonalisation of the generalised period matrix described
above corresponds to a particular degeneration of the spectral curve
$\Sigma$ occurring in the double-scaling limit. A simple intuitive
picture of the degeneration is that $\Sigma$ decomposes into a union
of the two Reimann surfaces $\Sigma_{-}$ and $\Sigma_{+}$ defined
above joined together by a long, thin tube as shown in Figure 5. 
A similar degeneration in the case $m=3$ was discussed in \cite{AD}. 
Such an interpretation must be treated with care as it refers 
to a particular choice of 
metric on the surface $\Sigma$. At this point we know the complex
structure of $\Sigma$ which only determines the
metric up to conformal transformations. Despite this, we will argue in
the next section that the picture of the Riemann surface degenerating
into two sectors linked by a long tube or throat is directly relevant
in the context of the string theory dual.       
\paragraph{}
The couplings in the F-term effective action reflect the splitting of
the period matrix described above. In particular, the three-point
coupling $\tilde{V}^{(3)}_{ijk}$ is only non-zero in the
double-scaling limit if all three indices
are constrained to take values less than or equal to $p=(m-1)/2$. Thus
we find that, 
\begin{equation}
\tilde{V}^{(3)}_{ijk}= \frac{1}{\Lambda^{\frac{m}{2}}\Delta} 
v_{\alpha\beta\delta}(\tilde{g}_{l})
\label{scr3}
\end{equation}
for $i=\alpha=1,2,\dots,p$, $j=\beta=1,2,\dots,p$,
$j=\delta=1,2,\dots,p$ and vanishes identically in the double-scaling
limit otherwise. Similar results apply to the four-point coupling 
$\tilde{V}^{(4)}$. 
\paragraph{}
It is also interesting to consider the behaviour of the glueball
superpotential $W_{eff}^{(0)}$, as given in (\ref{sp7}), in the 
double-scaling limit. This
depends explicitly on the integers $\tilde{N}_{l}$ and $\tilde{b}_{l}$ 
which determine our choice of vacuum.  
These may be evaluated by calculating the 
periods of the differential form $T$ on the new basis cycles 
$\tilde{A}_{l}$ and $\tilde{B}_{l}$. Alternatively, we can apply the
$Sp(2m;{\bf Z})$ transformation (\ref{sp2m}) to the corresponding
integers $N_{1}=N_{2}=\ldots=N_{m}=n$ and $b_{1}=b_{2}=\ldots=b_{m}=0$ 
in the original basis. By either method we find that  
$\tilde{N}_{l}=\tilde{b}_{l}=0$ for $l=1,2,\ldots, p$. In terms of the
string dual, this is equivalent to the statement that there is 
no RR flux through those three-cycles which vanish at the generalised
conifold singularity. Thus (\ref{sp7}) simplifies to become,  
\begin{eqnarray} 
W_{eff}^{(0)}   & = 
& 2\pi i \sum_{l=p+1}^{2p}\, \left(\tilde{N}_{l}\tilde{S}^{D}_{l} 
+ \tilde{b}_l \tilde{S}_{l}
\right)
+ 2\pi i\tilde N_{2p+1}\tilde{S}^{D}_{2p+1}  
+ 2\pi i ( \tilde \tau_0 + \tilde{b}_{2p+1} )\tilde{S}_{2p+1}
\label{spreduced}
\end{eqnarray}
\paragraph{}
In equation (\ref{spreduced}), 
the superpotential is expressed purely in terms of periods of $y
dx$ around one-cycles which have zero intersection with all the
vanishing one-cycles. In terms of the diagonal decomposition of the
period matrix illustated in Figure 5, these correspond to periods of 
the surface 
$\Sigma_{+}$. As one might expect, these have only weak dependence on
the moduli of the surface $\Sigma_{-}$. 
%This point is made precise in
%the Appendix, where we show explicitly that the partial derivatives 
%$\partial W^{(0)}_{eff}/\partial\tilde{S}_{l}$ vanish off-shell 
%in the double-scaling limit for $l\leq p$.         
This is illustrated by the behaviour of the two-point coupling, 
\begin{equation}
\tilde{H}^{(2)}_{ij}=\frac{\partial W^{(0)}_{eff}}{\partial
  \tilde{S}_{i}\partial \tilde{S}_{j}}
\label{hess2}
\end{equation}
which is given by,  
\begin{equation} 
\tilde{H}^{(2)}_{ij}=\frac{n}{\varepsilon{\cal U}^{\frac{1}{2}}} 
\, h^{(2)}_{AB}
\label{hess3}
\end{equation}
for $i=p+A=p+1,\ldots 2p+1$,  $j=p+B=p+1,\ldots 2p+1$ and vanishes
identically in the double scaling limit otherwise. Here $h^{(2)}_{AB}$
is a constant $q\times q$ matrix. Thus $\tilde{H}^{(2)}$ 
has precisely the opposite behaviour of the three-point
coupling, namely it is only non-zero in the $q\times q$ block
corresponding to the period matrix $\Pi^{+}$. 
\paragraph{}
With some 
additional assumptions, the vanishing of the complimentary $p\times p$ 
block corresponding to $\Pi_{-}$ indicates the emergence of $p$ 
massless glueballs in the double-scaling limit. In particular,
if we make the standard assumption that the glueball fields $\tilde{S}_{l}$
can be treated as independent fields with non-singular kinetic terms 
then we can infer that the theory contains $p$ neutral massless
chiral multiplets of ${\cal N}=1$ supersymmetry in the double-scaling
limit. The occurrence of a massless glueball in the case $m=2$ was
noted in \cite{ferr1}. Beyond this, one can also check that all higher
derivatives of the superpotential where even one of the derivatives is
with respect to a field ${\tilde S}_{l}$ with $l\leq p$ vanish in the 
double-scaling limit. This means that all the F-term interaction vertices
for these fields coming from the glueball superpotential 
vanish in the double-scaling limit.         
\paragraph{}
There is a simple way to summarise the results described in this 
section which is inspired by the intuitive picture of the
spectral curve $\Sigma$ degenerating into curves $\Sigma_{+}$ and
$\Sigma_{-}$ joined by a long thin tube as shown in Figure 5. 
The main points are, 

{\bf 1} The scaling of the couplings
in the F-term effective action are consistent with the existence of two
sectors ${\cal H}_{\pm}$ in the Hilbert space of the theory which
decouple in the large-$N$ double-scaling limit. 
One sector, ${\cal H}_{-}$, contains $p=(m-1)/2$ massless $U(1)$
vector multiplets of ${\cal N}=1$ SUSY, $\tilde{w}_{\alpha l}$, 
together with $p$ neutral
chiral multiplets $\tilde{S}_{l}$ for $l=1,2,\ldots, p$. 
Our result (\ref{hess3}) indicates that the chiral
multiplets also become massless in the double-scaling limit. Standard
large-$N$ reasoning suggests that this sector should also contain many 
additional massive degrees of freedom. 
These include for example, infinite towers
of massive single-particle states
created by the single trace operators $\tilde{S}_{l}$. 
The other sector ${\cal H}_{+}$ contains the remaining $q=m-p$
massless $U(1)$ vector multiplets as well as other massive degrees of 
freedom.    

{\bf 2} Interactions between colour-singlet states in the sector
${\cal H}_{-}$ are controlled by the coupling,  
\begin{equation}
\frac{1}{N_{eff}}\sim \frac{1}{\Delta}
\end{equation}
which is held fixed in the double-scaling limit\footnote{To make the coupling
dimensionless we could instead write $1/N_{eff}\sim 
\Lambda^{3-\frac{m}{2}}/\Delta$ where $\sim$ denotes equality up to an
unknown dimensionless constant of order $N^{0}$}. These interactions also
depend on the remaining double-scaling parameters $\tilde{g}_{l}$ for
$l=2,\ldots p$. Interactions between colour-singlet states in the sector
${\cal H}_{+}$ are suppressed by powers of $1/N$ as in a conventional
large-$N$ 't Hooft limit. 
Interactions between states in the sector ${\cal H}_{+}$ and those in
the sector ${\cal H}_{-}$ also go to zero as $N\rightarrow\infty$. 

{\bf 3} In the double-scaling limit, 
the F-term effective action for the fields in sector ${\cal H}_{-}$ is
determined by the reduced spectral curve $\Sigma_{-}$ defined in 
(\ref{sigmaminus}) above. In particular we can find a canonical basis of
one-cycles ${\cal A}_{l}$ and ${\cal B}_{l}$ for $l=1,2,\ldots p$ such
that,  
\begin{eqnarray}
\tilde{S}_{l}=2i\Lambda^{\frac{m}{2}}\Delta 
\,\oint_{{\cal A}_{l}}\, {\cal Y}d{\cal X} & \qquad{} & 
\tilde{S}^{D}_{l}=\frac{1}{2\pi i}\frac{\partial{\cal F}_{-}}
{\partial \tilde{S}_{l}}=  
2i\Lambda^{\frac{m}{2}}\Delta\,\oint_{{\cal B}_{l}}\, {\cal Y}d{\cal X}
\label{ssd3}
\end{eqnarray} 
which defines a reduced prepotential ${\cal
  F}_{-}[\tilde{S}_{1},\ldots,\tilde{S}_{p}]$. The complete F-term
  action for the sector ${\cal H}_{-}$ in the double-scaling limit 
can then be written as, 
\begin{equation}
{\cal L}^{-}_{F}= {\rm Im} \left[\int\, d^{2}\theta \,\sum_{k,l\leq p}
\frac{1}{2}\frac{\partial^{2}{\cal F}_{-}}
{\partial\tilde{S}_{k}\partial\tilde{S}_{l}}
\tilde{w}_{\alpha k}\tilde{w}^{\alpha}_{l} \,\,\right]
\label{result}
\end{equation}
Note, in particular that the effective superpotential for
the massless fields $\tilde{S}_{l}$, $l\leq p$, is exactly zero
reflecting the exact vanishing of the interaction vertices for these
fields described above.    
\paragraph{}
In this Section we have restricted our attention to the case of odd
$m$. The additional complication of the case of even $m$ stems from
the presence of an additional holomorphic differential which is 
log-normalisable at the singularity. In this case we find an
interacting sector of $p=m/2-1$ massless vectors and $p$ massless neutral
chirals\footnote{As above we define $p=[(m-1)/2]$ for all values of
  $m$.}. The non-interacting sector contains 
$q=m-p$ massless free vectors together with $q-1$ massive 
chirals. The theory also contains one extra massless chiral multiplet
which is also free. Note that the interacting sector has no massless
fields for the
special case $m=2$ which corresponds to the ordinary conifold. These results
will be reported in more detail elsewhere. 
    
\section{The Dual String Theory}
\paragraph{}
In this section we will interpret the results of our field theory
calculation in terms of the dual string theory. As a preliminary, 
we will begin by reviewing some standard facts about Type IIB string theory on
the generalised conifold geometry without flux.
  
\subsection{The generalised conifold without flux}
\paragraph{}
The generalised conifold geometry has the form, 
\begin{equation}
U^{2}+V^{2}+W^{2}+Z^{m}=0
\label{conifold1}
\end{equation} 
String amplitudes on this background are
singular at each order in string perturbation theory. 
The singularity shows up in a linear $\sigma$-model formulation of the 
world-sheet theory \cite{Phases}, 
where a new non-compact branch opens up apparently 
leading to an infinite number of massless states. When
non-perturbative effects are included the picture is somewhat
different. As we now review, the singularity can be understood in terms
of a finite number of massless states corresponding to D3 branes
wrapped on vanishing three-cycles \cite{Strom}.     
\paragraph{}
It is convenient to consider first the non-singular
geometry obtained when we slightly deform the complex structure of 
the Calabi-Yau (\ref{conifold1}) to\footnote{We are using conventions
  where the dimensionless space-time coordinates $U$, $V$, $W$ and $Z$ measure
  distance in units of the string length
  $\sqrt{\alpha'}$. Consequently the deformation parameters $\mu_{l}$
  are also dimensionless.},   
\begin{equation}
F(U,V,W,Z)=U^{2}+V^{2}+W^{2}+Z^{m}-\sum_{l=1}^{p}\mu_{l}Z^{l-1}=0
\label{conifold2}
\end{equation}
where $p=[(m-1)/2]$ as above. The
restriction to monomials $Z^k$ with $k\leq p$ 
corresponds to considering only normalisable deformations of the
geometry \cite{Shapere}. Each normalisable deformation leads to an ${\cal N}=2$
vector multiplet in the 
low-energy four-dimensional effective theory. Thus the low-energy theory has
gauge group $U(1)^{p}$. The effect of this deformation is to blow up various
three-cycles to finite size. The theory now contains contains massive 
BPS states corresponding to D3 branes wrapped on these cycles. 
These states carry electric and magnetic 
charges under the low-energy $U(1)^{p}$ gauge symmetry. 
Their BPS mass formula has the form,  
\begin{equation}
M_{\rm D3} \sim \frac{1}{g_{s}\sqrt{\alpha'}}\left|\int_{S^{3}} \Omega \right|
\label{dthrees}
\end{equation}
where 
\begin{equation}
\Omega=\frac{dU dV dW dZ}{dF}
\end{equation}
is the holomorphic three-form of the Calabi-Yau
geometry (\ref{conifold2}). By setting $\mu_{l}=0$, the wrapped 
D3 branes become massless and we return to the generalised conifold
singularity (\ref{conifold1}). It is believed that the effect of the
light D-branes on the target space geometry is to replace the region 
close to the singularity by an infinite throat with a linear
dilaton \cite{Comments, OV2d}. 
As the dilaton grows, the effective string coupling becomes
large and string perturbation theory breaks down. 
\paragraph{}
Another effect
related to the appearance of an infinite throat is that the strongly
interacting massless degrees of freedom which appear at the
singularity can be decoupled from gravity. In particular, taking the   
limit $g_{s}\rightarrow 0$ makes the ten dimensional bulk theory away from the
singularity free. States localised at the singularity can retain finite
interactions in this limit precisely 
because of the linear growth of the dilaton in the throat. 
The resulting interacting decoupled theory includes the $p$ massless vector
multiplets and the mutually non-local (for $m>2$) 
charged states corresponding to 
the wrapped D3 branes. In additional to these massless modes, the
theory also contains a full set of massive 
stringy excitations corresponding to bound states of the 
fundamental IIB string localised at the singularity. The interactions of
these localised strings are described by a 
non-critical string theory in four dimensions without gravity, also
known as a Little String Theory (LST). Like the IIB background we
started with, the LST has ${\cal N}=2$ supersymmetry in four
dimensions. Far below the string scale, only the massless degrees of 
freedom are relevant 
and the interactions of the vector multiplets and mutually non-local
charges are described by the Argyres-Douglas 
${\cal N}=2$ superconformal field theory of type $a_{m-1}$. 
\paragraph{}
As in other examples, the dynamics of LST can be usefully
studied using holographic duality. In general, non-local theories
without gravity of this type are dual to ``ordinary'' 
critical string theory on a linear dilaton background \cite{holog}.    
In the present case, the LST of the generalised conifold is
holographically dual to the IIB string propagating in the
infinite throat region described above. This description is still of
limited use because the effective string coupling becomes large in one
region of the spacetime. However, as suggested in \cite{GKP,GK}, 
one may eliminate the
strong-coupling region by an appropriate deformation of the
string world-sheet theory. In fact, the required deformation simply
corresponds to reintroducing the deformation parameters $\mu_{l}$.   
As above this has the effect of giving non-zero masses to the BPS
states corresponding to wrapped D3 branes. One then takes a
double-scaling limit where $g_{s}\rightarrow 0$ and 
$\mu_{l}\rightarrow 0$ in such a way that the D3 brane masses are held
fixed.  
\paragraph{}
To describe the double scaling limit of Giveon and Kutasov in more
detail, it is 
convenient to rescale the spacetime coordinates in (\ref{conifold2})
according to,
\begin{equation}
U=\tilde{U}\mu_{1}^{\frac{1}{2}} \qquad{}
V=\tilde{V}\mu_{1}^{\frac{1}{2}} \qquad{} 
W=\tilde{W}\mu_{1}^{\frac{1}{2}} \qquad{}
Z=\tilde{Z}\mu_{1}^{\frac{1}{m}} \qquad{}
\end{equation}
In this way we find that the masses (\ref{dthrees}) 
of wrapped D3 branes are given by,  
\begin{equation}
M_{\rm D3}\sim \frac{\mu_{1}^{\frac{m+2}{2m}}}{g_{s}\sqrt{\alpha'}}\, 
\left|\int_{S^{3}} \tilde{\Omega} \right|
\end{equation}
where $\tilde{\Omega}=d\tilde{U} d\tilde{V} d\tilde{W}
d\tilde{Z}/d\tilde{F}$ is the holomorphic three-form of the rescaled
geometry, 
\begin{equation}
\tilde{F}(\tilde{U},\tilde{V},\tilde{W},\tilde{Z})=\tilde{U}^{2}+
\tilde{V}^{2}+\tilde{W}^{2}+\tilde{Z}^{m}-\sum_{l=2}^{p}\tilde{\mu}_{l}
\tilde{Z}^{l-1} - 1=0
\label{conifoldtilde}
\end{equation}
with $\tilde{\mu}_{l}=\mu_{l}/\mu_{1}^{\frac{m-l+1}{m}}$ for
$l=2,3,\ldots, p$.
\paragraph{}
To keep the masses of the wrapped D3 branes constant as $g_{s}\rightarrow 0$, 
we take the limit $\mu_{l}\rightarrow 0$, with the parameter, 
\begin{equation}
\kappa^{-1}=\frac{\mu_{1}^{\frac{m+2}{2m}}}{g_{s}}
\label{kappa}
\end{equation}
held fixed. We also keep, $\alpha'$ and the rescaled parameters 
$\tilde{\mu}_{l}$ fixed. The resulting Double-Scaled Little String
Theory has an effective string coupling constant 
$\kappa\sim 1/(M_{D3}\sqrt{\alpha'})$ which remains constant in the limit. 
\paragraph{}
One case where the dual string theory is particularly well understood
is the ${\bf Z}_{m}$ symmetric case \cite{GK} where $\tilde{\mu}_{l}=0$ for
$l=2,3,\ldots, p$. In this case the string world-sheet theory has the
form, 
\begin{equation}
R^{3,1}\times \left(\frac{SL(2)_{k}}{U(1)}\times LG(W)\right) /{\bf
 Z}_{m}
\label{background2}
\end{equation}
Here $LG(W)$ denotes the Landau-Ginzburg model with four chiral
superfields $U$, $V$, $W$ and $Z$ and superpotential, 
\begin{equation}
{\cal W}=U^{2}+V^{2}+W^{2}+Z^{m}
\end{equation}
The non-compact coset $SL(2)_{k}/U(1)$ corresponds to a semi-infinite
cigar geometry with a linear dilaton, 
\begin{eqnarray}
\phi = - \frac{Q}{2} \sigma &  \qquad{} & 
Q^2 = \frac{2}{k\alpha'} = \frac{m+2}{m\alpha'} \nonumber
\end{eqnarray}
The level $k$ of the coset appearing in (\ref{background2}) is given by 
\begin{equation}
k = \frac{2m}{m+2}
\end{equation}
The effective string coupling has an upper bound given
by its value at the tip of the cigar of order 
\begin{equation}
g_{cigar} \sim \kappa
\end{equation} 
\paragraph{}
In the more general case, with $\tilde{\mu}_{l}\neq 0$ the string world-sheet
theory is not known. However, it will be useful to describe the 
low-energy effective action for the massless degrees of freedom. As
above these are $p$ ${\cal N}=2$ vector multiplets 
corresponding to the normalisable complex structure
moduli of the Calabi-Yau. In the double-scaling limit where gravity
decouples, the low-energy effective action is completely determined by
rigid special geometry. The rescaled geometry (\ref{conifoldtilde}) 
corresponds to an $a_{1}$ singularity fibered over a complex curve $\Gamma$, 
\begin{equation}
\tilde{Y}^{2}=\tilde{Z}^{m}- \sum_{l=2}^{p}\tilde{\mu}_{l}\tilde{Z}^{l-1}-1 
\label{gam}
\end{equation}
The curve defines a double-cover of the complex $Z$ plane
corresponding to a Riemann surface of genus $p$. Choosing a 
basis of one-cycles $\alpha_{l}$ and $\beta_{l}$, for
$l=1,2,\ldots,p$, with the canonical intersection properties, the
vector multiplets $a_{l}$ and their duals can be expressed as, 
\begin{eqnarray}
a_{l}=\frac{1}{\kappa\sqrt{\alpha'}}\, \oint_{\alpha_{l}}\,
\tilde{Y}\,d\tilde{Z}
 & \qquad{} &  
a^{D}_{l}=\frac{1}{2\pi i}\frac{\partial {\cal G}}{\partial a_{l}}
=\frac{1}{\kappa\sqrt{\alpha'}}\, \oint_{\beta_{l}}\, {\tilde
  Y}\,d{\tilde Z}
\label{aad}
\end{eqnarray}
where the second relation defines the prepotential ${\cal G}$. 
The BPS mass formula for a wrapped D3 brane with electric/magnetic
charges $(\vec{n}_{E},\vec{n}_{M})$ is, 
\begin{equation}
M_{D3}=|n_{E}^{l}a_{l}+n_{M}^{l}a^{D}_{l}| 
\end{equation}
The low energy effective action for the vector multiplets is 
written as an F-term in ${\cal N}=2$ superspace, 
\begin{equation}
{\cal L}_{eff}={\rm Im} \left[ \int \, d^{4}\theta \,\, {\cal
    G}(a_{l}) \, \right]
\label{leffn2}
\end{equation}   
\paragraph{}
Finally, it will also be useful to rewrite the above results in the
language of ${\cal N}=1$ supersymmetry. In particular, we can
decompose the ${\cal N}=2$ vector multiplet $a_{l}$, into an ${\cal N}=1$
vector multiplet ${\cal W}_{\alpha l}$ and a neutral massless chiral multiplet 
$\varphi_{l}$ for $l=1,2,\ldots,p$. 
The low energy effective action (\ref{leffn2}), contains both D-terms and
 F-terms when rewritten in ${\cal N}=1$ superspace. 
After integrating out half the Grassmann coordinates we find,
\begin{equation}
{\cal L}_{eff}={\rm Im} \left[ \int \, d^{2}\theta \,\, \frac{1}{2}
\sum_{k,l\leq p}\frac{\partial^{2}{\cal
    G}(\varphi_{j})}{\partial \varphi_{k}\partial\varphi_{l}}\, 
{\cal W}_{\alpha k}{\cal W}^{\alpha}_{l} \, \right] \,\,\,\,\, +
\,\,\,\,\, {\rm D-terms}
\label{leffn3}
\end{equation}   
Note that a non-zero superpotential for the chiral multiplets $\varphi_{l}$
is forbidden by ${\cal N}=2$ supersymmetry. 
\subsection{A duality proposal}
\paragraph{}
We now return to consider the 
IIB geometry dual to the four-dimensional gauge theory with
superpotential (\ref{sp1}). In a generic vacuum this has the form, 
\begin{eqnarray}
u^{2}+v^{2}+w^{2}+\varepsilon^{2}\left(x^{m}-\sum_{l=1}^{m}g_{l}x^{l-1}
\right)^{2}-4\varepsilon^{2}\Lambda^{2m} & = &
\nonumber \\u^{2}+v^{2}+w^{2}+
\varepsilon^{2}\prod_{l=1}^{m}(x-x_{+}^{(l)})(x-x_{-}^{(l)}) & = & 0 
\label{geom9}
\end{eqnarray}
At the critical point the $m$ branch-points $x^{(l)}_{-}$ collide and
we find the singular geometry, 
\begin{equation}
u^{2}+v^{2}+w^{2}+\varepsilon^{2}x^{m}(x^{m}-4\Lambda^{m})=0
\label{conifold3}
\end{equation}
which has a generalised conifold singularity at the origin. 
\paragraph{}
There are two main differences with the standard discussion of the
generalised conifold given in the previous section. First, the
singularity appears as part of a larger geometry which also includes
non-trivial cycles of finite size of order $\Lambda$. 
Secondly there is RR flux through these cycles. Specifically
there is $n$ units of flux through the cycles whose projections on the
$x$-plane surround the cuts running from $x^{(l)}_{-}$ to
$x^{(l)}_{+}$ for $l=1,2,\ldots m$. The effect of this flux is to 
introduce a non-trivial superpotential which fixes the moduli of the
Calabi-Yau. As discussed by Aganagic, Mari\~no and Vafa \cite{AMV}, one
way to make contact with the flux-free case of the previous section 
is to take a  particular 
$\Lambda\rightarrow \infty$ limit where the non-vanishing cycles 
become very large and the RR field-strength goes to zero everywhere. 
%the limit discussed in \cite{AMV} is 
%$\Lambda\rightarrow \infty$, $\varepsilon\rightarrow 0$ and 
%$g_{1}\rightarrow \infty$ with $\alpha=-4\varepsilon^{2}\Lambda^{m}$ and 
%$\delta=g_{1}-2\Lambda^{m}$ held fixed. In the ${\bf Z}_{m}$ symmetric
%case with $g_{l}=0$ for $l>2$, this yields the generalised conifold
%geometry,  
%\begin{equation}
%u^{2}+v^{2}+w^{2}+\alpha (x^{m}-\delta)=0
%\label{conifold4}
%\end{equation}  
%with zero flux. 
\paragraph{}
In the following we will suggest a quite different limit which also makes
contact with the generalised conifold without flux. Specifically, we 
will consider the large-$N$ double-scaling limit of the dual field theory
defined in (\ref{dscale}) above. As in the standard 't Hooft limit, we
take $N\rightarrow \infty$ while holding fixed the dynamical scale
$\Lambda$. The new
ingredient is that the superpotential couplings are scaled in such a
way that the mass of the lightest $\hat{G}$-charged state is held
fixed. As always, the mapping
to string theory is such that $g_{s}$ scales like $1/N$. The
tension of the confining string is identified with that of the 
IIB string. We also know that the charged dibaryons and
magnetic monopoles of the field theory are identified with D3
branes wrapped on cycles which vanish at the critical point. 
Thus field theory double-scaling therefore corresponds to a limit in the dual
string theory where 
$g_{s}\rightarrow 0$ with $\alpha'$ fixed, and the moduli of the
Calabi-Yau are tuned to their critical values in such a way that the
masses of wrapped branes are also held fixed. This is precisely the
same type of limit considered by Giveon and Kutasov in the context of the
generalised conifold without flux. 
\paragraph{}
We also argued above that only one
sector of the field theory, denoted ${\cal H}_{-}$, 
retains finite interactions in the
double-scaling limit, while the remaining states become free. This
feature closely resembles the decoupling in the Giveon-Kutasov limit
of the interacting LST describing 
states localised at the singularity from the free
theory in the bulk. 
On the other hand, the double-scaling limit described above keeps
$\Lambda$ fixed, so the 
size of the non-vanishing cycles in the geometry stays fixed in string
units. This means that the dual string background still involves RR
field-strengths which remain large in the limit. 
Despite this we would like to propose that 
the interacting sector of the four-dimensional field theory 
is correctly described by the double-scaled Little 
String Theory of the generalised conifold without flux. 
For brevity we will refer to the latter theory as DSLST. 
In the rest of this Section, we will explain why such a correspondence 
should hold. 
\paragraph{}
The first observation is that the massless degrees of freedom 
in the interacting sector ${\cal H}_{-}$ and their F-term interactions
precisely match those of 
DSLST. In terms of ${\cal N}=1$ multiplets, both theories contain 
$p=(m-1)/2$ $U(1)$ vector multiplets and $p$
neutral massless chiral multiplets. The F-term effective action for
these degrees of freedom is given for the field theory in Eqs
(\ref{sigmaminus}, \ref{ssd3}, \ref{result}) and for
DSLST in Eqs (\ref{gam}, \ref{aad}, \ref{leffn3}). The two sides of the
correspondence match exactly if we identify the fields according to, 
\begin{eqnarray}
\tilde{w}_{\alpha l}\leftrightarrow {\cal W}_{\alpha l} & \qquad{} \,\,\,\,
& \tilde{S}_{l}\leftrightarrow K\varphi_{l} 
\label{ident}
\end{eqnarray}
for $l=1,2,\ldots, p$ where
$K=2i\Lambda^{\frac{m}{2}}\kappa\sqrt{\alpha'}$ will be identified below.  
We must also identify the 
double-scaling parameters on both sides as  
$\tilde{g}_{l} \leftrightarrow \tilde{\mu}_{l}$ for $l=2,3,\ldots,p$. 
This ensures that the reduced spectral curve $\Sigma_{-}$ matches the
curve $\Gamma$ which governs the F-terms of DSLST.  
\paragraph{}
It is straightforward to reinterpret this agreement 
in terms of the dual string geometry. 
To do so we rescale the spacetime
coordinates in (\ref{geom9}) as follows, 
\begin{equation}
u=\tilde{u}\delta^{\frac{1}{2}}\alpha^{\frac{1}{2}} \qquad{} 
v=\tilde{v}\delta^{\frac{1}{2}}\alpha^{\frac{1}{2}} \qquad{} 
w=\tilde{w}\delta^{\frac{1}{2}}\alpha^{\frac{1}{2}} \qquad{} 
x=\tilde{x}\delta^{\frac{1}{m}}
\label{rescaleb}
\end{equation}
where $\delta=g_{1}-2\Lambda^{m}$ and 
$\alpha=-4\varepsilon^{2}\Lambda^{m}$. 
We now take the double-scaling
limit (\ref{dscale}) while keeping the rescaled coordinates fixed. 
Keeping only the leading terms, (\ref{geom9}) becomes, 
\begin{equation}
\tilde{u}^{2}+\tilde{v}^{2}+\tilde{w}^{2}+\tilde{x}^{m}-
\sum_{l=2}^{p}\tilde{g}_{p}\tilde{x}^{l-1}-1=0
\label{rescale2}
\end{equation}
The resulting geometry can be thought of as an $a_{1}$-fibration of
the reduced spectral curve $\Sigma_{-}$ and also coincides with the
rescaled generalised conifold (\ref{conifoldtilde}) with
identification $\tilde{g}_{l}=\tilde{\mu}_{l}$. 
\paragraph{}
With the identifications (\ref{ident}), we see that the two 
theories coincide exactly at the level of F-terms. On the other hand, 
an obvious discrepancy is that the DSLST has ${\cal N}=2$ supersymmetry in four
dimensions while the field theory apparently has only ${\cal N}=1$
SUSY. Note however that the chiral multiplets 
$\tilde{S}_{l}$ become exactly massless in the double-scaling
limit. Further the superpotential for these degrees of freedom
vanishes exactly in this limit. These facts are consistent with the
proposal that the supersymmetry of the interacting sector of the field 
theory is enhanced from ${\cal N}=1$ to ${\cal N}=2$ in the
double-scaling limit. In particular, our large-$N$ counting arguments 
imply that the glueballs of the interacting sector have couplings of
order one to the massive $\hat{G}$-charged states. Normally such
interactions would generate non-zero masses for the glueballs unless
some symmetry forbids it. In the present case, there are no obvious
candidates apart from an enhanced ${\cal N}=2$ supersymmetry. 
\paragraph{}
In the dual string theory, the superpotential and the consequent
breaking of ${\cal N}=2$ SUSY down to ${\cal N}=1$ are directly
associated with the presence of non-zero RR fluxes. As we have
emphasized above, the full string geometry certainly does contain strong
RR fluxes. On the other hand, 
the superpotential for the fields in the interacting
sector vanishes in the double-scaling limit suggesting that this
sector of the theory is unaffected by the flux.  
The heuristic picture which reconciles these apparently contradictory
features is as follows.  
The generalised conifold singularity involves
the same set of light charged states wrapped on the vanishing
three-cycles whether or not there is flux through the remaining 
non-vanishing cycles. Thus we expect the same infinite throat to form in
the region near the singularity in both cases. As discussed in Section
5, the degeneration of the spectral curve $\Sigma$ in the double
scaling limit is consistent 
with two curves $\Sigma_{+}$ and $\Sigma_{-}$ joined by
a long tube as shown in Figure 5. As the dual spacetime
geometry is an $a_{1}$-fibration of $\Sigma$, we are essentially 
suggesting that the spacetime metric in the directions along $\Sigma$
is similar to the one implied in Figure 5.  
\paragraph{}
Given the presence of a throat region near the origin, the resulting 
linear growth of the dilaton provides a natural explanation of 
the breakdown of the $1/N$ expansion in the dual field theory. 
It also suggests an explanation for the puzzle of the RR flux raised
above. Even though RR flux is present in the bulk geometry, 
as there are no RR charges in the throat, the RR field-strength
should decay as we move down the throat. In the limit 
where the asymptotic string coupling goes to zero, the only states
which retain finite interactions are localised 
infinitely far down the throat. These states should therefore 
be unaffected by the presence of flux in the bulk. 
With this interpretation, the rescaled spacetime coordinates 
defined in (\ref{rescaleb}), which are held fixed in the double-scaling
limit, correspond to coordinates for the region far down the throat 
where the effective string coupling remains fixed as $g_{s}\rightarrow 0$. 
In the next Section we will discuss a related model where these  
points can actually be made quite precise. 
\paragraph{}
To make our proposal more concrete, we should specify the relation
between the parameters of the two theories. We begin by considering
the ${\bf Z}_{m}$ symmetric theory with 
$\tilde{g}_{l}=\tilde{\mu}_{l}=0$  
The double-scaled Little String Theory of the generalised conifold
singularity is characterised by the renormalised string coupling 
$\kappa$ and the fundamental string tension $1/2\pi \alpha'$. 
In field theory, these correspond to the effective three-point 
glueball coupling and the confining string tension respectively. 
We will mainly be interested in the case $\kappa<<1$, where string
perturbation theory is valid. In this case we propose the identifications, 
\begin{eqnarray}
\kappa=\frac{1}{N_{eff}} & = & 
\frac{\Lambda^{3-\frac{m}{2}}}{\Delta}F_{1}(\tilde{\varepsilon},\Lambda,
M_{UV}) 
\nonumber \\ 
\alpha' &= & \frac{1}{\Lambda^{2}}F_{2}(\tilde{\varepsilon},\Lambda,
M_{UV})
\label{ident6} 
\end{eqnarray}
where $F_{1}$ and $F_{2}$ are two unknown dimensionless functions 
of the parameters which remain fixed in double scaling limit. 
These include $\tilde{\varepsilon}=\varepsilon/N$,$\Lambda$, and the UV
cut-off $M_{UV}$. Thus we find the constant $K$ appearing in
(\ref{ident}) is given as $K=2i\Lambda^{2}F_{1}\sqrt{F_{2}}$. 
Away from the ${\bf  Z}_{m}$ symmetric point the DSLST also has 
$p-1$ additional parameters $\tilde{\mu}_{l}$ associated with the normalisable
deformations of the geometry. As above, comparison of F-term
observables implies that these are identified with the corresponding 
field theory parameters as $\tilde{g}_{l}=\tilde{\mu}_{l}$ for
$l=2,3,\ldots p$.  

\section{The $\beta$-deformed Theory and its AdS Dual}
\paragraph{}
As emphasized in the introduction, partially confining phases of the
type studied above are generic to ${\cal N}=1$ 
supersymmetric theories with adjoint matter. In this Section we will 
study a very special model in this class: the $\beta$-deformation of
${\cal N}=4$ SUSY Yang-Mills. 
The matter content of this model is 
an ${\cal N}=1$ vector multiplet $W_{\alpha}$ together
with three adjoint chiral superfields, $\Phi_{1}$, $\Phi_{2}$ and
$\Phi_{3}$ with superpotential, 
\begin{equation}
W={\rm Tr}_{N}\left[e^{+i\beta/2}\Phi_{1}\Phi_{2}\Phi_{3}-
e^{-i\beta/2}\Phi_{1}\Phi_{3}\Phi_{2}\right]
\label{supA2}
\end{equation}
We will consider the case of gauge group $G=SU(N)$. This theory is
distinguished because it corresponds to an exactly marginal deformation of 
${\cal  N}=4$ SUSY Yang-Mills \cite{LS1} and inherits some of
special properties of the latter theory. In particular as the
deformation is marginal the resulting theory shares the exact
conformal invariance of ${\cal N}=4$ SUSY Yang-Mills. It also inherits 
an exact electric-magnetic duality related to the S-duality 
of the ${\cal N}=4$ theory \cite{DHK}. Finally, proximity to ${\cal N}=4$ SUSY
Yang-Mills means that the deformed theory can easily be studied in the
context of the AdS/CFT correspondence using IIB supergravity. 
\paragraph{}
Despite these special features we will see that the $\beta$-deformed
theory has very similar behaviour to the model (\ref{sp1}). In
particular it has a partially confining branch which exhibits many of
the features we discussed above, including a critical point where charged
states become massless. The key point is that, for this particular
model, the physics of the critical point can be studied quantitatively 
using the AdS/CFT correspondence. This Section is mainly a review of
the results of \cite{nd1,nd2}, interpreted in the context of our more
general study.         
\paragraph{} 
The model with superpotential (\ref{supA2}) corresponds to 
a two complex parameter family of
conformal field theories labelled by the complexified gauge coupling 
$\tau=4\pi i/g^{2}+\theta/2\pi$ and the deformation parameter
$\beta$. The ${\cal N}=4$ theory with gauge
group $SU(N)$ is obtained by setting $\beta=0$. Turning on the
deformation breaks the ${\cal N}=4$ superconformal
invariance down to an ${\cal N}=1$ sub-algebra. 
At linear order, the $\beta$-deformation corresponds to adding a certain 
chiral primary operator to the ${\cal N}=4$ Lagrangian. Via the
AdS/CFT correspondence, the deforming operator is mapped to a 
dual supergravity field in $AdS_{5}$. The field in question is 
a particular mode of the complexified
three-form field-strength, $G^{(3)}=F^{(3)}_{RR}+\tau H^{(3)}_{NS}$, of 
IIB supergravity compactified on $S^{5}$. 
Starting from the string dual of the ${\cal N}=4$ theory 
on $AdS_{5}\times S^{5}$, we can introduce the $\beta$-deformation 
by turning on an appropriate source for $G^{(3)}\sim \beta$ on the 
boundary of $AdS_{5}$. As the deformed theory is conformally invariant, the
resulting string background has the form $AdS_{5}\times
\tilde{S}_{5}$, where $\tilde{S}_{5}$ is a deformation of the
five-sphere. The corresponding geometry was constructed to second
order in a  perturbation series in $\beta$ in \cite{AKY}. 
More recently an exact 
supergravity solution, valid for all values of $\beta$, has been
obtained by Lunin and Maldacena \cite{lunin}.    
\paragraph{}
A surprising aspect of the $\beta$-deformed theory is that 
it has a much richer vacuum structure than
that of the ${\cal N}=4$ theory. The latter theory has a conformal
invariant phase and also a Coulomb branch where the gauge symmetry is
spontaneously broken to its Cartan subalgebra.  
For generic values of $\beta$,
the deformed theory also has a conformal point and a Coulomb branch. 
However, for special values of $\beta$, it also has branches of vacua 
in various Higgs, confining and oblique confining phases. These
branches were studied in detail in \cite{nd1,nd2,DH} by solving the 
corresponding Dijkgraaf-Vafa matrix model in the planar limit. In the
following we will focus on a particular branch of vacua which exhibits
partial confinement of the type discussed in the previous section. 
In particular we will review several results from \cite{nd2} and refer
the reader to this paper for further details. 
\paragraph{}
We will consider the $\beta$-deformed theory with gauge group
$G=SU(N)$ where $N=mn$ for some integers $m$ and $n$ and, for
simplicity, we set the vacuum angle to zero so that $\tau=4\pi
i/g^{2}$. If we choose the 
special value $\beta=8\pi^{2}i/g^{2}n$, the theory has a branch of
vacua  with low-energy gauge group $\hat{G}=U(1)^{m-1}$.  
Along this branch, the lowest components 
of the three chiral superfields $\Phi_{i}$ acquire non-zero 
vacuum expectation values (VEVs). A gauge invariant description of
these VEVs is,    
\begin{eqnarray}
\frac{1}{N}\left\langle{\rm Tr}_{N} \Phi_{i}^{k} \right\rangle  
& \, \, = \,\, & \beta^{(l)}_{i} \qquad{} {\rm for} \,\,k=ln \nonumber
\\ & \,\, =\,\, & 0 \qquad{} \,\,\,\,\, {\rm otherwise} \nonumber
\end{eqnarray}
The resulting branch is parametrised by $3m$ complex moduli 
$\beta^{(l)}_{i}$, with $l=1,2,\ldots,m$ and $i=1,2,3$. For generic 
points on this branch the low-energy gauge group is
$\hat{G}=U(1)^{m-1}$ and the massless spectrum includes $m-1$ abelian 
vector multiplets and $3m$ $\hat{G}$-neutral chiral multiplets
corresponding to fluctuation of the moduli $\beta_{i}^{(l)}$. 
\paragraph{}
The branch described above exhibits partial 
confinement of the original gauge group, $G=SU(N)$, down to the 
low-energy gauge group $\hat{G}=U(1)^{m-1}$. More precisely, we will 
see that the confinement index, $t$, introduced above 
takes the value $n$. This result can be established in several
different ways. As in \cite{nd1}, we may use the exact S-duality of
the theory to map the confining branch at $\beta=8\pi^{2}i/g^{2}n$ to 
a Higgs branch which occurs for $\beta=2\pi/n$. The dual phenomenon,
in which magnetic charges are partially confined by the formation of 
flux tubes can then be studied using semiclassical methods. An alternative 
approach, developed in \cite{DH}, is to find the root of the confining 
branch using an exact description of the Coulomb branch as the moduli
space of a certain complex curve (see also \cite{ben}). The root of
the new branch occurs at a special point on the Coulomb branch 
where additional magnetic
degrees of freedom become massless. These degrees of freedom condense
on the new branch leading to partial confinement of electric charges. 
As usual, confinement can be demonstrated explicitly 
using an abelian effective Lagrangian which is valid near 
the root. Finally, as we now review, the physics of the confining
branch can also be analysed using the AdS/CFT correspondence \cite{nd2}. 
\paragraph{}                        
As mentioned above the AdS-dual of the $\beta$-deformed theory in its
conformal phase is a spacetime of the form $AdS_{5}\times
\tilde{S}^{5}$ with a source for $G^{(3)}\sim \beta$ on the boundary 
of $S^{5}$. Here we have the usual identifications $g^{2}=4\pi g_{s}\sim
1/N$ and $\lambda=g^{2}N={\cal R}^{4}/\alpha'^{2}$ where ${\cal R}$ is
the radius of the geometry. The supergravity description is valid provided 
$\lambda=g^{2}N>>1$. Choosing the slightly stronger condition
$g^{2}n>>1$, we find that the special value $\beta=8\pi^{2}i/g^{2}n$ 
can be made small and the dual background can be thought of as a small 
perturbation of $AdS_{5}\times S^{5}$. Under these conditions, a
string dual for the new branch was identified in
\cite{nd2}. Specifically a generic vacuum on the partially confining 
branch is dual to a configuration of $m$ NS fivebranes in\footnote{At
  lowest order in $\beta$ we can neglect the deformation of $S^{5}$}
$AdS_{5}\times S^{5}$ with the following properties: 

{\bf 1:} Each fivebrane is located at a definite radial position 
$r=\bar{r}_{l}$ in  $AdS_{5}$, for $l=1,2,\ldots,m$, and fills the 
four dimensions parallel to the boundary. The radii $\bar{r}_{l}$ are 
adjustable moduli of the configuration determined by the field theory 
VEVs\footnote{The precise relation between the radii $\bar{r}_{l}$ and
  the moduli $\beta^{(l)}_{i}$ can be deduced from the results of
  \cite{nd2} but will not be needed here.}. 
Without loss of generality we can choose $\bar{r}_{1}
\leq \bar{r}_{2}\leq \ldots\leq \bar{r}_{l}$. 

{\bf 2:} Each fivebrane is wrapped on a toroidal subspace of
$S^{5}$, $T^2_{(l)}\subset S^{5}$. The remaining moduli 
of the configuration correspond to the shape of each torus and its 
orientation on $S^{5}$. 

{\bf 3:} Each fivebrane carries exactly $n$ units of D3 brane charge, 
which can be realised as $n$ units of magnetic flux of the 
six-dimensional world-volume gauge field through the torus
$T^{2}_{(l)}$. 

The NS fivebrane configuration is stabilised by the presence of a 
non-zero field
strength for $H^{(3)}_{NS}\sim 1/n$ corresponding to 
$\beta=8\pi^{2}i/g^{2}n$. The Neveu-Schwarz threeform couples
magnetically to the NS5 branes and renders the toroidally wrapped
branes marginally stable. This is a version of Myers dielectric
effect \cite{myers} which arises in the analysis of ${\cal N}=1^{*}$ SUSY
Yang-Mills \cite{PS}. 
\paragraph{}
Several features of the partially confining 
branch can be understood in terms of the
brane configuration described above. First, the low energy gauge group 
of the dual field theory is visible in the worldvolume theory on the
NS5-branes. In particular, each fivebrane has an abelian gauge field living 
on its six-dimensional worldvolume. After dimensional reduction on the
torus $T^{2}_{(l)}$, we obtain a $U(1)$ gauge field in the four
non-compact dimensions on each brane. As in \cite{PS}, we expect that
the central $U(1)$ subgroup of $U(1)^{m}$ is lifted by its coupling to the 
bulk fields, leaving the low-energy gauge group $\hat{G}=U(1)^{m-1}$.   
\paragraph{}
To analyse the confinement of electric charges, we introduce external 
charges in the ${\bf N}$ and $\bar{\bf N}$ representations of 
$SU(N)$, separated by a distance $R$. By standard arguments, 
these charges correspond to the endpoints of a fundamental string,
separated by the same distance $R$, on the boundary of $AdS_{5}$. In a
conformally-invariant vacuum, the energetically preferred
configuration is one where the string droops into the interior of 
$AdS_{5}$, corresponding to the spreading out of electric flux lines
in the dual gauge theory. The result is a Coulomb potential between
the charges, $V(R)\sim 1/R$, as dictated by conformal invariance 
\cite{mal2}. 
\paragraph{}
As explained in \cite{PS}, the situation is modified by the presence 
of NS5 branes in the bulk. A key fact about Neveu-Schwarz fivebranes is that
fundamental strings cannot end on them. Instead the fundamental string
can form a boundstate with the NS5, corresponding to a solitonic
string of finite tension $\sigma$ in its worldvolume gauge theory. 
The tension of these objects is set by the scale of the VEVs and 
obeys the standard 't Hooft scaling, $\sigma\sim N^{0}$. 
In the present case the string can droop into the bulk geometry as far
as one of the NS fivebranes and smoothly join onto a segment of  
solitonic strings which extends a distance $R$ parallel to the
boundary. The result is the onset of a 
confining linear potential $V(R)\sim \sigma
R$ between the external charges in the gauge theory at large
distances. However, this is not quite the whole story: 
the theory also contains $m$ distinct baryon vertices, labelled by an
integer $l=1,2,\ldots,m$, each corresponding to
a D3 brane whose spatial worldvolume fills a three-manifold
$M^{(l)}_{3}$ whose boundary is the torus $T^{2}_{(l)}$ on which the
$l$'th NS5 is wrapped. As each NS5 brane carries $n$ units of
worldvolume magnetic through the torus there there is a non-trivial 
selection rule \cite{PS} which demands that 
$n$ fundamental strings must end on the D3 brane. The resulting
configuration therefore corresponds to a baryon vertex which can
screen external charges in the 
$n$-fold product of the fundamental representation of $SU(N)$. 
Thus we conclude that the confinement
index $t$ takes the value $t=n$.        
\paragraph{}
The AdS analysis described above corresponds to a 't Hooft limit of
the $\beta$-deformed theory where $N=mn\rightarrow \infty$ with 
$m$, $\lambda=g^{2}N$ and the  moduli $\beta^{(l)}_{i}$ held fixed.  
This is the same type of large-$N$ limit considered in Section 3 above, 
and it is interesting to compare the spectrum of the 
$\beta$-deformed theory with the general discussion given there. 
From the dual AdS description, we find the following states,   

{\bf 1:} The massless spectrum includes $m-1$ photons and their 
${\cal N}=1$ superpartners as well as 3m additional neutral 
chiral multiplets. 

{\bf 2:} As we have gauge theory flux tubes of tension 
$\sigma\sim N^{0}$, we expect to find a spectrum of glueballs with 
masses of order $N^{0}$ corresponding to the excitation spectrum of
the corresponding closed strings. These states are neutral under the 
low-energy gauge group. 

{\bf 3:} It is straightforward to identify dibaryonic states carrying 
electric charges $(+n,-n)$ with respect to $U(1)_{r}\times U(1)_{s}
\subset \hat{G}$. These states correspond to D3 branes wrapped on
3-chains $C_{rs}$ of topology $T^{2}\times I$ where $I$ is a closed 
interval. The two boundaries of the chain are the torii $T^{2}_{(r)}$ 
and $T^{2}_{(s)}$ on which the $r$'th and $s$'th NS fivebranes are
wrapped. The masses of these states go like 
${\rm Vol}(B_{rs})/g_{s}\alpha'^{2}$. The inverse power of $g_{s}$
implies that the masses of these states scale like $N$ as expected for
baryons. 

{\bf 4:} States carrying magnetic charges $(+1,-1)$ under 
$U(1)_{r}\times U(1)_{s}\subset \hat{G}$ correspond to D-strings 
stretched between the $r$'th and $s$'th NS fivebranes with masses of
order $D_{rs}/g_{s}\alpha'$ where $D_{rs}$ is the proper distance
between the branes. As before, the inverse power of $g_{s}$
implies that the masses of these states scale like $N$. 
These states correspond to unconfined magnetic monopoles.     
 
Thus, the large-$N$ spectrum is in qualitative agreement with the
field theory analysis of Section 2. The only significant 
difference is the presence of massless chiral multiplets in generic
vacua. This reflect the conformal invariance of the theory. Indeed
conformal invariance and other global symmetries are spontaneously
broken on the partially confining branch and some of the massless
scalars living in the chiral multiplets can be interpreted as
Goldstone bosons.   

\subsection{The critical point}  
\paragraph{}
So far we have discussed generic points on the partially confining
branch of the $\beta$-deformed theory. The string dual to these vacua 
is a configuration of $m$ NS fivebranes at distinct radial positions 
$r=\bar{r}_{l}$ in $AdS_{5}$, for $l=1,2,\ldots, m$ wrapped on
distinct tori $T^{2}_{(l)}$. In this case, the 
spectrum of the theory includes massive states corresponding to 
D-strings and toroidally wrapped D3 branes stretched between 
the NS five-branes. In the previous Section, we identified these
states as magnetic and electric dibaryons 
respectively. In generic vacua, the masses of these states scale like 
$g^{-1}_{s}\sim N$. However, there are also special points in the
moduli space where two or more NS5 branes coincide and some of these
states become massless. In this subsection we will discuss
interpretation of these special points in, both in field theory and in
the string dual.   
\paragraph{}
We will focus on the maximal critical point where all $m$
NS-fivebranes become coincident. The new massless states include
$m(m-1)$ stretched D-strings and, as for NS fivebranes in flat space, 
we expect the low-energy gauge group to be enhanced from 
$\hat{G}=U(1)^{m-1}$ to $\hat{H}=SU(m)$. We also find $m(m-1)$ new
massless states corresponding to stretched D3 branes which are
magnetically charged under $\hat{G}$. As both electric and magnetic 
charges are becoming light simultaneously, we expect the 
theory in these special vacua to flow to an interacting CFT in the
IR. For $m$ coincident toroidally-wrapped 
NS5 branes in flat space, the resulting CFT would be ${\cal N}=4$ SUSY
Yang-Mills with gauge group $\hat{G}=SU(m)$. As shown in \cite{nd2}, 
the ${\cal N}=4$
theory also arises in the present case where the fivebranes live in
an asymptotically-$AdS$ space.   
 \paragraph{}
In the dual field theory, the
corresponding submanifold of the partially confining branch is 
parametrised by three complex numbers 
$\alpha_{1}$, $\alpha_{2}$ and $\alpha_{3}$, defined by  
\begin{eqnarray}
\frac{1}{N}\left\langle{\rm Tr}_{N} \Phi_{i}^{k} \right\rangle  
& \, \, = \,\, & \alpha^{k}_{i} \qquad{} {\rm for} \,\,k=ln \nonumber
\\ & \,\, =\,\, & 0 \qquad{} \,\,\,\,\, {\rm otherwise} \nonumber
\label{vaccr}
\end{eqnarray}
Equivalently, in terms of the moduli $\beta^{(l)}_{i}$ introduced 
above we have $\beta^{(l)}_{i}=\alpha_{i}^{ln}$. As before, these
vacua only occur for the special value $\beta=8\pi^{2}i/g^{2}n$. 
The low energy effective theory in these vacua is ${\cal N}=4$ SUSY
Yang-Mills with gauge group $\hat{G}=SU(m)$ and gauge coupling 
$\hat{g}^{2}=g^{2}n$. These conclusions can be checked by using the exact 
description of the vacuum structure of the theory presented in 
\cite{DH}, to verify the low energy gauge group, gauge coupling and 
moduli space near the corresponding root.   
\paragraph{}
A notable feature of the above results is that the rank $m$ of the IR
gauge group and the corresponding gauge coupling 
$\hat{g}^{2}=16\pi^{2}/g^{2}n$ remain fixed when we take the 
't Hooft limit of the $U(N)$ theory. To understand the implications 
of this, it is useful to consider the correlation functions of single-trace
operators. We will focus on $L$-point correlation
function of the operator $\hat{O}={\rm Tr}_{N}[F_{\mu\nu}F^{\mu\nu}]$ 
appearing in (\ref{scaling}). The large-distance behaviour of this
correlation function will be 
dominated by the fluctuations of the unconfined components 
of the gauge field. In this case the non-abelian field strength 
$F_{\mu\nu}\in {\rm Lie}(G)$ can be replaced by its unconfined component 
$f_{\mu\nu}\in  {\rm Lie}(\hat{G})$ inside the trace and the
resulting correlations calculated in the low-energy effective theory. 
\paragraph{}
For generic vacua on the partially confining branch we have 
$\hat{G}=U(1)^{m}$. The low-energy theory is abelian and hence free,
implying that all connected correlation functions apart from 
the two-point function vanish. The normalization of the two-point
function is determined by the low-energy gauge coupling 
$\hat{g}^{2}=16\pi^{2}/g^{2}n$. The resulting correlations functions trivially 
obey the bound, 
\begin{eqnarray}
\langle \hat{O}(x_{1})\hat{O}(x_{2})\ldots\ldots
\hat{O}(x_{L})\rangle & 
\qquad{} \leq 
\qquad{} N^{2-L}
\label{bound}
\end{eqnarray}
implied by standard large-$N$ scaling. Now consider the theory at the
critical point, which has $SU(m)$ gauge interactions with a coupling
which remains fixed in the large-$N$. Perturbation theory in 
$\hat{g}^{2}$ yields contributions limit to $L$ point functions of 
order $N^{0}$ for any $L$. Thus the bound (\ref{bound}) is badly 
violated at the critical point. This indicates that the 't Hooft large-$N$
expansion considered above breaks down at the critical point.  

\subsection{The gravity dual}
\paragraph{}  
The string dual of the vacuum (\ref{vaccr}) was studied in detail in
\cite{nd2}. The relations between string and gauge theory parameters
are simplest when one of the three moduli (say $\alpha_{3}$) is set
to zero. In this case the dual configuration consists of 
$m$-coincident NS fivebranes located at fixed radial distance 
$r=\bar{r}=\sqrt{r_{1}^{2}+r^{2}_{2}}$ with $r_{i}=|\alpha_{i}|(2\pi
\alpha')$ for $i=1,2$.   
The NS5 branes are wrapped on a rectangular torus $T^{2}\subset \tilde{S}^{5}$
and carry a total of $N=mn$ units of D3 brane charge.  
\paragraph{}
Following Polchinski and Strassler \cite{PS}, one can find a smooth 
supergravity solution corresponding to the brane configuration described above
which is valid provided $g^{2}n>>m$. The
resulting geometry interpolates between two regions: 

{\bf Region I:} Sufficiently far away from the branes, the effect of
$N$ units of D3 branes charge dominates over that of the NS5 branes. 
Specifically, for $v=|r-\bar{r}|>>v_{cr}$ with, 
\begin{equation}
v_{cr}=\sqrt{\frac{mr_{1}r_{2}}{g_{s}n}}
\label{vcr}
\end{equation}
the dual geometry can be approximated by the one produced by the
toroidal distribution of $N$ D3 branes only. This leads to a 
warped $AdS_{5}$ metric corresponding to a point on the Coulomb branch 
of ${\cal N}=4$ SUSY Yang-Mills. 

{\bf Region II:} Very close to the toroidally wrapped branes, the 
gravitational effect of the NS5 branes dominates over that of the D3
branes. The dual geometry can be approximated by the the near-horizon
geometry of $m$ NS fivebranes compactified to four dimensions on a
torus. This is valid for $v<<v_{cr}$. 

The full supergravity solution interpolating between these two 
regions was given in \cite{nd2}. For the present purposes we will only
need the metric in region II, which can be written in the form, 
\begin{equation}
ds^{2}=\eta_{AB}dY^{A}dY^{B}+d\sigma^{2}+\alpha' m d\Omega_{3}^{2}
\label{metric}
\end{equation}
This is the standard infinite throat geometry of the NS fivebrane. 
The $Y_{A}$, with $A=0,1,2,\ldots 5$ are coordinate on the brane
world-volume and $\eta_{AB}$ is the flat metric on six dimensional 
Minkowski space. Compactification to four dimensions on $T^{2}$ is
implimented by the identifications, 
\begin{eqnarray}
Y_{4}\sim Y_{4}+2\pi L_{1} & \quad{} \qquad{} &  Y_{5}\sim Y_{5}+2\pi
L_{2} \nonumber 
\end{eqnarray}
with 
\begin{eqnarray}
L_{1}=\left(\frac{g_{s}n\alpha'}{r_{1}r_{2}}\right)^{\frac{1}{2}}r_{1}
& \quad{} \qquad{} &
L_{2}=\left(\frac{g_{s}n\alpha'}{r_{1}r_{2}}\right)^{\frac{1}{2}}
r_{2} \nonumber 
\end{eqnarray}
\paragraph{}
It is important to note that that energies of string states in the
geometry (\ref{metric}) are measured relative to the time 
coordinate $Y_{0}$ on the fivebrane worldvolume. On the other hand
energies in gauge theory are naturally measured relative to the time coordinate
$x_{0}$ on the boundary of $AdS_{5}$. The relation between $Y_{0}$ and
$x_{0}$ is fixed uniquely by matching the solutions in region I and
region II \cite{nd2}. The result is, 
$Y_{0}=(T_{\rm IIB}/T_{\rm QFT})^{-\frac{1}{2}}x_{0}$ where, 
\begin{eqnarray}
T_{\rm IIB}=\frac{1}{2\pi \alpha'} & \qquad{} \qquad{} & 
T_{\rm QFT}=\frac{8\pi^{2}|\alpha_{1}||\alpha_{2}|}{g^{2}n} \nonumber
\end{eqnarray}       
The upshot of this rescaling is that the fundamental IIB string of
tension $T_{\rm IIB}$, propagating in the throat geometry
(\ref{metric}) is identified with a gauge theory flux tube 
of tension $T_{\rm QFT}$. The above formula for the gauge theory flux
tube tension can also be checked explicitly via S-duality to the Higgs
branch where the corresponding object emerges as a classical soliton.   
\paragraph{}
The four directions transverse to the brane are labelled by a radial
coordinate $\sigma=-\sqrt{m\alpha'}\log(g_{s}v_{cr}/v)$ and three 
polar angles on the  3-sphere. The IIB dilaton 
$\phi$ grows linearly along the radial direction, 
\begin{eqnarray}
\phi=-\frac{Q}{2}\sigma & \qquad{} \qquad{} &
Q=\frac{2}{\sqrt{m\alpha'}} \nonumber 
\end{eqnarray}
Hence the effective string
coupling $g^{eff}_{s}=\exp(\phi)$ grows in the direction of increasing
$\sigma$, becoming equal to unity at $\sigma=0$. Consequently string
perturbation theory breaks down for states propagating in the region 
$\sigma>0$. This corresponds to the breakdown of the $1/N$ expansion in the
dual field theory discussed in the previous section.  
\paragraph{} 
As we have $m$ NS5 branes the solution also includes $m$
units of Neveu-Schwarz threeform flux through the transverse
$S^{3}$. However, note that the region II solution, valid for
$v<<v_{cr}$, contains no background Ramond-Ramond fields. 
The full geometry contains both three-form and five-form RR 
field-strengths, but these die off rapidly as we enter the 
fivebrane throat \cite{nd2}. The throat geometry (\ref{metric}) corresponds 
to an exact solution of IIB string theory and has 
a worldsheet description \cite{CHS} where
the radial direction is realised as a free scalar with an appropriate
background charge and the $S^{3}$ factor becomes an $SU(2)$ WZW model 
of level $m$. This description is useful for analysing states 
propagating in the region where the effective string coupling is
small. 

\subsection{The double-scaling limit}
\paragraph{}    
The near-horizon geometry of parallel NS fivebranes is the throat 
geometry described in the previous section. String theory in this
background is hard to analyse because of the growth of the 
dilaton. As discussed in the introduction above, we can cure this
problem by defining an appropriate
double-scaling limit.  
\paragraph{}
Following Giveon and Kutasov \cite{GK}, 
we consider a configuration where the
NS5 branes are no longer coincident.  
Seperating the NS5 branes corresponds to moving away from the critical 
point on the partially confining branch. In particular, we 
can produce a circular 
distribution of $m$ NS fivebranes by introducing a VEV for 
the chiral superfield $\Phi_{3}$, 
\begin{equation}
\frac{1}{N}\langle{\rm Tr}_{N} \Phi_{3}^{N} \rangle = \bar{\mu}^{N}
\end{equation}
The VEVs of $\Phi_{1}$ and $\Phi_{2}$ are given in terms of moduli 
$\alpha_{1}$ and $\alpha_{2}$ as in (\ref{vaccr}). As we will
eventually take a limit where $\bar{\mu}\rightarrow 0$, we will 
choose $|\bar{\mu}|<< |\alpha_{1}|$, $|\alpha_{2}|$. 
\paragraph{}
As we move away from the critical point we find a spectrum of massive
states corresponding to stretched D-strings. 
As the NS5 branes are wrapped on tori, we also find states
corresponding to stretched wrapped D3 branes. As explained above, 
these states are identified with magnetic and electric dibaryons
respectively. The deformation can also be understood as moving onto the Coulomb
branch of the low-energy theory which is ${\cal N}=4$ SUSY
Yang-Mills with gauge group $SU(m)$ and gauge coupling
$\hat{g}^{2}=g^{2}n$. This low energy theory has an S-dual formulation 
with coupling $16\pi^{2}/\hat{g}^{2}=16\pi^{2}/g^{2}n$.  
In the latter formulation, the electric 
dibaryons are realised as BPS saturated W-bosons while the magnetic 
dibaryons are BPS monopoles. There will also be dyonic dibaryons
carrying both electric and magnetic charges. 
The S-duality of the low-energy theory permutes these states.  
\paragraph{}
The double scaling limit involves taking $g_{s}$ to zero while holding
the masses of stretched D-strings and the fundamental string tension
fixed. In field theory this means taking $N\rightarrow\infty$ and 
$\bar{\mu}\rightarrow 0$ with $\lambda=g^{2}N$ and $m=N/n$ held fixed. In
addition we also hold the moduli $\alpha_{1}$, $\alpha_{2}$ and 
$\rho=\bar{\mu} N$ fixed. In this limit the Region II solution becomes 
the cigar geometry of Giveon and Kutasov, with two directions
compactified on a torus. This corresponds to IIB string theory on,       
 \begin{equation}
R^{3,1}\times T^{2}\times \left(\frac{SL(2)}{U(1)}\times 
\frac{SU(2)}{U(1)} \right)/{\bf Z}_{m}
\label{gk2}     
\end{equation}
where both cosets are at level $m$ as above. 
The effective string tension $T_{\rm QFT}$  is the tension of 
the gauge theory flux tubes studied in \cite{nd1,nd2}. 
The maximum value, $g_{cigar}$ of the effective string coupling 
$g_{s}^{\rm eff}=\exp(\phi)$ is attained at the tip of the cigar. 
In terms of gauge theory parameters we have,  
\begin{eqnarray}
T_{\rm QFT}=\frac{8\pi^{2}m|\alpha_{1}||\alpha_{2}|}{\lambda} 
& \qquad{} \qquad{} & g_{cigar}\sim \frac{2\pi}{\rho}
\sqrt{\lambda m|\alpha_{1}||\alpha_{2}|} \label{final1} 
\end{eqnarray}
The radii of the torus are, 
\begin{eqnarray}
R_{1}=\frac{\lambda}{8\pi^{2}m|\alpha_{1}|}  & \qquad{} \qquad{} &
R_{2}=\frac{\lambda}{8\pi^{2}m|\alpha_{2}|} \label{final2}
\end{eqnarray}
\paragraph{}
For large but finite $N$, the dual string background consists of the
asymptotically-AdS Region I and the cigar-like Region II described
above. The two regions are joined by a tube or throat of proper length 
$L \sim \log N$. The Hilbert space of the dual theory has two sectors, 
one corresponding to string states localised in Region I which have
interactions controlled by $1/N\sim g_{s}$ and the other to states 
localised near the tip of the cigar which have interactions controlled
by $1/N_{eff}\sim g_{cigar}$. In the large-$N$ limit these two sectors
decouple as discussed in the Introduction.  

%confining theory discussed in Section 2. In both cases the confinement
%index $t$ is non-zero and the unconfined
%gauge group $\hat{G}$ is a product of $r$ $U(1)$s. The nature of the
%relevant large-$N$ limit in the two theories is also similar. 
%In both cases $t$ scales linearly with $N$ while $r$ remains fixed.       
%There are also some important differences. The spectrum of the 
%$\beta$-deformed theory contains massless scalar fields which are
%absent in the model of Section 2. In addition the UV physics of the
%two models is quite different. The $\beta$-deformed theory flows to a
%strongly-coupled superconformal field theory in the UV, while the
%model of Section 2 is asymptotically free. 

\subsection{Summary}
\paragraph{}
To conclude this Section we will sumarise the common features of the
two models studied in detail in this paper. For brevity we will refer
to the model with superpotential (\ref{sp1}) as Model A and the
$\beta$-deformed theory as Model B. 

{\bf 1} Model A (B) has a phase where the gauge group
$G=U(N)$ ($SU(N)$) 
is partially confined down to an abelian subgroup 
$\hat{G}=U(1)^{m}$ ($U(1)^{m-1}$) with $N=mn$. 
The screening of external charges 
is characterised by the
existence of $m$ ($m-1$) different baryon vertices and 
the confinement index of \cite{CSW} takes the value $t=n$ in the
relevant vacua of both models. 

{\bf 2} In generic vacua both models have a standard 't Hooft limit 
where $N=mn \rightarrow \infty$ with $m$ fixed leading to a weakly
coupled closed string dual. The large-$N$ spectrum
includes states carrying electric and magnetic charges under
$\hat{G}$. The masses of these states scale linearly
with $N$. In the dual string theory, these states are realised as
wrapped (stretched) D-branes for Model A (B).  

{\bf 4} Both models have critical points where electric and
magnetic charges simultaneously become massless leading to an interacting 
conformal field theory. Model B at its critical point flows to
${\cal N}=4$ SUSY Yang-Mills with gauge group $SU(m)$ in the IR. It has been
proposed in \cite{eguchi} that, for the critical values of the
parameters, Model A flows to the  Argyres-Douglas 
${\cal N}=2 $superconformal fixed point of type $a_{m-1}$ in the IR. 

{\bf 5} In both models, the critical point corresponds to the
appearance of a singularity in the dual string background. 
For Model A, the vanishing of certain intersecting
three-cycles in a Calabi-Yau three-fold leading to the generalised
conifold singularity. 
For Model B the singularity involved coincident NS5 branes. 
In both cases we argued that standard $1/N$ expansion 
breaks down at the 
critical point. 
\paragraph{}
We also found some differences between the two theories. In
particular, 
Model B has some additional special features due to its close relation
to ${\cal N}=4$ SUSY Yang-Mills. The partially confining phase of
Model B is realised on a branch with moduli corresponding to the
expectation values of massless scalar fields. In contrast for Model A
we found isolated vacua. This distinction is related to the underlying
superconformal invariance of Model B which could be removed by adding
further terms to the superpotential. 
\paragraph{}
In our study of Model A, the critical point lead to singular behaviour
in the low-energy effective action corresponding to the divergence of
the sum of planar diagrams. For Model B, the corresponding terms in
the F-term effective action actually vanish identically because of
enehanced ${\cal N}=4$ SUSY in the IR (which is already present at
generic points on the partially confining branch). For example, the
exact low-energy gauge coupling is $\tau/n$ at all points on the
partially confining branch and receives no corrections from planar diagrams. 
On the other hand, in one corner of the Model B parameter space, 
we can study the critical point directly using Type IIB supergravity.  
\paragraph{}     
Although we studied the two models using very different methods, we
were lead to a very similar picture of the critical point in both
cases. The main points are, 

{\bf 1} In both models we propose that the region near the singularity
in the spacetime of the dual string theory is replaced by an infinite
throat with a linear dilaton. The growth of the dilaton corresponds
directly to the breakdown of the $1/N$ expansion in the dual field
theory. 

{\bf 2} In both models we defined a double-scaling limit 
where $N\rightarrow \infty$ and the parameters/moduli of the theory 
are tuned to their critical values in such a way that the masses $M_{B}$ of
the $\hat{G}$-charged states are held fixed. As in the conventional 't
Hooft limit, the tension, $T$, of the confining string is also held
fixed. The characteristic
feature of the double-scaled theory in both cases was the emergence of
a decoupled sector with interactions controlled by a parameter
$1/N_{eff}\sim \sqrt{T}/M_{B}$ which is held fixed in the limit.  

{\bf 3} In both cases we have proposed that the decoupled sector is
described by a double-scaled Little String Theory (DSLST). For Model A the
relevant DSLST lives in four-dimensions and has ${\cal N}=2$
supersymmetry. For Model B we obtained a description in terms of a
six-dimensional DSLST with sixteen supercharges compactified to
four-dimensions on a torus. 

{\bf 4} In both cases there is an equivalent description of the
DSLST in terms of IIB string theory on a particular 
linear dilaton background including a semi-infinite cigar. 
The background is pure NS and the effective string coupling is
controlled by the field theory parameter $1/N_{eff}$
   
\section{Discussion}
\paragraph{}
In this paper we have presented two examples of a new type of duality between 
double-scaled confining field theories and non-critical superstrings. The main
result is a limit in which the spectrum and S-matrix of the
interacting sector of the field
theory can be calculated using standard worldsheet methods. In
particular, the glueball spectrum of the field theory is identified
with the discrete spectrum of string states localised near the tip of
the cigar in the dual string background. As discussed in \cite{nd2},
the pre-exisiting results \cite{GK,AGK} on the string side indicate
some of the expected features of large-$N$ gauge theory such as 
a Hagedorn density of states and an S-matrix with linear Regge
trajectories. Other features such as the continuous spectrum of plane-wave
normalisable states \cite{GK} are unexpected
\footnote{See however \cite{Pol} for earlier discussion of the
  possibility of a continuous spectrum for large-$N$ QCD.} 
but not contradictory: 
the continuum should be replaced
  by a discrete spectrum at finite $N$, with mass splittings which go
  to zero as $N\rightarrow \infty$ \cite{nd2}.             
\paragraph{}
One of the most interesting aspects of the results described above is
the close relation between the four-dimensional double-scaling limit
and the related phenomena in zero-dimensional matrix models. 
A key feature of the zero-dimensional case is the emergence 
of universality classes associated with particular critical points. 
In that context, the Feynman diagrams of the matrix model
correspond to a triangulation of the string world-sheet and the
universal behaviour is obtained in the continuum limit 
of this discrete world-sheet theory. In four-dimensional theories
considered here a similar type of universality emerges in the
double-scaling limit. In particular, the interacting double-scaled
theory only depends on the behaviour near the singularity of the dual
string background. Many details of the microscopic theory reside in
the bulk region which decouples from the theory in the throat 
in the double-scaling limit. In this way many different models should
yield the same double-scaled theory. 
The model with superpotential (\ref{sp}) provides a striking example
of this. This theory is non-renormalisable and our analysis was
performed with a fixed UV cut-off $M_{UV}$ in place. Nevertheless, the
large-$N$ double-scaling limit yields a continuum theory where all
dependence on $M_{UV}$ is absorbed in the definitions (\ref{ident6})   
of the two parameters $\kappa$ and $\alpha'$. A decoupled continuum
theory with a fixed mass scale, is therefore obtained {\em without}
taking the limit $M_{UV}\rightarrow \infty$. A very similar phenomenon
was also observed in \cite{nd2}, where the $\beta$-deformed model was
interpreted as a lattice regularisation of a six-dimensional gauge
theory. This type of behaviour is very unexpected from a field theoretic 
point of view and is probably of more general interest.      
\paragraph{}
The original motivation for this work was to find examples where the
't Hooft large-$N$ limit of a four-dimensional confining field theory is dual
to a pure Neveu-Schwarz string background. We were only able to
accomplish this at the cost of considering 
an alternative large-$N$ double-scaling limit. An obvious question is 
why this was necessary and what is the connection between the
double-scaling limit and the absence of RR flux? The close parallels
to the double-scaled limit of zero-dimensional matrix models suggest
an answer to this question. In this context, the double-scaling limit 
is a continuum limit on the world-sheet of the string. Before taking
this limit we really have a discrete theory where the world-sheet has
holes. As discussed in the introduction, this is in line with the
emergence of a discrete spin chain description of the 
$AdS_{5}\times S^{5}$ string \cite{Min}. The ideas of 
\cite{OV1,OV2} also indicate the appearance of holes in the closed
string world-sheet due to the presence of background RR flux. 
In this case, the holes correspond to bubbles
of a new phase in a linear $\sigma$-model formulation of the string
worldsheet theory. Note that, in this formulation, the 
underlying string $\sigma$-model is
still formulated on a continuous worldsheet and the holes are described as 
large fluctuations of the worldsheet
fields. There is no contradiction here, as these large 
fluctuations mean that this formulation inevitably becomes 
strongly-coupled as soon
as a significant number of holes appear. In this case, it is plausible
that the appropriate (ie weakly-coupled) effective description is in terms of a
discretised world-sheet. If RR fields really lead to a discrete
worldsheet theory, then it is natural that the 
appropriate double-scaling limit, which can be thought of as the
corresponding continuum limit, also leads a pure NS background. 
  
ND is supported by a PPARC Senior Fellowship. The authors would like
to thank Ofer Aharony for many useful discussions. ND would like to thank
the particle theory groups at the Weizmann Institute and at Hebrew
University for hospitality during the completion of this work. 
GB would like to thank the particle theory group at DAMTP,
University of Cambridge for hospitality.

%%%%%%%%%%%%%%%%%%%

\section*{Appendix}

In this appendix, we are going to analyze 
the double-scaling limit of the couplings
$\tilde{V}^{(2)}_{ij}, \tilde{V}^{(3)}_{ijk}$
and the Hessian matrix $\tilde{H}^{(2)}_{ij}$
which were discussed in Section 5.
\paragraph{}
The F-terms of the theory are controlled 
by the matrix model spectral curve $\Sigma$ which is given by
\begin{equation}
y^2 = W'(x)^2 - f(x) =
\varepsilon^2 \prod_{l=1}^m \left( x - x_+^{(l)} \right)   
\left( x - x_-^{(l)} \right)
\label{curveApp}\end{equation}
where 
\begin{equation}
W'(x) = \varepsilon \left(
x^m - \sum_{l=1}^{m} g_l x^{l-1} 
\right) \ ,
\qquad
f(x) = \sum_{l=1}^m \kappa_l x^{l-1} \ .
\label{W&fApp}\end{equation}
The curve $\Sigma$ is hyperelliptic, 
a double cover of the complex plane with $2m$ branch points, $x_\pm^{(l)},
l=1, \ldots, m$.
The Argyres-Douglas singularities correspond to
$g_{l}=0, \kappa_l=0, l \geq 2, \kappa_1 = 4 \varepsilon^2 \Lambda^{2m},
g_1 = {\cal U} = \pm 2 \Lambda^m$. Let us set
\begin{equation}
x_{\pm}^{(l)}=  e^{\frac{2\pi i l}{m}} \, \left({\cal U}\pm 
2 \Lambda^m \right)^{\frac{1}{m}} \ .
%\label{branch}
\end{equation} 
\noindent 
We are going to focus on the singularity at ${\cal U}=2\Lambda^m$,
where the roots $x^{(l)}_-$ collide.
\paragraph{}
The coupling constants of the low-energy $U(1)^m$ theory 
are governed by the spectral curve $\Sigma$ 
via the generalized period matrix $\Pi$ (\ref{vijk})(\ref{vp}) 
\begin{equation}
W^{(2)}_{eff} = \frac{1}{2} \sum_{k,l=1}^m \Pi_{k l} 
\, \omega_{\alpha k} \, \omega^\alpha_l \ ,
\qquad  
\Pi_{k l} =  \frac{\partial S^D_k}{\partial S_l}
= \frac{\partial^2 {\cal F}}{\partial S_k \partial S_l} \ ,
\end{equation}
\begin{equation}
\int d^2 \theta \, W^{(2)}_{eff} \supset 
V^{(2)}_{kl} f^{k}_{\alpha\beta} f^{\alpha\beta\, l}
\ , 
\qquad
V^{(2)}_{kl} = \left\langle
\frac{\partial^2 {\cal F}}{\partial S_k \partial S_l}
\right\rangle \ . 
\end{equation}
In the new basis
\begin{equation}
W^{(2)}_{eff} = \frac{1}{2} \sum_{k,l=1}^m \tilde \Pi_{k l} 
\, \tilde \omega_{\alpha k} \, \tilde \omega^\alpha_l \ ,
\qquad  
\tilde \Pi_{k l} =  \frac{\partial \tilde S^D_k}{\partial \tilde S_l}
= \frac{\partial^2 {\cal \tilde F}}{\partial \tilde S_k \partial \tilde S_l} \ .
\end{equation}
The new period matrix $\tilde \Pi$ is related to 
$\Pi$ as follows. Using
\begin{equation}
\left(
\begin{array}{c}
\tilde B \\
\tilde A \\
\end{array} 
\right) = 
M \cdot 
\left(
\begin{array}{c}
B \\
A \\
\end{array} 
\right) 
=
\left(
\begin{array}{cc}
X & Y \\
W & Z \\
\end{array} 
\right) 
\left(
\begin{array}{c}
B \\
A \\
\end{array} 
\right) 
\qquad M \in Sp(2m; {\bf Z}) \ ,
\label{Msymplectic}\end{equation}
we find 
\begin{equation}
\tilde S^D_k = X_{kl} S^D_l + Y_{kl} S_l \ ,
\qquad
\tilde S_k = W_{kl} S^D_l + Z_{kl} S_l \ ,
\label{tildeSDtildeS}\end{equation}
which in turn implies 
\begin{equation}
\tilde \Pi = \left( X \, \Pi + Y \right) \left( W \Pi + Z \right)^{-1} \ .
\label{tildePi}\end{equation}
For instance, the symplectic transformation that connects the basis of cycles
shown in Fig.(\ref{fig2}) to the basis shown in Fig.(\ref{fig3})
is given by 
\begin{equation}
\left(
\begin{array}{c}
\tilde{B}_1 \\
\tilde{B}_2 \\
\tilde{B}_3 \\
\tilde{A}_1 \\
\tilde{A}_2 \\
\tilde{A}_3 \\
\end{array}
\right)
=
\left(
\begin{array}{cccccc}
1 & 0 & -1 & 0 & 0 & 0 \\
0 & 1 & -1 & 1 & 1 & 1 \\
1 & -1 & 1 &-1  &0  &-1  \\
-1 & 1 & 0 & 1 & 0 & 0 \\
1 &-1  & 0 & 0 &1  & 0 \\
0 & 0 & 0 & 1 & 1 & 1 \\
\end{array} 
\right) 
\left(
\begin{array}{c}
{B}_1 \\
{B}_2 \\
{B}_3 \\
{A}_1 \\
{A}_2 \\
{A}_3 \\
\end{array}
\right) \ .
\label{Mm=3}\end{equation}
\paragraph{}
However, in order to find the behaviour of $\tilde \Pi$
close to the Argyres-Douglas singularity, we will
consider 
\begin{equation}
M_{ij} = \frac{\partial \tilde S^D_j}{\partial \kappa_i} 
= - \frac{1}{2} 
\oint_{\tilde B_j} \frac{x^{i-1}}{y}\, dx 
\ ,
\qquad
N_{ij} = \frac{\partial \tilde S_j}{\partial \kappa_i} 
=
- \frac{1}{2} 
\oint_{\tilde A_j} \frac{x^{i-1}}{y}\, dx \ ,
\label{MijNij}\end{equation}
which determine the period matrix via 
\begin{equation}
\tilde \Pi_{ij} = \frac{\partial \tilde S^D_j}{\partial \kappa_k} 
\frac{\partial \kappa_k}{\partial \tilde S_i} 
= \left( N^{-1} \right)_{ik} M_{kj} \ .
\label{PiN-1M}\end{equation}
\paragraph{}
Near the singular point, the curve (\ref{curveApp}) 
can be approximated by
\begin{equation}
y^2 \approx - 2 \,\varepsilon^2 \, {\cal U}
\left(
x^m - \sum_{l=2}^m g_l x^{l-1} - \delta 
\right) \ ,
\qquad
\delta = {\cal U} - 2 \Lambda^m 
\ .
\end{equation} 
Using the above expression, we can easily find the
behaviour of $M_{ij}, N_{ij}$ close to the AD point. 
First of all, let us focus on $N_{ij}$ where $i=\alpha=1,\ldots,p$,
$j=\beta=1,\ldots,p$. 
Recalling that $\tilde A_\beta = A^{(\beta)}_-$ is
a vanishing cycle, we find 
$$
N_{\alpha\beta} = \frac{\partial S^{(\beta)}_-}{\partial \kappa_\alpha} 
\sim 
\int_{x_-^{(l)}}^{x_-^{(l')}} \frac{x^{\alpha-1}}{y}\, dx
\sim 
\frac{1}{\varepsilon\,{\cal U}^{1/2}} \ \delta^{\frac{\alpha}{m}-\frac{1}{2}}
\int_{\tilde x_-^{(l)}}^{\tilde x_-^{(l')}} \frac{\tilde x^{\alpha-1}}{\tilde y}\, d \tilde x  
$$
\begin{equation}
\sim 
\frac{1}{\varepsilon\,{\cal U}^{1/2}} \ \delta^{\frac{\alpha}{m}-\frac{1}{2}}
\, f_{\alpha\beta} \left( \tilde g_l \right) \ ,
\label{N--}\end{equation}
where we defined
\begin{equation}
x = \delta^{\frac{1}{m}} \tilde x \ , 
\qquad
\tilde y^2 = \tilde x^m - \tilde g_l \tilde x^{l-1} - 1 \ ,
\qquad
\tilde g_l = \frac{g_l}{\delta^{\frac{m-l+1}{m}}} \ .
\end{equation}
Similarly
\begin{equation}
M_{\alpha\beta} = \frac{\partial S^{(\beta)}_{D,-}}{\partial \kappa_\alpha} 
\sim 
\frac{1}{\varepsilon\,{\cal U}^{1/2}} \ \delta^{\frac{\alpha}{m}-\frac{1}{2}}
\, f^D_{\alpha\beta} \left( \tilde g_l \right) \ .
\label{M--}\end{equation}
We see that, since $\alpha < m/2$, 
both $N_{\alpha\beta}$ and $M_{\alpha\beta}$ diverge in the limit
$\delta \to 0$.
Then, let us consider $N_{ij}$ where $i= A = p+1,\ldots,2p+1$,
$j=\beta=1,\ldots,p$. 
In this case
$$
N_{A\beta} = \frac{\partial S^{(\beta)}_-}{\partial \kappa_A} 
\sim 
\int_{x_-^{(l)}}^{x_-^{(l')}} \frac{x^{A-1}}{y}\, dx
\sim 
\frac{1}{\varepsilon\,{\cal U}^{1/2}} \ \delta^{\frac{A}{m}-\frac{1}{2}}
\int_{\tilde x_-^{(l)}}^{\tilde x_-^{(l')}} \frac{\tilde x^{A-1}}{\tilde y}\, d \tilde x  
$$
\begin{equation}
\sim 
\frac{1}{\varepsilon\,{\cal U}^{1/2}} \ \delta^{\frac{A}{m}-\frac{1}{2}}
\, f_{A\beta} \left( \tilde g_l \right) \ ,
\end{equation}
which vanishes in the limit $\delta \to 0$. The same 
result holds for $M_{A\beta}$.
\paragraph{}
The behaviour of the periods along the cycles 
$\tilde{A}_l,\tilde{B}_l, l > [m/2]$,
which correspond to the cycles $A_+,B_+,A_\infty,B_\infty$,
is markedly different. 
Indeed, both $N_{\alpha B},M_{\alpha B}$ and $N_{AB},M_{AB}$, 
$B=p+1,\ldots,2p+1$, are finite in the limit $\delta \to 0$
and analytic.
One can do a Taylor expansion of the integrand in powers of 
${\cal U} - 2 \Lambda^m, \kappa_l, l\geq 2$ and find that
the coefficients are finite. 
Therefore, the periods along the cycles $\tilde{A}_l, \tilde{B}_l, l > [m/2]$ 
are manifestly analytic at the Argyres-Douglas point.
This is a property of the new basis we will exploit again
in the following. 
\paragraph{}
In summary, in the limit $\delta \to 0$, the matrices
$N$ and $M$ will have the following block structure
\begin{equation}
N \to \left(
\begin{array}{cc}
N_{--} & N_{-+}^{(0)} \\
0 & N^{(0)}_{++} \\
\end{array} 
\right) 
\qquad
M \to \left(
\begin{array}{cc}
M_{--} & M_{-+}^{(0)} \\
0 & M^{(0)}_{++} \\
\end{array} 
\right) \ ,
\end{equation}
where by $-$ or $+$ we denote indices in the 
ranges $\{1, \ldots, p\}$ and $\{p+1,\ldots,2p+1\}$
respectively. 
In order to evaluate the generalized period matrix,
we also need the inverse of $N$, which is given by
\begin{equation}
N^{-1}
\to \left(
\begin{array}{cc}
N^{-1}_{--} & {\cal N} \\
0 & \left( N^{(0)}_{++} \right)^{-1} \\
\end{array} 
\right) \ ,
\qquad 
{\cal N} = - N_{--}^{-1} \, N_{-+}^{(0)} \, 
\left( N_{++}^{(0)} \right)^{-1} \ . 
\label{Ninv}\end{equation}
Finally
\begin{equation}
\tilde \Pi = N^{-1} \, M \to 
\left(
\begin{array}{cc}
N^{-1}_{--} M_{--} & \quad N_{--}^{-1} M_{-+}^{(0)}
+ {\cal N} M_{++}^{(0)}
\\
0 & \quad \left( N^{(0)}_{++} \right)^{-1} M_{++}^{(0)} \\
\end{array} 
\right) \ .
\label{Pi1}\end{equation}
Let us look more closely at the structure of each
block in the above matrix.
First of all, by (\ref{N--}) %and (\ref{M--})
\begin{equation}
\left( N_{--} \right)^{-1}_{\alpha\beta} 
\sim 
\varepsilon\,{\cal U}^{1/2} \ \delta^{ \frac{1}{2} - \frac{\beta}{m} }
\, f^{-1}_{\alpha\beta} \left( \tilde g_l \right) \ ,
\label{Ninv--}\end{equation}
which, by (\ref{M--}), implies  
\begin{equation}
\left( N_{--} \right)^{-1}_{\alpha\beta} \, 
\left( M_{--} \right)_{\beta\gamma}
= 
f^{-1}_{\alpha\beta} \left( \tilde g_l \right) 
f^D_{\beta\gamma} \left( \tilde g_l \right) 
= \Pi^{-}_{\alpha\gamma} \left( \tilde g_l \right)
\ .
\label{Piminus}\end{equation}
Furthermore, since $N_{--}^{-1}$ vanishes 
in the limit $\delta \to 0$, we find that
\begin{equation}
N_{--}^{-1} M_{-+}^{(0)}
+ {\cal N} M_{++}^{(0)} =
N_{--}^{-1} \,M_{-+}^{(0)}
- N_{--}^{-1} \,N_{-+}^{(0)} 
\left( N_{++}^{(0)} \right)^{-1} 
M_{++}^{(0)} 
\rightarrow 0 \ .
\end{equation}
Therefore, the period
matrix has the following block-diagonal form
in the double scaling limit
\begin{equation}
\tilde \Pi \to 
\left(
\begin{array}{cc}
\Pi^{-} & 0
\\
0 & \Pi^{+} \\
\end{array} 
\right) \ .
\label{Pifinal}\end{equation} 
\paragraph{}
The upper block $\Pi^-$ is actually
the period matrix of the {\it reduced spectral curve} $\Sigma_-$ 
\begin{equation} 
\tilde y^2 = \tilde x^m - \sum_2^m \tilde g_l \tilde x^{l-1} - 1 \ .
\label{reducedApp}\end{equation} 
To realize this, note that the cycles 
$A^{(\alpha)}_-, B^{(\alpha)}_-$
form a standard homology basis for $\Sigma_-$ and that
\begin{equation}  
\frac{\tilde x^{\beta-1}}{\tilde y} d \tilde{x}\ , \qquad
\beta = 1 , \ldots, p \ ,
\end{equation}
span a basis of holomorphic 1-forms on $\Sigma_-$.
Therefore, by (\ref{N--})(\ref{M--}) and (\ref{Piminus}),
we can conclude that $\Pi^-$ is the standard
period matrix of $\Sigma_-$.

\paragraph{}
Let us now consider the cubic coupling
$$
\tilde{V}^{(3)}_{ijk} = 
\frac{\partial^3 \tilde F}{\partial \tilde S_i \partial \tilde S_j  \partial \tilde S_k} 
$$
$$
%\frac{\partial^3 \tilde F}{\partial \tilde S_i \partial \tilde S_j  \partial \tilde S_k}
= \frac{\partial \kappa_r}{\partial \tilde S_i}
\left(
\frac{\partial^2 \tilde S^D_k}{\partial \kappa_r  \partial \kappa_p}
- \frac{\partial^2 \tilde S_s}{\partial \kappa_r  \partial \kappa_p}
\frac{\partial \kappa_q}{\partial \tilde S_s}
\frac{\partial \tilde S^D_k}{\partial \kappa_q}
\right)
\frac{\partial \kappa_p}{\partial \tilde S_j}
$$
\begin{equation}
=
N^{-1}_{ir} 
\left(
\frac{\partial^2 \tilde S^D_k}{\partial \kappa_r  \partial \kappa_p}
- \frac{\partial^2 \tilde S_s}{\partial \kappa_r  \partial \kappa_p}
\tilde \Pi_{sk}
\right)
( N^{-1} )^T_{pj} \ .
\label{V3}\end{equation}
Recall that in general $\tilde{V}^{(3)} \sim \varepsilon^{-1}$.
Therefore, only terms that are divergent in the limit
$\delta \to 0$ can survive in the double scaling limit.
On the other hand, such terms can only come from the
second derivatives of $\tilde{S}^D_k$ and $\tilde{S}_k$
for $k \leq p = [m/2]$, since the period integrals along the   
cycles $\tilde{A}_k, \tilde{B}_k, k > [m/2]$ are analytic
in $\Delta \kappa_1 = \kappa_1-\kappa_1^{cr}, \kappa_l, l \geq 2$. 
\paragraph{}
Eq.(\ref{V3}) can be recast in matrix form 
\begin{equation}
\tilde{V}^{(3)}_{ijk} =
N^{-1}_{ir} 
\Delta_{rp}^k \,
( N^{-1} )^T_{pj} \ . 
\label{N-DeltaN-}\end{equation}
In order to estimate the above expression for
$k < [m/2]$, we use the fact that 
$$
\Delta^k_{rp} \sim \oint_{\tilde A_k}
\frac{x^{r+p-2}}{y^3}\,dx
\sim
\frac{1}{\varepsilon^3\,{\cal U}^{3/2}} 
\int_{x_-^{(k)}}^{x_-^{(k')}} \frac{x^{r+p-2}}{
\left( \sqrt{x^m - g_l x^{l-1} - \delta} \right)^3
}\, dx
$$
$$
\sim 
\frac{1}{\varepsilon^3\,{\cal U}^{3/2}} 
\frac{\partial}{\partial \delta}
\int_{x_-^{(k)}}^{x_-^{(k')}} \frac{x^{r+p-2}}{
\sqrt{x^m - g_l x^{l-1} - \delta}
}\, dx
\sim
\frac{1}{\varepsilon^3\,{\cal U}^{3/2}} 
\frac{\partial}{\partial \delta}
\left(
\delta^{\frac{r+p-1}{m}-\frac{1}{2}}
F( \tilde{g}_l )
\right) \ ,
$$
where
$$
F( \tilde{g}_l ) =
\int_{\tilde{x}_-^{(k)}}^{\tilde{x}_-^{(k')}} \frac{\tilde{x}^{r+p-2}}{
\sqrt{\tilde{x}^m - \tilde{g}_l \tilde{x}^{l-1} - 1}
}\, d\tilde{x} \ ,
\qquad
\tilde g_l = \frac{g_l}{\delta^{\frac{m-l+1}{m}}} \ .
$$
Then, since $F ( \tilde{g}_l )$ is an analytic function
of $\tilde {g}_l$ and these parameters are
kept finite in the double scaling limit, we find 
\begin{equation}
\Delta^k_{rp} \sim
\frac{1}{\varepsilon^3\,{\cal U}^{3/2}} \,
\delta^{\frac{r+p-1}{m}-\frac{3}{2}} 
\, G \left( \tilde{g}_l
\right) \ ,  
\label{Deltakrp}\end{equation}
where $G(\tilde{g}_l)$ is analytic as well.
\paragraph{}
Then, Eq. (\ref{N-DeltaN-}) is equivalent to
$$
\left(
\begin{array}{cc}
N^{-1}_{--} & {\cal N} \\
0 & \left( N^{(0)}_{++} \right)^{-1} \\
\end{array} 
\right) 
\left(
\begin{array}{cc}
\Delta^k_{--} & \Delta^k_{-+} \\
\Delta^k_{+-} & \Delta^k_{++}  \\
\end{array} 
\right) 
\left(
\begin{array}{cc}
( N^{-1}_{--} )^T & 0 \\
{\cal N}^T & \left( N^{(0)}_{++} \right)^{-T} \\
\end{array} 
\right) \ .
$$
A careful inspection shows that the most relevant term
is actually
\begin{equation}
N^{-1}_{--} \, \Delta^k_{--} \, ( N^{-1}_{--} )^T \ ,
\end{equation}
namely, in the double scaling limit
$$
N^{-1} \, \Delta^k \, (N^{-1})^T 
\sim 
\left(
\begin{array}{cc}
N^{-1}_{--} \, \Delta^k_{--} \, ( N^{-1}_{--} )^T & 0 \\
0 & 0 \\
\end{array} 
\right) \ .
$$
In fact, using 
$$
\left( N_{--} \right)^{-1}_{\alpha\beta} 
= N^{-1}_{\alpha\beta} 
\sim 
\varepsilon\,{\cal U}^{1/2} \ \delta^{ \frac{1}{2} - \frac{\beta}{m} }
$$
we find
\begin{equation}
N^{-1}_{\alpha\beta} \, \Delta^k_{\beta\gamma} \, N^{-1}_{\rho\gamma} 
\sim
\frac{1}{\varepsilon\,{\cal U}^{1/2}} \,
\delta^{ \frac{1}{2} - \frac{\beta}{m} } \,
\delta^{\frac{\beta+\gamma-1}{m}-\frac{3}{2}} \,
\delta^{ \frac{1}{2} - \frac{\gamma}{m} }
= \frac{1}{\varepsilon\,{\cal U}^{1/2}} \,
\delta^{-\frac{m+2}{2m}} \ .
\end{equation}
\paragraph{}
In conclusion, the only surviving terms in the double scaling
limit are
\begin{equation}
\tilde{V}^{(3)}_{ijk} \sim
%\frac{1}{\varepsilon\,{\cal U}^{1/2}} \,
%\delta^{-\frac{m+2}{2m}} \ ,  
\frac{1}{\Lambda^{m/2} \, \Delta} v_{\alpha\beta\gamma} 
\left( \tilde{g}_l \right) \ ,
\label{VijkFinal}\end{equation}
for $i=\alpha=1,\ldots,p, \,
j=\beta=1,\ldots,p, \, k=\gamma=1,\ldots,p$.

\paragraph{}
Finally, let us turn our attention to the Hessian matrix
\begin{equation}
\tilde{H}^{(2)}_{ij} =
\frac{\partial^2 W_{eff}^{(0)}}{\partial \tilde{S}_i \partial \tilde {S}_j}
\sim N^{-1}_{ik}
\left(
\frac{\partial^2 W_{eff}^{(0)}}{\partial \kappa_k  \partial \kappa_l}
\right)
( N^{-1} )^T_{lj}
= 
N^{-1}_{ik} \,
\Delta^H_{kl} \,
( N^{-1} )^T_{lj}
\end{equation}
The effective superpotential in the new basis reads
\begin{eqnarray} 
W_{eff}^{(0)}   & = 
& \sum_{l=1}^{2p}\, 
\left(
\tilde{N}_{l}\frac{\partial{\tilde{\cal F}}}{\partial 
\tilde{S}_{l}} 
+ 2\pi i \tilde{b}_l \tilde{S}_{l}
\right)
+ \tilde N_{2p+1} \frac{\partial{\tilde{\cal F}}}{\partial 
\tilde{S}_{2p+1}}  
+ 2\pi i ( \tilde \tau_0 + \tilde{b}_{2p+1} )
\tilde{S}_{2p+1} \ .
\end{eqnarray}
However, since there is no flux through 
the $A^{(\alpha)}_- , B^{(\alpha)}_-$ cycles
\begin{equation}
\oint_{A_-^{(\alpha)}} T(x) dx = 
\oint_{B_-^{(\alpha)}} T(x) dx = 0 \ ,
\end{equation}
the expression simplifies to
$$
W_{eff}^{(0)}
= 
\sum_{\alpha=1}^{p}\, 
\left(
\tilde{N}^+_{\alpha}
\frac{\partial{\tilde{\cal F}}}{\partial S^{(\alpha)}_{+}} 
+ 2\pi i \tilde{b}^+_\alpha \,S^{(\alpha)}_{+}
\right)
+ \tilde N_{2p+1} \frac{\partial{\tilde{\cal F}}}{\partial 
\tilde{S}_{2p+1}}  
+ 2\pi i ( \tilde \tau_0 + \tilde{b}_{2p+1} )
\tilde{S}_{2p+1} \ ,
$$
which is manifestly analytic at the Argyres-Douglas point.
In fact, as we remarked before, $S^{D,(\alpha)}_{+}$, $S^{(\alpha)}_{+}$,
$\tilde{S}^D_{2p+1}$ and $\tilde{S}_{2p+1}$ are all analytic functions
of the moduli $\kappa_l$ close to the critical point.
Therefore, the matrix of second derivatives $\Delta^H$ will be non-singular 
\begin{equation}
\Delta^H_{kl} \sim \frac{n}{ \varepsilon^3 \, {\cal U}^{3/2} } \, \delta^0
\quad .
\end{equation}
Then, using 
\begin{equation}
N^{-1}
\to \left(
\begin{array}{cc}
0 & 0 \\
0 & \left( N^{(0)}_{++} \right)^{-1} \\
\end{array} 
\right) \ ,
\label{Ninv2}\end{equation}
we find that 
\begin{equation}
\tilde{H}^{(2)}_{ij} 
= \frac{n}{\varepsilon\,{\cal U}^{1/2}}
\left(
\begin{array}{cc}
0 & 0 \\
0 & h_{++} \\
\end{array} 
\right) \ .
\label{H2}\end{equation}
In conclusion, the only Hessian matrix elements that survive
in the double scaling limit are
\begin{equation}
\tilde{H}^{(2)}_{ij} 
= \frac{n}{\varepsilon\, \Lambda^{m/2}} \,
h^{(2)}_{AB} \ , 
\end{equation}
where $i=p+A=p+1,\ldots,2p+1, \,j=p+B=p+1,\ldots,2p+1$.

\end{document}